\documentclass[numberedappendix,twocolumn,iop]{emulateapj} 
\usepackage{times}
\usepackage{natbib}

\usepackage[section]{placeins}
\usepackage{amssymb,amsmath}
\usepackage{graphicx,subfigure}
\usepackage{color}
\usepackage{rotating}
\usepackage{ulem} 
\usepackage{float}
\usepackage[citecolor=blue,hypertexnames=false]{hyperref}
\hypersetup{
    colorlinks=true,       
    linkcolor=blue,                 
    filecolor=magenta,      
    urlcolor=blue
}

\newcommand{\be}{\begin{equation}}
\newcommand{\ee}{\end{equation}}

\newcommand{\gray}{$\gamma$-ray}

\newcommand{\hi}{H~{\sc i}}
\newcommand{\hii}{H~{\sc ii}}

\newcommand{\fermi}{{\it Fermi}}

\newcommand{\GP}{{\sc GALPROP}}

\newcommand{\galprop}{{\sc GALPROP}}

\shorttitle{Time-Dependent Models for the Non-Thermal Interstellar Radiation from the Milky Way}
\shortauthors{Porter~et~al.}

\begin{document}

%\linenumbers


\title{Deciphering Residual Emissions: Time-Dependent Models for the Non-Thermal Interstellar Radiation from the Milky Way}

\author{
T.~A.~Porter\altaffilmark{1,$\dagger$},
G.~J\'ohannesson\altaffilmark{2,3}, and 
I.~V.~Moskalenko\altaffilmark{1}
}
\altaffiltext{$\dagger$}{email: tporter@stanford.edu}
\altaffiltext{1}{W. W. Hansen Experimental Physics Laboratory and Kavli Institute for Particle Astrophysics and Cosmology, Stanford University, Stanford, CA 94305, USA}
\altaffiltext{2}{Science Institute, University of Iceland, IS-107 Reykjavik, Iceland}
\altaffiltext{3}{AlbaNova Univ. Center Nordita, Roslagstullsbacken 23, SE-106 91 Stockholm, Sweden }

\begin{abstract}
Cosmic rays (CRs) in the Galaxy are an important dynamical component of the interstellar medium (ISM) that interact with the other major components (interstellar gas and magnetic and radiation fields) to produce broadband interstellar emissions that span the electromagnetic spectrum.
The standard modelling of CR propagation and production of the associated emissions is based on a steady-state assumption, where the CR source spatial density is described using a smoothly varying function of position that does not evolve with time. 
While this is a convenient approximation, reality is otherwise where primary CRs are produced in and about highly localised regions, e.g., supernova remnants, which have finite lifetimes. 
In this paper, we use the latest version of the \GP\ CR propagation code to model time-dependent CR injection and propagation through the ISM from a realistic 3D discretised CR source density distribution, together with full 3D models for the other major ISM components, and make predictions of the associated broadband non-thermal emissions.
We compare the predictions for the discretised and equivalent steady-state model, finding that the former predicts novel features in the broadband non-thermal emissions that are absent for the steady-state case.
Some of the features predicted by the discretised model may be observable in all-sky observations made by {\it WMAP} and {\it Planck}, the recently launched {\it eROSITA}, the \fermi-LAT, and ground-based observations by HESS, HAWC, and the forthcoming CTA. 
The non-thermal emissions predicted by the discretised model may also provide explanations of puzzling anomalies in high-energy \gray{} data, such as the \fermi-LAT north/south asymmetry and residuals like the so-called ``Fermi bubbles.''
\end{abstract}

\keywords{
astroparticle physics ---
Galaxy: general ---
Gamma rays: general ---
Gamma rays: ISM ---
(ISM:) cosmic rays ---
radiation mechanisms: nonthermal
}

%%%%%%%%%%%%%%%%%%%%%%%%%%%%%%%%%%%%%%%%%%%%%%%%%%%%%%%%%%%%%%%%%%%%

\section{Introduction}
\label{sec:intro}
\setcounter{footnote}{0}

The Galactic emission is dominated over a broad range of wavelengths by the radiation from CR particles that interact with the gas, radiation, and magnetic fields in the ISM.
These non-thermal interstellar emissions present a strong foreground for detection of point sources, diffuse signals of exotic origin (e.g., the dark matter), as well as all manner of extragalactic phenomena.
They present also a valuable tool for understanding CR sources, the CR injection and propagation through the ISM, and the spatial distributions of the other components of the diffuse ISM.

High-quality data tracing the non-thermal emissions now exist, or will be available shortly, over the electromagnetic spectrum. 
The {\it Wilkinson Microwave Anisotropy Probe} \citep[WMAP][]{2013ApJS..208...20B} and {\it Planck} \citep{2018arXiv180706206P,2018arXiv180706207P} instruments have provided comprehensive all-sky data at microwave frequencies, while the recently launched {\it eROSITA} mission \citep{2016SPIE.9905E..1KP} will make a sensitive all-sky survey at X-ray energies.
Observations by these instruments allow tracing of the CR electrons/positrons over a broad range of energies and spatial scales, ranging from across the Galaxy to close to the injection regions, via the synchrotron emissions by their interactions with the magnetic fields in the ISM.

The \fermi\ Large Area Telescope \citep[\fermi-LAT;][]{2009ApJ...697.1071A}, launched in 2008, has provided the most comprehensive view of high-energy \gray{s} from the Galaxy over $\sim$30~MeV to hundreds of GeV energies and higher.
Similar to the microwave/X-ray observations, the \fermi-LAT energy coverage spans the range from where the smoothly varying CR ``sea'' is dominating the production of \gray{} emissions to where the localised individual CR source regions are thought to make more intense contributions, and it traces the interactions of CR nuclei and leptons with the gas and radiation fields across the Galaxy.

At TeV energies, the ground-based array Milagro detected hard-spectrum interstellar emission from the Galactic plane~\citep{2008ApJ...688.1078A}.
The recent release of the High-Energy Stereoscopic System (HESS) Galactic plane survey \citep{2018A&A...612A...1H} shows localised, extended regions embedded in lower-intensity broadly distributed emissions -- similar to but not exactly the same as the \fermi-LAT data at lower energies.
The High Altitude Water Cherenkov (HAWC) experiment extends coverage of the Galactic plane to energies $\gtrsim$10~TeV, revealing that the emissions are dominated by hard-spectrum localised regions \citep{2017ApJ...843...40A}, some of which have identifications with objects detected at lower energies \citep{2018Natur.562...82A,2017Sci...358..911A,2019arXiv190806658J}.
The complementarity of these multi-wavelength data for probing the CRs and sources across the Galaxy has been anticipated by different authors \citep[e.g.,][]{1980A&A....92..175P,1997JPhG...23.1765P,2000A&A...362..937A}.

Modelling the non-thermal interstellar emissions relies on determination of the CR distribution throughout the Galaxy.
The standard approach employs a steady-state assumption for the CR source spatial density described as a smoothly varying function of position that does not evolve with time.
This readily facilitates CR propagation calculations using analytical and numerical methods.
For the $\sim$~GeV~nucleon$^{-1}$ energy data that are used to constrain model parameters (diffusion coefficient, halo size, etc.), the long CR residence times are thought to provide sufficient mixing to effectively erase individual contributions of the CR sources.
This mixing leads to a smooth spatial distribution for the CR densities -- the so-called `sea' -- motivating the initial assumption for the source density.

However, reality is otherwise, with primary CRs produced in and about highly localised regions, e.g., supernova remnants (SNRs), which have finite lifetimes.
The injection and propagation in the ISM of CRs produced by discrete (time/space) sources has, in fact, been investigated since the late 1960s~\citep[e.g.,][]{1969Natur.224.1182L,1973ICRC....1..621L,1971ICRC....1..377L,1970PhRvL..24..913R,1979ApJ...229..424L}.
A ``standard'' picture based on these ideas has emerged for interpreting local CR fluxes, particularly for the high-quality CR electron/positron data now available.
Building on the early work of \citet{1970ApJ...162L.181S} and \citet{1971ApL.....9..169S} and, later, \citet{1979ApJ...228..297C}, \citet{1979ICRC....1..488N}, and \citet{1995PhRvD..52.3265A}, the contributions of individual ``nearby'' sources ($\lesssim 1$~kpc) on top of a smoothly distributed background of more distant sources can be used to explain the broadband electron spectral intensity \citep[e.g.,][]{2004ApJ...601..340K}.

Statistical descriptions based on a Galaxy-wide ensemble of discrete sources have also been used to interpret the electron data and CR nuclei fluxes \citep[][]{2003ApJ...582..330H,2004ApJ...609..173T,2011JCAP...02..031M,2012A&A...544A..92B,2015RAA....15...15L,2015A&A...573A.134M,2017A&A...600A..68G,2018JCAP...11..045M,2006AdSpR..37.1909P}.
Unsurprisingly, given the lack of directionality for the CR fluxes and the large parametric uncertainty for such descriptions, a unique distribution of sources explaining the CR data, particularly for the high energies where spectral features are most evident, has proved elusive.
The respective contributions by different source classes across the Galaxy, such as SNRs, pulsars and pulsar wind nebulae (PWNe), and OB associations; the effect of the diffusive properties of the magnetised ISM; and the locations and properties of the nearest individual sources are ``known'' unknowns.

Because the non-thermal emissions data provide all-sky directional spectral intensities and probe far beyond the relatively limited horizon of the local CR observations, they can be useful information to assist in resolving these issues and provide better context for how the local CRs are representative of the Galaxy-wide populations.
So far, the steady-state models for the propagation and interactions of CRs in the ISM have been employed for explaining features of the multi-wavelength diffuse spectrum of the Galaxy; for reviews, see \citet{2007ARNPS..57..285S} and \citet{2015ARA&A..53..199G}.
However, there are many unexplained residual features, particularly for high-energy \gray{s} \citep[e.g.,][]{2012ApJ...750....3A}, which may be signatures of the discrete CR sources.
But modelling for time/space-discretised CR sources and their associated non-thermal emissions has been relatively unexplored.
An earlier version of the \GP\ code was used for preliminary calculations \citep{2001AIPC..587..533S,2001ICRC....5.1964S,2003ICRC....4.1989S}, but the quality of data available at the time and computational limitations meant that the modelling provided limited insight.

In this paper, we take the next steps to address this issue using the latest release of the \GP\ CR propagation code\footnote{\url{http://galprop.stanford.edu}} \citep{2017ApJ...846...67P,2018ApJ...856...45J}, which has been enhanced with new capabilities to facilitate efficient modelling for time/space-discretised CR sources and propagation in the ISM.
Two smooth CR source density models are used for this investigation: an axisymmetric disk-only distribution (SA0) for all of the injected CR power and a 3D distribution (SA50) that has an axisymmetric disk and spiral arm components with 50\% of the CR luminosity injected by each.
The SA50 source density model is used together with the new sampler facility (see below) to determine a time/space-discretisation for the CR injection regions in the Galaxy.
The solutions for both the SA50 steady state (hereafter SS$_{_{\rm SA50}}$) and time-dependent discretised of the same (hereafter TDD) are determined for the CR intensities, and the corresponding broadband non-thermal emissions are calculated.
We then compare the SS$_{_{\rm SA50}}$ and TDD solution predicted observables (CR spectra, non-thermal photon intensity maps), which test equivalent CR source models where the only ``variable'' is the time/space discretisation.
Because analysis of actual data almost universally employs the smooth/steady-state formalism to produce a baseline emissions model, the comparison between the SS$_{_{\rm SA50}}$ and TDD solution identifies features of the discretised CR injection and propagation that may be searched for in, e.g., high-energy \gray{} residual sky maps.

However, the ``true'' (3D) CR source distribution is not known, and the majority of analyses of data employ axisymmetric averaged steady-state solutions for the CR intensities; see, e.g., \citet{2012ApJ...750....3A} for high-energy \gray{s} and \citet{2013MNRAS.436.2127O} for synchrotron emissions.
We employ the SA0 distribution to investigate the case of smooth CR density mismodelling for the baseline.
The SA0 distribution is used as an alternative (incorrect) source density to determine a steady-state solution (hereafter SS$_{_{\rm SA0}}$) and its observables calculated and compared with the TDD predictions.
The evaluation of TDD solution observables, together with those for the SS$_{_{\rm SA50}}$ and SS$_{_{\rm SA0}}$ baselines, therefore enables the determination of features associated with discretised CR injection and propagation when the functional form for the CR distribution is known (SS$_{_{\rm SA50}}$) or mismatched (SS$_{_{\rm SA0}}$).

\section{Modelling Setup}
\label{sec:setup}

Theoretical understanding of CR propagation in the ISM is the 
framework that the \GP\ code is built around.
The key idea is that all CR-related data, including direct measurements, \gray{s}, synchrotron radiation, etc., are subject to the same physics and must therefore be modelled simultaneously.
The \GP\ code numerically solves the system of time-dependent partial 
differential equations describing the particle transport 
with a given source distribution and boundary conditions for all species of CRs.
Propagation is described using the advection--diffusion--reacceleration equation, which has proven to be remarkably successful at modelling transport processes in the ISM. 
The processes involved include diffusive reacceleration and, for nuclei, nuclear spallation, secondary particle production, radioactive decay, electron capture and stripping, electron knock-on, and electron K-capture, in addition to energy loss from ionisation and Coulomb interactions. 
For CR electrons and positrons, the important processes are the energy losses due to ionisation, Coulomb scattering, bremsstrahlung (with the neutral and ionised gas), inverse Compton (IC) scattering, and synchrotron emission.

Galactic properties on large scales, including the diffusion coefficient, halo size,
Alfv\'en  velocity and/or advection velocity, as well as
the mechanisms and sites of CR acceleration, can be probed by measuring stable and radioactive secondary CR nuclei. 
The ratio of the halo size to the diffusion coefficient can be constrained by measuring the abundance of stable secondaries, such as boron.
Radioactive isotopes of beryllium, aluminium, chlorine, and manganese ($^{10}$Be, $^{26}$Al, $^{36}$Cl, and $^{54}$Mn) then allow the resulting degeneracy to be lifted \citep[e.g.,][]{1998A&A...337..859P,1998ApJ...509..212S,1998ApJ...506..335W,2001ICRC....5.1836M}.
However, the interpretation of the peaks observed in secondary-to-primary ratios, such as boron/carbon (B/C), around energies of a few GeV~nucleon$^{-1}$ remains model-dependent.

The CR propagation in the heliosphere is described by the \citet{1965P&SS...13....9P} equation.
Spatial diffusion, convection with the solar wind, drifts, and adiabatic cooling are the main processes that determine transport of CRs to the inner heliosphere.
The resulting modified (modulated) fluxes significantly differ from the interstellar spectra below energies of $\sim$20-50 GeV~nucleon$^{-1}$, but correspond to the ones actually measured by balloon-borne and spacecraft instruments.
These effects have been incorporated into realistic (time-dependent, 3D) codes \citep[e.g.,][]{2003JGRA..108.1228F,2006ApJ...640.1119L,2004AnGeo..22.3729P,2018AdSpR..62.2859B}.
In particular, one of the most advanced codes, the HelMod\footnote{\url{http://www.helmod.org}} modelling package for heliospheric propagation, has been extensively used with GALPROP to derive local interstellar spectra of CR species based on the latest {\it AMS-02} data \citep{2017ApJ...840..115B,2018ApJ...854...94B,2018ApJ...858...61B,2019arXiv191103108B}.
The ``force-field'' approximation that is often used \citep{1968ApJ...154.1011G}, instead characterises the modulation effect as it varies over the solar cycle using a single parameter: the ``modulation potential'' that describes only the result of adiabatic cooling.
Despite having no predictive power, the force-field approximation is a useful low-energy parameterisation
of the modulated spectrum for a given interstellar spectrum. 

We have made further enhancements to the \GP\ code to enable more efficient time-dependent CR propagation and interstellar emissions modelling\footnote{These will be provided with a forthcoming point release for the latest version of \GP\ (v56) available. In addition, the full set of configuration files used for the calculations made in this paper will be made available, as usual, from the dedicated website.}.
The ``discrete sampler'' is a new facility that produces a spatial and temporal discretised list of CR source regions from a user-supplied smooth CR spatial density distribution and time interval.
This new code feature enables direct comparison of CR intensity distributions and interstellar emission intensity maps for a steady-state and equivalent time/space-discretised distribution realisation from the same smooth CR density model.
The discrete sampler uses an acceptance/rejection method with a pseudo-random number generator, which allows full reproducibility of the discretisation of the smooth density model for different luminosity evolutionary scenarios for the CR sources in the time-dependent case.

In addition, the \GP\ code now includes an option to use non-equidistant grids that allow for increased spatial resolution over user-specified regions of the calculation volume.
This \GP\ enhancement is inspired by the Pencil Code\footnote{See 
  \url{http://pencil-code.nordita.org/doc/manual.pdf}, Section 5.4.}
\citep{BrandenburgDobler:2002}, where the usage of analytic functions can have advantages in terms of speed and memory usage compared to purely numerical implementations for non-uniform grid spacing.
As for the recent work of \citet{2019ApJ...879...91J}, the calculations in this paper use the grid function for each coordinate $X,Y,Z,$
\begin{equation}
  X(\zeta_{_X}) = \frac{\epsilon_{_X}}{a_{_X}}\tan\left[ a_{_X}\left( \zeta_{_X} - \zeta_0 \right)  \right] + X_0
  \label{eq:gridFunction}
\end{equation}
where $\epsilon_{_X}, a_{_X}, \zeta_{0}, X_{0}$ are parameters.
This function maps from the linear grid $\zeta$ to the nonlinear grid for each of $X,Y,Z$ using (possibly) different parameter sets for the individual coordinate transformations.
The transport equations are solved on the $\zeta$ grid accounting for the change in first and second derivatives.

Without loss of generality, we use the SA50 model \citep{2017ApJ...846...67P,2018ApJ...856...45J} for the underlying source density distribution from which the discretised injection regions are sampled. 
The SA50 density model assumes that 50\% of the injected CR luminosity is provided by an axisymmetric disk component following \citet{2004A&A...422..545Y} and the other 50\% is from the four-arm spiral of \citet{2012A&A...545A..39R}.
The CR source scale height perpendicular to the Galactic plane is taken to be 200~pc.
The R12 interstellar radiation field (ISRF) model developed by \citet{2017ApJ...846...67P} is used for the CR electron energy losses and \gray{} production via IC scattering.
The ISRF model sampling grid differs from that employed for the CR calculations (see below), and trilinear (3D) interpolation is used to determine the ISRF spectral intensity over the \GP{} spatial grid.
Meanwhile, CR electron synchrotron losses/radiation production use the bisymmetric spiral model of \citet{2011ApJ...738..192P} (hereafter PBSS) for the regular component of the magnetic field, together with a uniformly distributed 4$~\mu$G random component.

The propagation model parameters are as determined by \citet{2019ApJ...879...91J} using the 3D neutral gas (atomic and molecular) distribution model described by \citet{2018ApJ...856...45J} with 90\% hydrogen and 10\% helium by number and the ionised gas distribution described by the hybrid \hii{} model included in the \GP\ code, which is based on the NE2001 model of \citet{2002astro.ph..7156C} and the work of \citet{2008PASA...25..184G}.
The tuning procedure follows that of \citet{2017ApJ...846...67P} and \citet{2018ApJ...856...45J}, where the size of the CR halo is set to 6~kpc and the parameters are adjusted to reproduce recent CR data.
Solar modulation is accounted for by using the force-field approximation, one
modulation potential value for each observation period.
This simplified approach is justified because the main objective here is to
study the effect of the time-dependent source injection compared with the steady-state approach, rather than obtain an update for the local interstellar spectra or accurate determination of the
propagation parameters \citep[see, e.g.,][for more advanced treatment of the solar modulation with the \GP/HelMod framework]{2017ApJ...840..115B,2018ApJ...858...61B,2018ApJ...854...94B,2019arXiv191103108B}.

The propagation parameters are determined by first fitting the models to the observed spectra of CR nuclei: Be, B, C, O, Mg, Ne, and Si.
These are kept fixed, and the injection spectra for electrons, protons, and He nuclei are then fitted together to the data.
The procedure is then iterated until convergence.
(Iteration is required because the proton spectrum affects the normalisation of the heavier species and hence the propagation parameters.)
The only difference to \citet{2017ApJ...846...67P} and \citet{2018ApJ...856...45J} are the data: elemental spectra on Be, C, and O from {\it AMS-02} has been added to the fit, as well as elemental spectra from {\it ACE}-CRIS, while data from {\it HEAO}-C3 and PAMELA has been removed \citep[for the full list of CR data used see Table~1 of][]{2019ApJ...879...91J}.

Our model calculations use a 3D right-handed spatial grid with the solar system on the positive $X$-axis and $Z=0$~kpc defining the Galactic plane employing the IAU-recommended $R_S = 8.5$~kpc~\citep{1986MNRAS.221.1023K} for the distance from the Sun to the Galactic centre (GC).
The coordinate transformation for the spatial grid is given by Eq.~(\ref{eq:gridFunction}), where the parameters\footnote{The $X/Y$ offsets ($X_0/Y_0$) in the respective forms for Eq.~(\ref{eq:gridFunction}) differ because of the solar system location.} of the transforms for the $X$ and $Y$ coordinates are chosen so that the resolution near the solar system is $\sim$50~pc, increasing to $\sim$0.5~kpc at the boundary of the Galactic disk, which is 20~kpc from the GC.
In the $Z$-direction, the resolution is 25~pc in the plane, increasing to 0.5~kpc at the boundary of the grid at $|Z_{\rm halo}| = 6$~kpc.
The momentum grid is logarithmic from 100~MeV to 10~TeV with 32 planes.
This range largely covers the CR energies that are responsible for producing the non-thermal emissions detected by the \fermi-LAT and current Cherenkov instruments, as well as radio/microwave/X-ray synchrotron emissions.
The upper limit is set due to physical considerations related to the size of the discrete CR injection regions (see below); extension to higher energies of prediction for Galaxy-wide non-thermal emissions, including accounting for the distance-dependent pair absorption on the ISRF \citep{2006ApJ...640L.155M,2018PhRvD..98d1302P}, will be addressed elsewhere. 

The simulation epoch for the TDD solution is set to 600~Myr, which is a factor of $\sim$10 longer than the dominant (propagation) time scale around a few GeV determined using the steady-state parameters.
Using such a time span ensures that the discretised solution intensities reach the steady-state limit.
The size of an individual CR injection volume is set to be 50~pc in $X$, $Y$, and $Z$ coordinates, with frequency 0.01~yr$^{-1}$ and active time $10^5$~yr with a constant luminosity over this time and using the same spectral parameters as the smooth density distribution.
(Parametric variations for the definition of the region properties are, obviously, many, and we give the rationale for our choices below.
For the purposes of this paper, it is sufficient to use a single set that is representative.)
A fixed timestep of 5~kyr is used for the TDD solution, which is small enough to capture the propagation and energy losses at the upper boundary of the energy grid described above.

The propagation parameters for the TDD solution are assumed to be the same as derived for the steady-state case.
Recent work by \citet{2019ApJ...879...91J} showed that propagation parameters (diffusion coefficient, etc.) obtained for a time-independent model that included ``slow diffusion'' regions around the CR sources in the Galactic disk were practically the same as for the homogeneous case.
For the time-dependent case, we expect that this is also a reasonable assumption to make because the active times of the injection regions (100~kyr) are much shorter than the millions of years confinement time of CRs in the Galaxy.
Short-term perturbations by the localised injection regions will have a limited effect on the global diffusive properties of CRs in the ISM.

The size of the CR injection volumes is chosen to approximate the physical dimension where the CR propagation becomes `ISM-like'; for smaller sizes, the propagation is likely characterised by local effects about the true CR sources rather than in the general ISM~\citep[see, e.g.,][]{2008AdSpR..42..486P,2013ApJ...768...73M,2016MNRAS.461.3552N}.
Indeed, recent observations of the extended TeV emission around Geminga and the PSR~B0656+14 PWN by the HAWC experiment \citep{2017Sci...358..911A} show evidence for inhomogeneous diffusion properties that may be explained by the non-linear propagation phenomena \citep[e.g.,][]{2018PhRvD..98f3017E}.
But including such sub-grid scale effects is beyond the scope of the present work\footnote{It should be noted that there is no agreed phenomenology describing the propagation of CRs from the source injection and the local environment to `interstellar' space. The CR `source' spectra that are used in this paper include the localised propagation from the real sources before the propagation is governed by interstellar conditions, and not those directly injected by individual CR sources.}.

The self-generated wave picture may also be applicable to describing propagation of the CR particles throughout the extended halo \citep[e.g.,][]{2018PhRvL.121b1102E,2019PhRvL.122e1101B}, while the latter itself can be nonuniform or have a rigidity-dependent height above the plane \citep[e.g.,][]{2015PhRvD..92h1301T}.
  But such models involving possible CR feedback processes on the magnetic turbulence throughout the general ISM are also beyond the scope of our current investigation.

The frequency and active time are chosen as characteristic of SN-related sources (so, encompassing SNRs, pulsars, and others) in the Galaxy without making an identification with a specific class of objects.
Canonical estimates for the Galactic SN rate give one every 30--100~yr \citep[e.g.,][]{2006Natur.439...45D}, but statistical modelling by \citet{2018JCAP...11..045M} suggests that an even lower rate (one every $\sim$500~yr) may be necessary to explain the CR electron spectral measurements $\gtrsim$1~TeV.
The rate that we use (0.01~yr$^{-1}$) is within this range.

Meanwhile, the whole process of acceleration by sources and injection in the ISM is not entirely understood, with a combination of effects leading to injection of high-energy particles at earlier times and low-energy particles at later times spanning $\sim$30--300~kyr \citep[e.g.,][]{2012JCAP...01..010B}.
Consequently, the choice of a single active time of $\sim$100~kyr (within the likely time span for individual CR sources injecting particles) without any source spectral time evolution is reasonable for the calculations made in this paper to enable comparison with the continuous source density solution.

Only protons and electrons are used for the comparison between the steady-state and discretised CR injection modelling solutions, because they are also the primary CR species of major interest for producing also secondary electromagnetic emissions\footnote{The addition of CR He will produce $\sim 30-40$\% more $\pi^0$-decay \gray{s} over the pure proton case, but the qualitative results will be the same. Because computational time and storage for the TDD solution with only protons and electrons is already significant, resource requirements are minimised by not including the heavier nuclei.}.
For the discrete injection regions, using a constant luminosity over their individual active times and the same spectral parameters as determined for the steady-state distribution enables an equivalent comparison between the two solutions\footnote{Formal correspondence between CR fluxes obtained for steady state and discretised models has been shown using analytic solutions \citep[e.g.,][]{2012A&A...544A..92B}.}.
The discrete sampler also allows for more sophisticated luminosity and spectral evolution models to be used for individual source regions, but this usage is beyond the scope of the investigation described by this paper.

\section{Results}
\label{sec:results}

\subsection{Cosmic Rays}
\label{sec:CR}

\begin{figure*}[htb!]
  \subfigure{
    \includegraphics[scale=0.85]{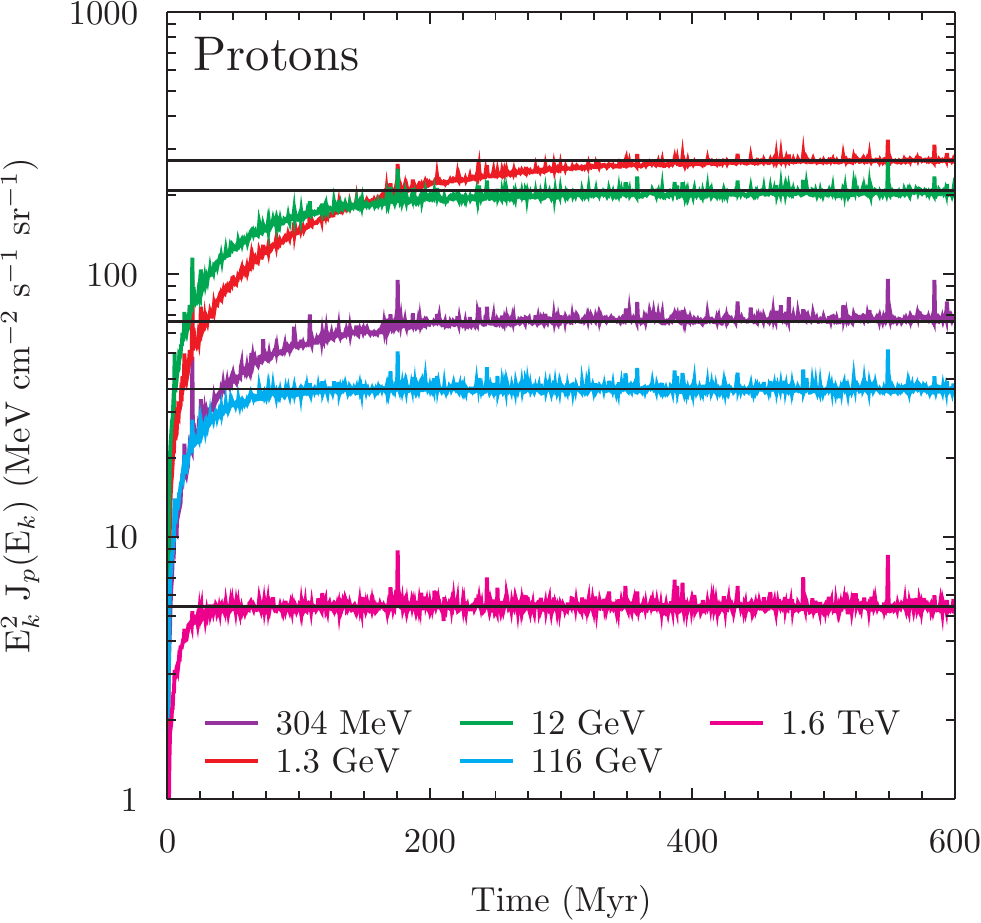}
    \includegraphics[scale=0.85]{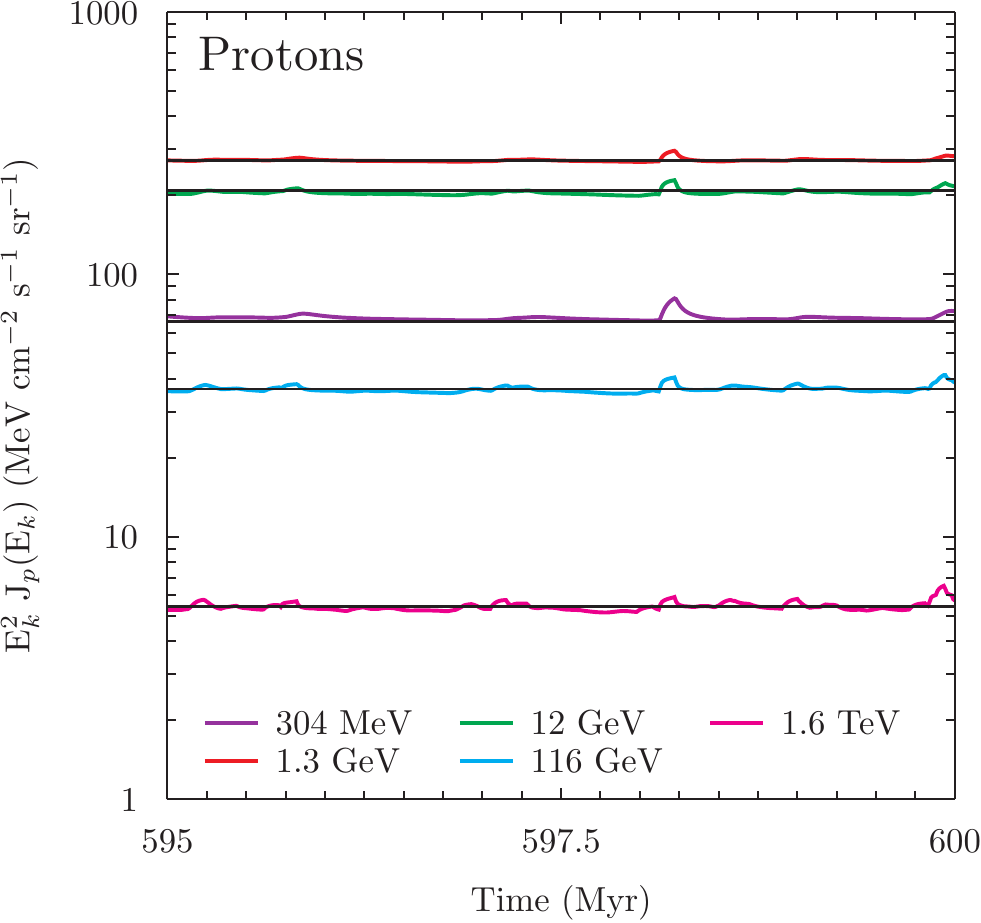}
  }\\
  \subfigure{
    \includegraphics[scale=0.85]{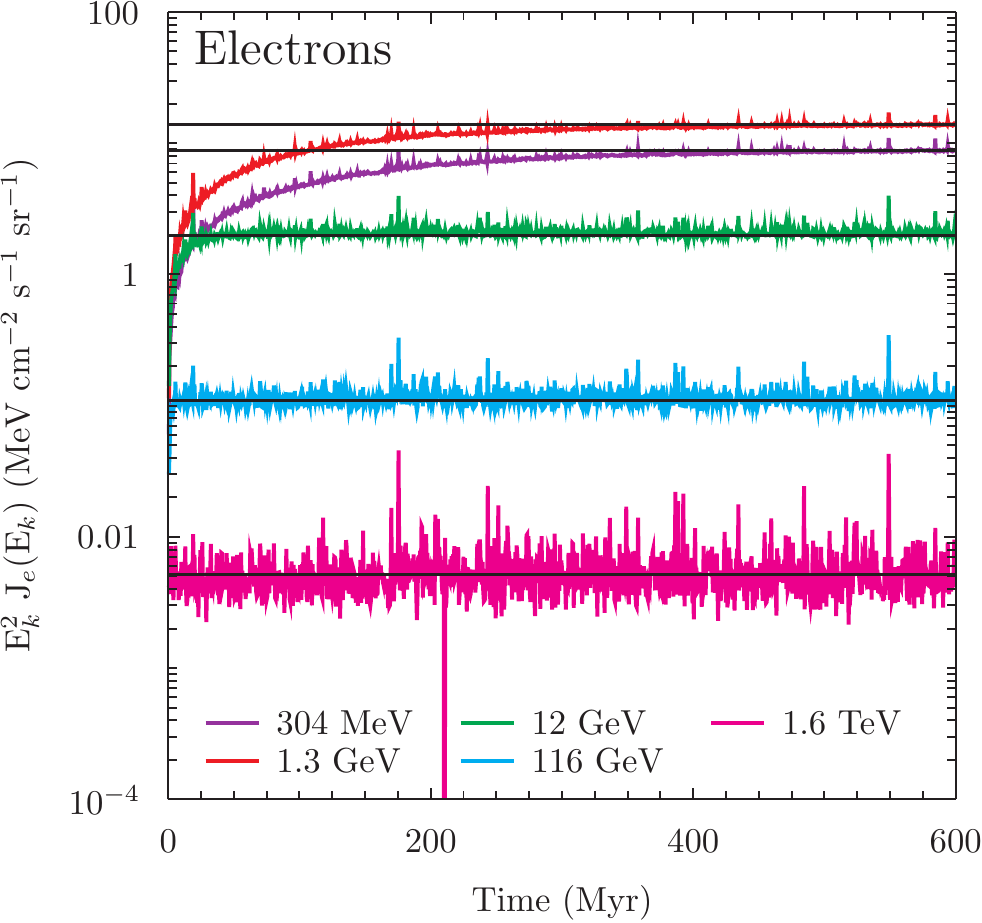}
    \includegraphics[scale=0.85]{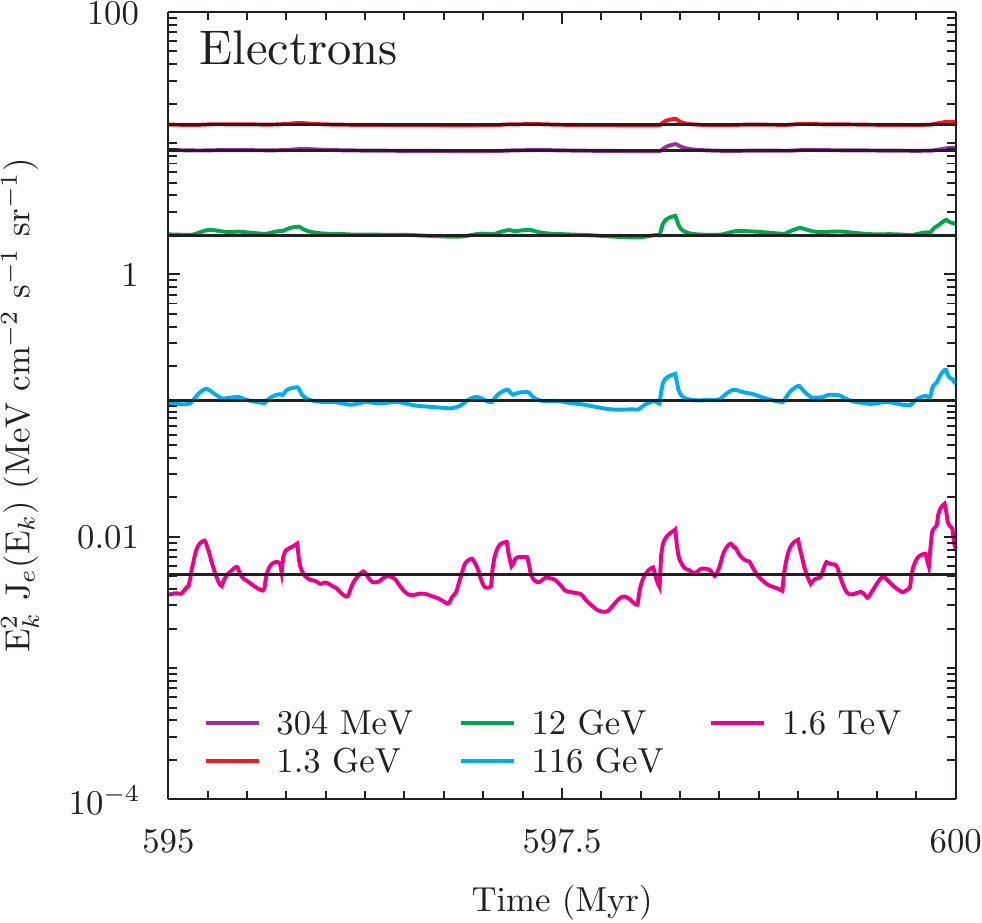}
  }
  \caption{Time series of TDD solution CR proton (upper) and electron (lower) intensities compared with SS$_{_{\rm SA50}}$ intensities at selected energies for the solar system location.
    Line colours: purple, 304~MeV; red, 1.3~GeV; green, 12~GeV; cyan, 116~GeV; magenta, 1.6~TeV; black, steady state at corresponding energies.
    Left panels show the total time series for the intensities over the 600~Myr of the simulation epoch sample at 500~kyr intervals.
    Right panels show a zoom of the last 5~Myr of the run sampled at 10~kyr intervals.
    \label{fig:crtimeseries}
  }
\end{figure*}

Figure~\ref{fig:crtimeseries} shows the TDD solution CR intensity time series (protons upper panels, electrons lower panels) at selected energies, together with the corresponding SS$_{_{\rm SA50}}$ intensities for the same energies at the solar system location.
The left panels show the evolution over the entire simulation epoch, while the right panels show the last 5~Myr of the run at higher sampling. 
(The sampling period for the left panels is 500~kyr, while for the right panels, it is 10~kyr.)
The normalisation for the steady-state solution is made to local CR data at 100~GeV for protons and 35~GeV for electrons, well above the energies that are affected by the solar modulation.
The TDD solution is normalised to the same data at the conclusion of the simulation epoch.
The average over the last 10 evenly spaced 50~kyr samples\footnote{Half of the active region lifetime.} for the CR intensities at the solar system is used to minimise fluctuation biases.
All earlier samples for the TDD solution CR intensities are then scaled according to the normalised ones from the end of the simulation epoch.
%due to nearby short-term active injection regions.
\begin{figure*}[tb!]
  \subfigure{
    \includegraphics[scale=0.9]{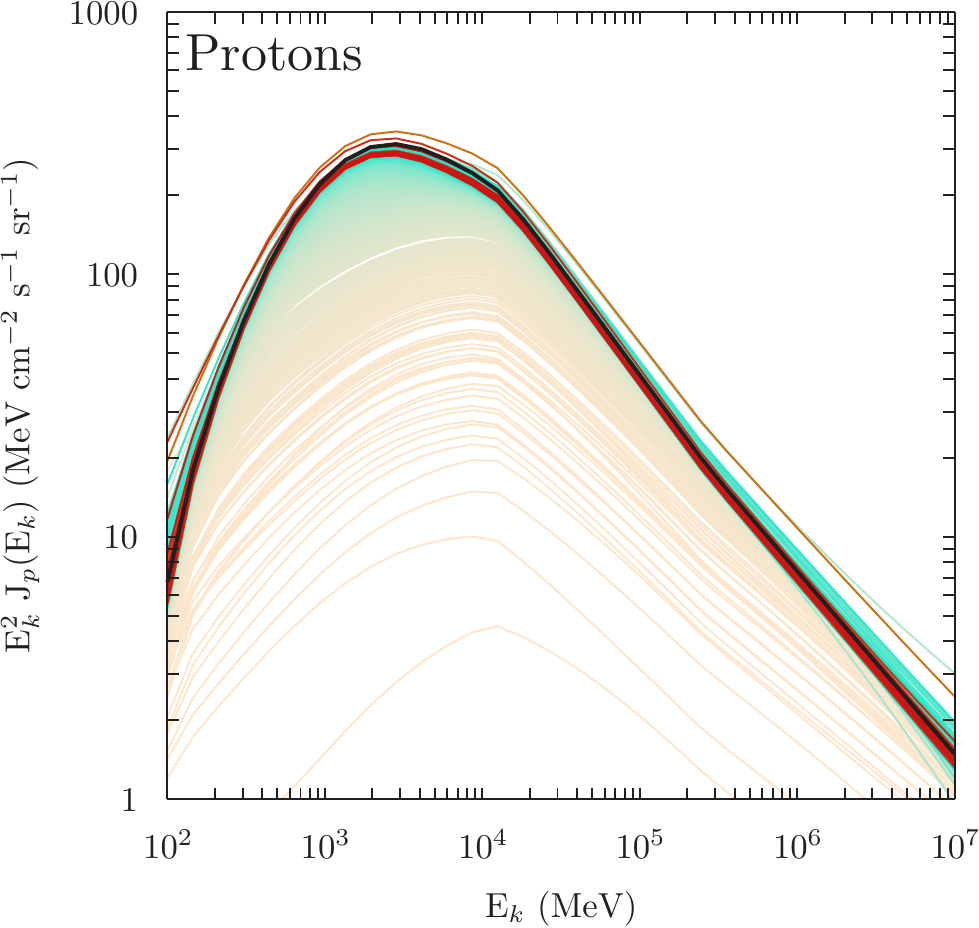}
    \includegraphics[scale=0.9]{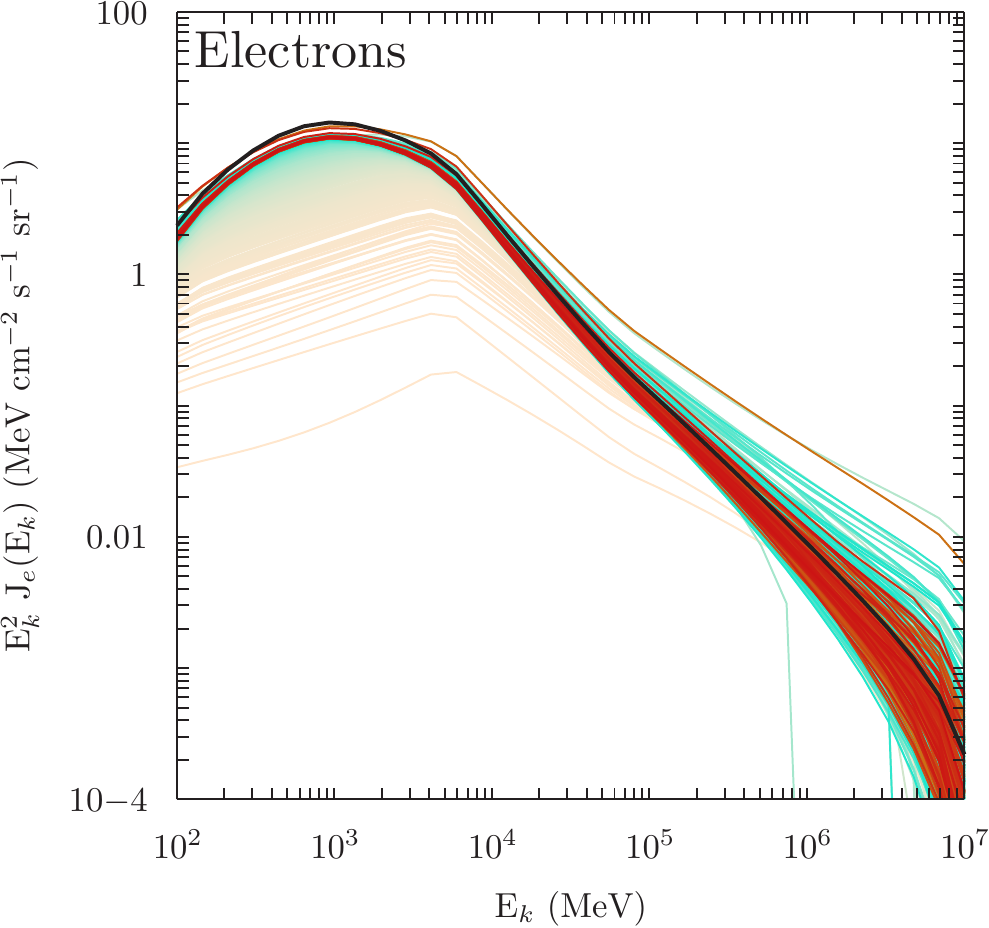}
  }
  \caption{Spectral intensities for protons (left) and electrons (right) at the solar system location. The solid black line shows SS$_{_{\rm SA50}}$ solution. Coloured lines show time evolution for the TDD solution from early ($\lesssim$100--200~Myr, light tan) to late ($\gtrsim$450~Myr, red) times of the simulation epoch (600~Myr). Normalisation of the spectral intensities is made to local proton and electron data, as described in the text.
    \label{fig:localspectra}
  }
\end{figure*}

The TDD solution starts with an initially empty Galaxy and evolves with time, eventually approaching the steady-state intensities.
For protons $\gtrsim$100~GeV the intensities are comparable to the steady state at relatively early times compared to lower energies because of the energy-dependent propagation.
For the electrons, the much more rapid energy losses that determine the relevant time scale $\gtrsim$10~GeV mean that the TDD solution approaches that of the steady state even quicker than the high-energy protons.
Only for energies $\lesssim$10~GeV does the TDD solution need comparably long times to approach the steady-state intensities, as is expected because the dominant time scale is determined by the diffusion and halo size.

The steady-state solution effectively produces a time-averaged intensity, whereas the TDD solution has fluctuations that are energy-dependent.
Generally, the latter are most evident for electrons energies $\gtrsim$100~GeV, but even the protons show some effect when nearby localised source regions can noticeably influence the intensity at the solar system location.
An example of this can be seen in the right panels around $\sim$598.2~Myr where the contribution of a recently active nearby injection region produces an enhanced intensity for both protons and electrons\footnote{The injection spectrum is generally harder than that following propagation and energy losses/gains. The CR nuclei and electron injection spectra are not the same; hence, the effect of the nearby region produces a different energy-dependent intensity enhancement.}.
The balance of fluctuations about the average intensity is generally more one-sided (up) for long simulation times $\lesssim$1~TeV, reflecting contributions to the intensity by the nearest source regions on top of an established CR background.
For electrons at higher energies, this is, however, not the case where the fluctuations about the average intensity are distributed about it and are larger in magnitude, showing that at $\gtrsim$1~TeV the CR electron intensity is strongly dependent on the historical distribution of the nearby CR injection regions.
The origin of the significant `spikes' (several) and the `dip' (around 210~Myr) that can be seen in the lower left panel are due to the presence of very close (spikes) or the absence of (dip) nearby injection regions.
For the upward fluctuations, it only takes one or a few nearby injection regions, while the $\gtrsim 1$~TeV downward fluctuation (dip) comes from a deficit of nearby regions over several hundred kyr.

%\begin{figure*}[tb!]
%  \subfigure{
%    \includegraphics[scale=0.9]{protons_local_spec_v2.pdf}
%    \includegraphics[scale=0.9]{electrons_local_spec_v2.pdf}
%  }
%  \caption{Spectral intensities for protons (left) and electrons (right) at the solar system location. The solid black line shows SS$_{_{\rm SA50}}$ solution. Coloured lines show time evolution for the TDD solution from early ($\lesssim$100--200~Myr, light tan) to late ($\gtrsim$450~Myr, red) times of the simulation epoch (600~Myr). Normalisation of the spectral intensities is made to local proton and electron data, as described in the text.
%    \label{fig:localspectra}
%  }
%\end{figure*}

Figure~\ref{fig:localspectra} shows the time evolution for the TDD solution over the full CR energy grid at the solar system location, together with that for the  SS$_{_{\rm SA50}}$.
The broadband spectral intensities at the same sampling interval are shown for protons (left) and electrons (right), with the SS$_{_{\rm SA50}}$ solution shown by the solid black line for either species and the TDD for the same from early (light tan) to late (red) times of the simulation epoch. 
For both protons and electrons, the spectral intensities generally steepen as the solution evolves forward in time because the energy losses and energy-dependent propagation shift particles to populate the lower energies in the spectrum.
At late times, the TDD proton spectral shape follows reasonably closely that of the steady-state solution, and its normalisation is slightly different, as described above.

\begin{figure*}[htb!]
  \subfigure{
    \includegraphics[scale=0.9]{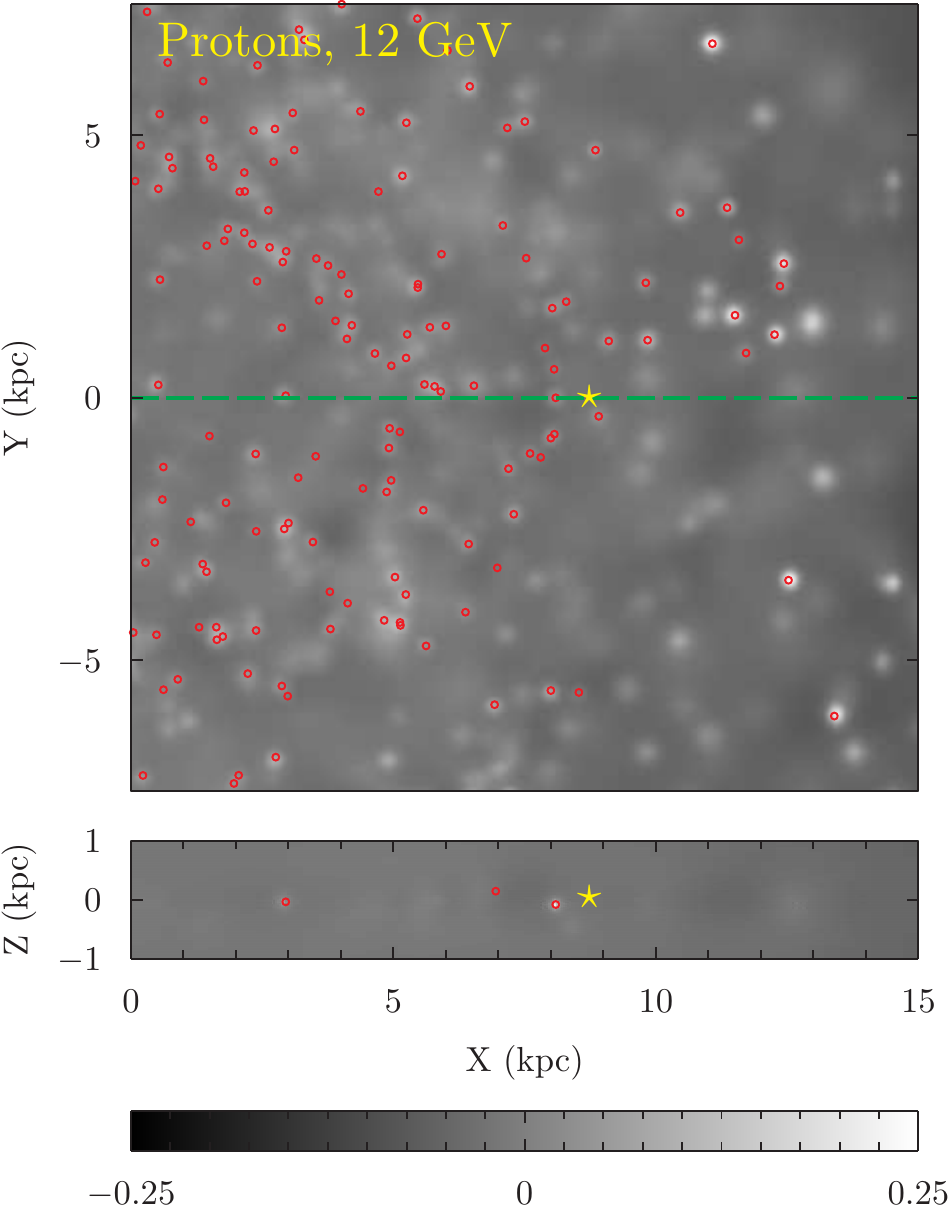}
    \includegraphics[scale=0.9]{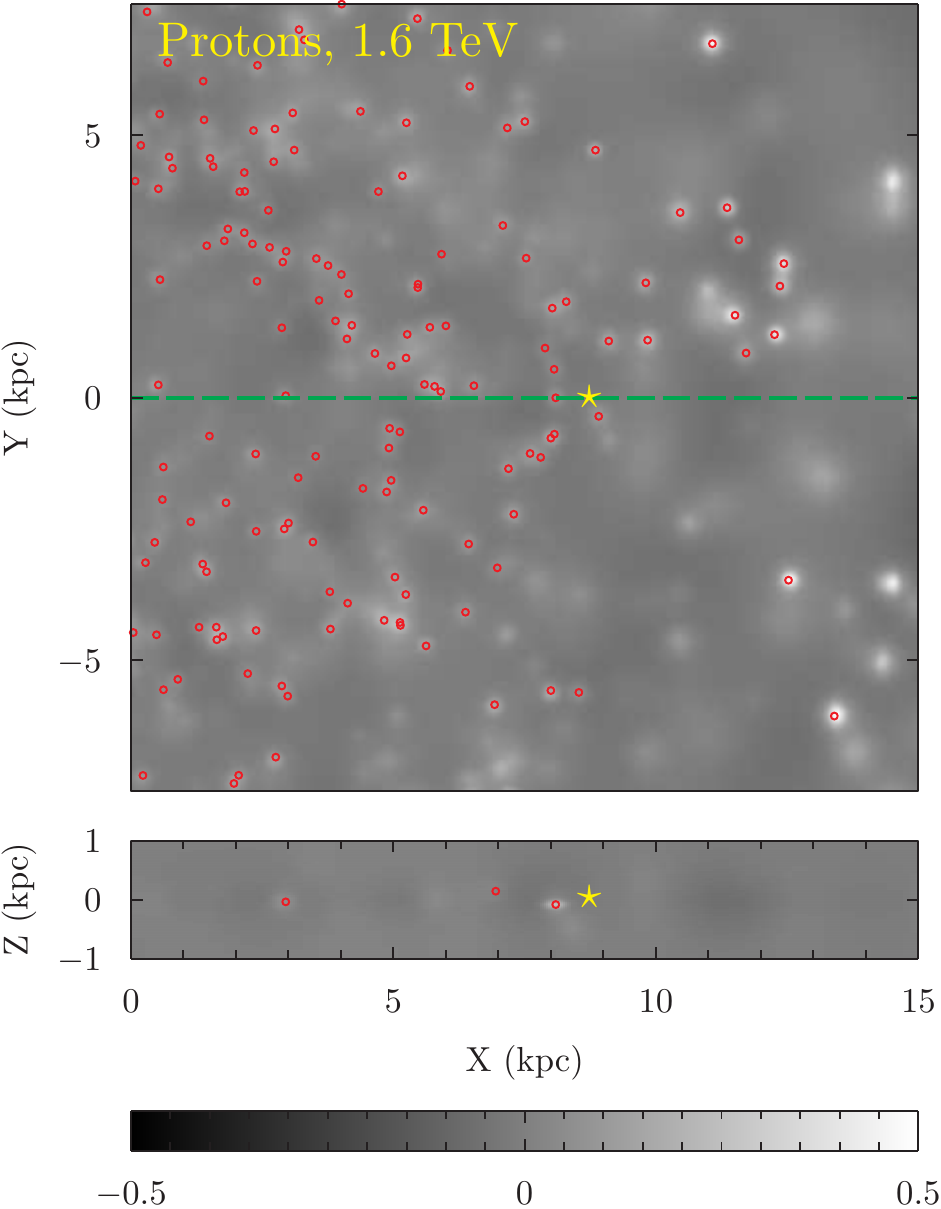}
  }\\
  \subfigure{
    \includegraphics[scale=0.9]{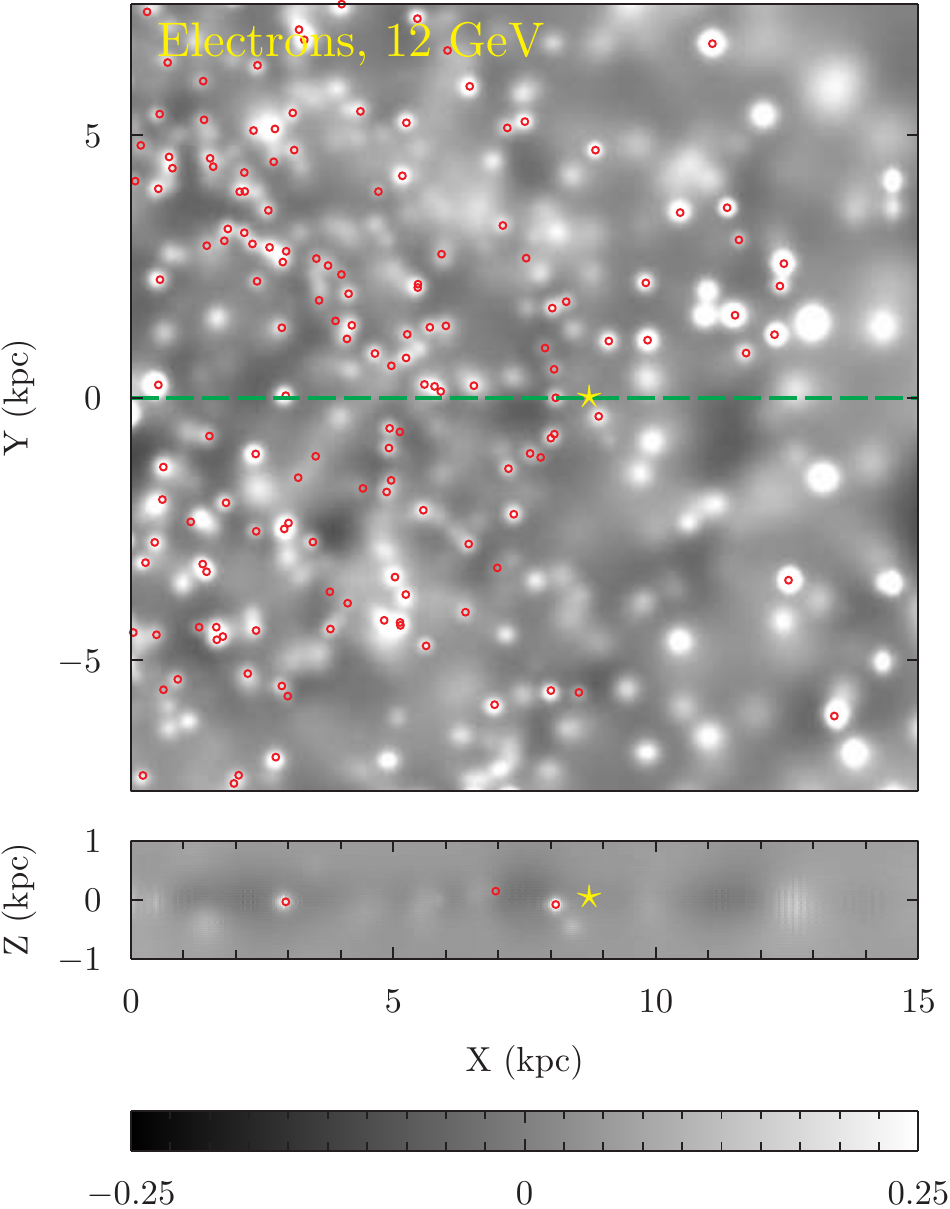}
    \includegraphics[scale=0.9]{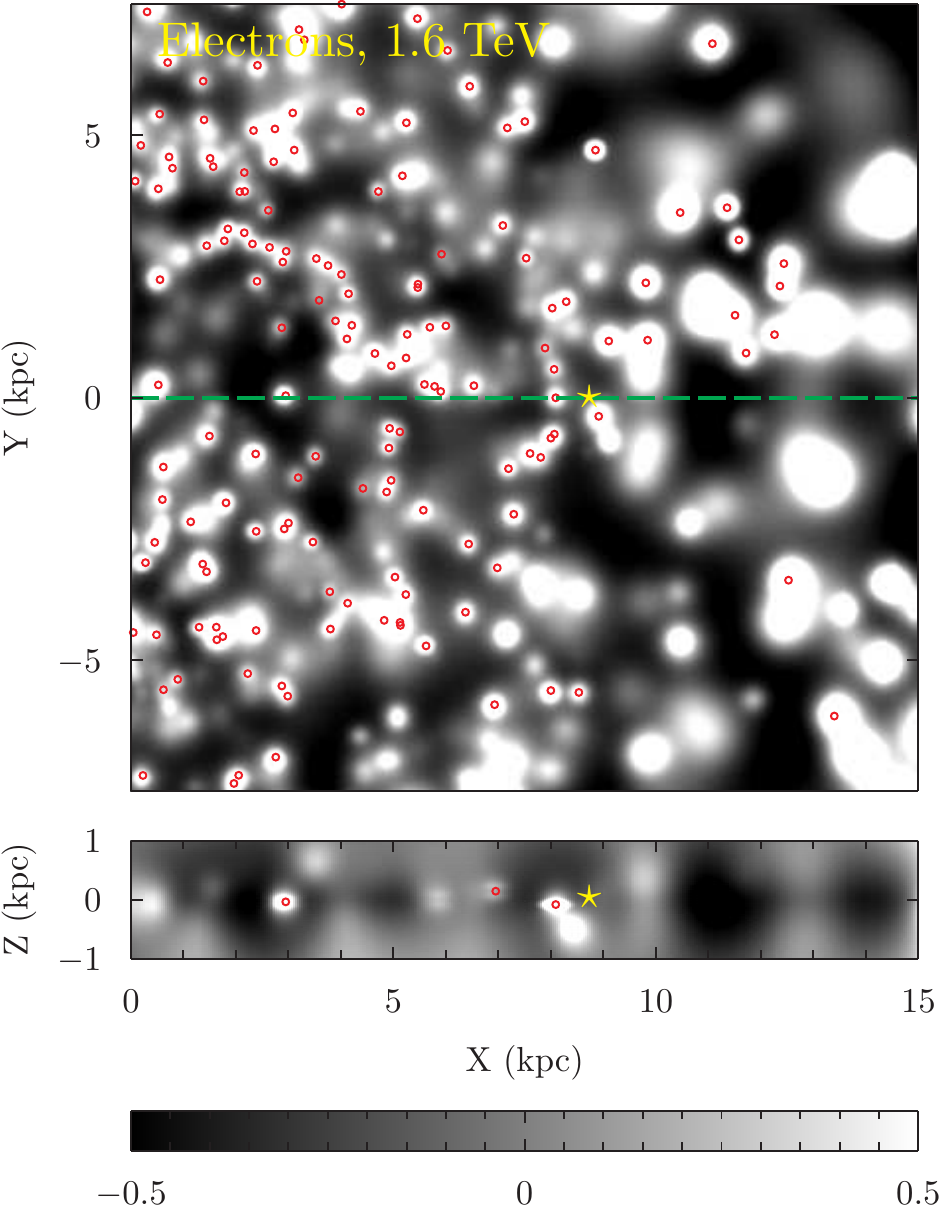}
  }
  \caption{The CR proton (top) and electron (bottom) intensity fractional residuals for the TDD solution with respect to the SS$_{_{\rm SA50}}$ case at selected energies. The solar system location is marked by a yellow star. The dashed line in each $X/Y$ panel shows the location of the corresponding slice in the $X/Z$ plane shown below it for each particle species and energy. The TDD solution sample is taken at 599.5~Myr into the simulation epoch (600~Myr). Overlaid on the residuals are the locations (red circles) of the discretised injection regions active within the last 100~kyr whose centres are within $\pm$50~pc of the $X/Z = 0$ coordinate for the respective plane. Note that the fractional scale is different for the left and right panels, with range $\pm$25\% and $\pm$50\%, respectively. \label{fig:edenfracres}
  }
\end{figure*}

For the electron spectrum, the TDD solution takes a somewhat shorter time to approach that of the steady-state case for energies $\lesssim$50~GeV.
However, for higher energies at late times of the simulation epoch, there is an envelope for the electron intensity and spectral shape that is not convergent to the steady state, and instead fluctuates around it.
The spectral envelope is due to the fresh injection of particles by relatively nearby active regions that can produce a hardening of the spectral shape, and (if there is very little nearby recent injection) the rapid cooling for CR electrons $\gtrsim$50~GeV that generally steepens the spectrum\footnote{The aqua solid line that is cutoff around $\sim$1~TeV in the right panel for this figure is the spectrum corresponding to the intensity `dip' in Fig.~\ref{fig:crtimeseries} around $\sim$210~Myr. Note that there is no spectral cutoff for the electron injection spectrum, so that seen in this figure is solely due to the paucity of nearby active regions prior to the sample.}.
Generally, over the energy range where the normalisation to the CR data is made ($\sim$10--100~GeV) the spectral shape for both protons and electrons is fairly constant, with the effects of the nearby injection activity becoming visible at lower and higher energies.

The variations of the spectral intensity exhibited by the TDD solution seen at the solar system are also present at other locations throughout the Galaxy.
Compared to the SS$_{_{\rm SA50}}$ solution, which is smoothly distributed, the TDD solution shows fluctuations of differing magnitudes over all energies.
Examples of these fluctuations are shown in Fig.~\ref{fig:edenfracres} for protons and electrons toward the end of the simulation epoch.
They are the fractional residuals of the TDD solution using the SS$_{_{\rm SA50}}$ for the baseline, and, while the 599.5~Myr sample is being used for illustration, similar fluctuations for the TDD CR intensity distributions are seen over the entire run.

%\begin{figure*}[htb!]
%  \subfigure{
%    \includegraphics[scale=0.9]{protons_eden_fracres_12GeV_ts119900.pdf}
%    \includegraphics[scale=0.9]{protons_eden_fracres_1561GeV_ts119900.pdf}
%  }\\
%  \subfigure{
%    \includegraphics[scale=0.9]{electrons_eden_fracres_12GeV_ts119900.pdf}
%    \includegraphics[scale=0.9]{electrons_eden_fracres_1561GeV_ts119900.pdf}
%  }
%  \caption{The CR proton (top) and electron (bottom) intensity fractional residuals for the TDD solution with respect to the SS$_{_{\rm SA50}}$ case at selected energies. The solar system location is marked by a yellow star. The dashed line in each $X/Y$ panel shows the location of the corresponding slice in the $X/Z$ plane shown below it for each particle species and energy. The TDD solution sample is taken at 599.5~Myr into the simulation epoch (600~Myr). Overlaid on the residuals are the locations (red circles) of the discretised injection regions active within the last 100~kyr whose centres are within $\pm$50~pc of the $X/Z = 0$ coordinate for the respective plane. Note that the fractional scale is different for the left and right panels, with range $\pm$25\% and $\pm$50\%, respectively. \label{fig:edenfracres}
%  }
%\end{figure*}

\begin{figure*}[tb!]
  \subfigure{
    \includegraphics[scale=0.8]{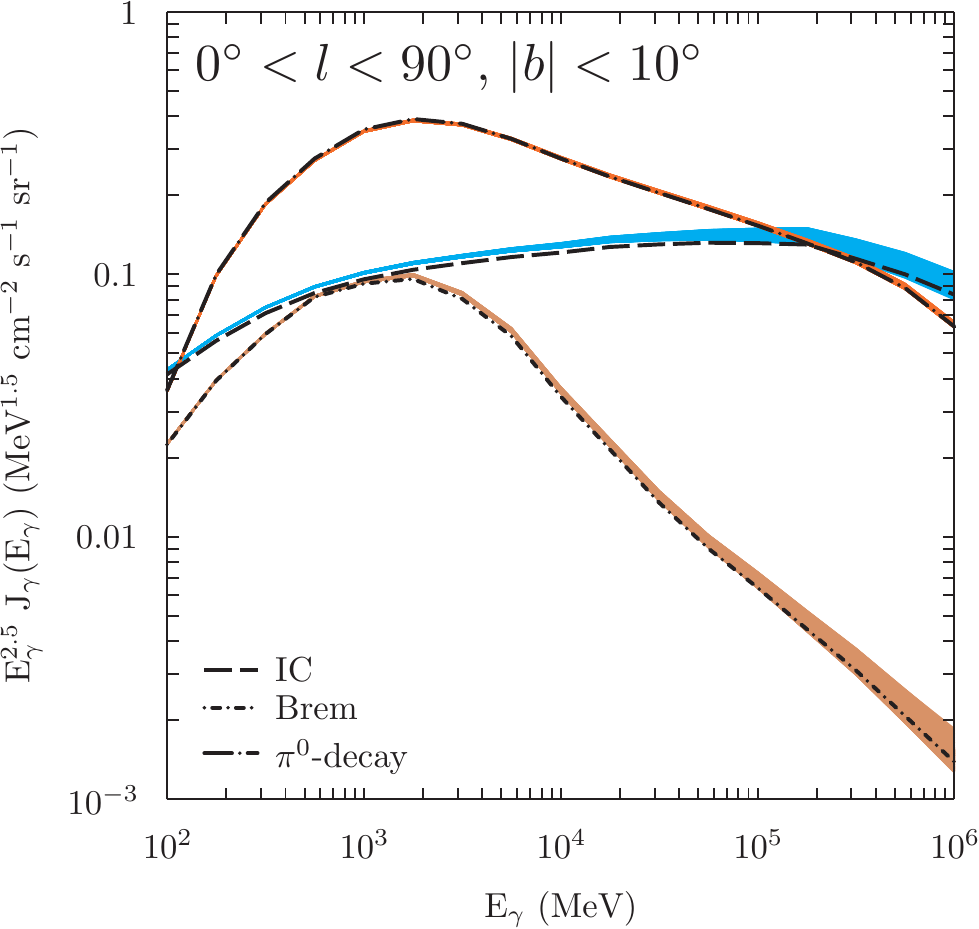}
    \includegraphics[scale=0.8]{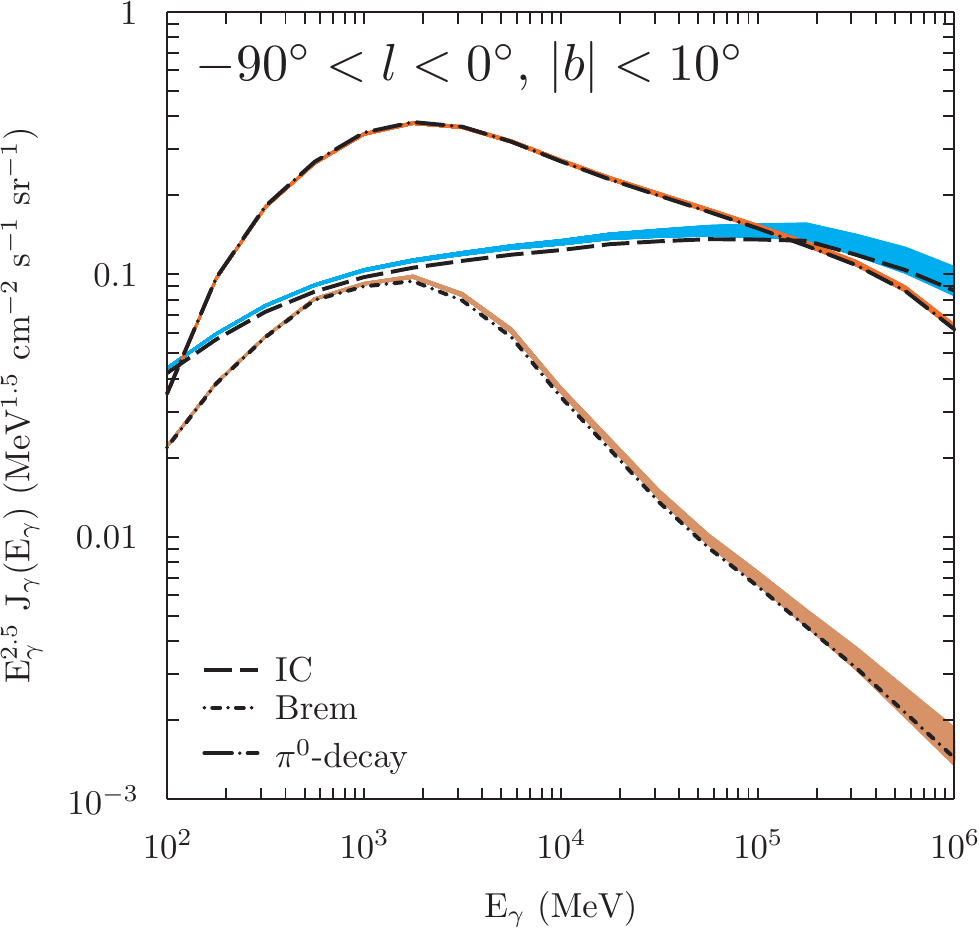}
  }\\
  \subfigure{
    \includegraphics[scale=0.8]{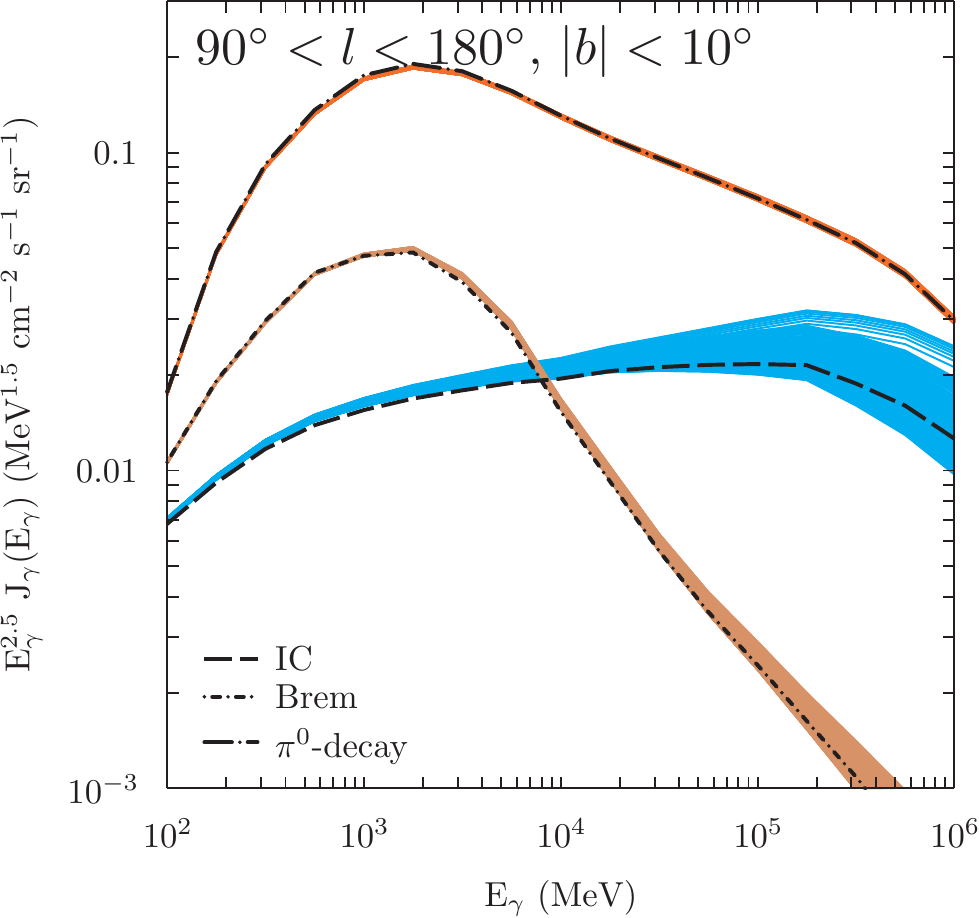}
    \includegraphics[scale=0.8]{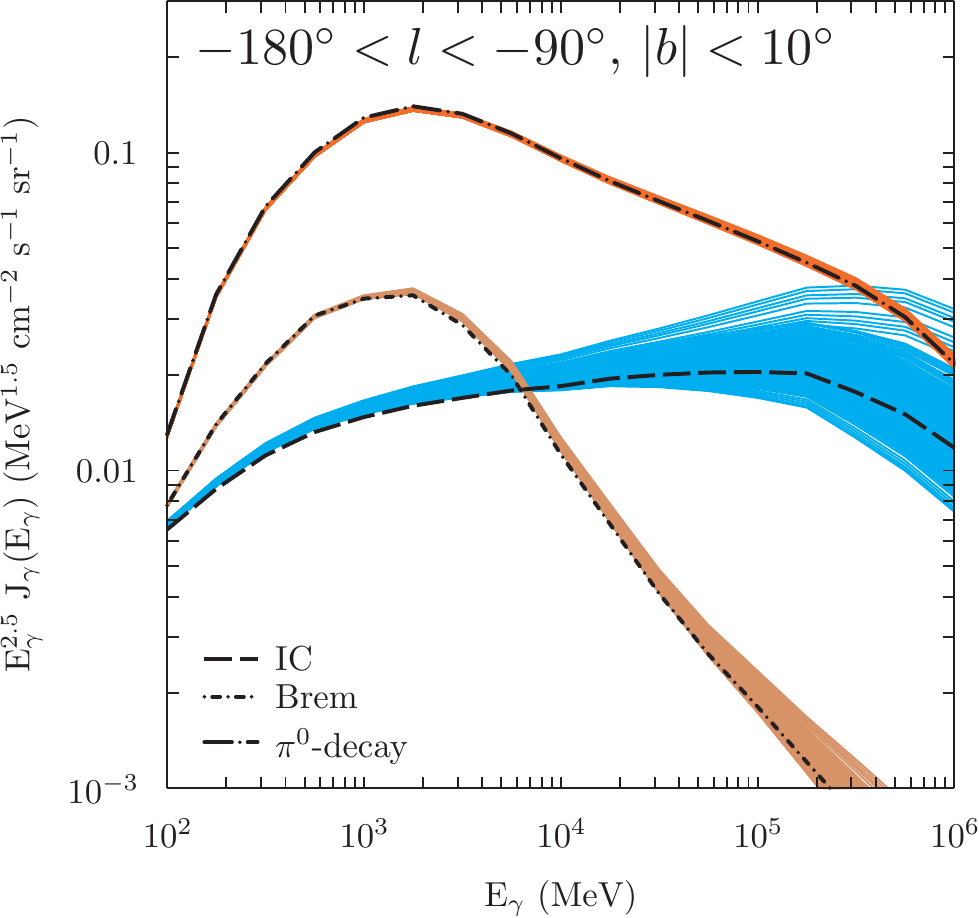}
  }\\
  \subfigure{
    \includegraphics[scale=0.8]{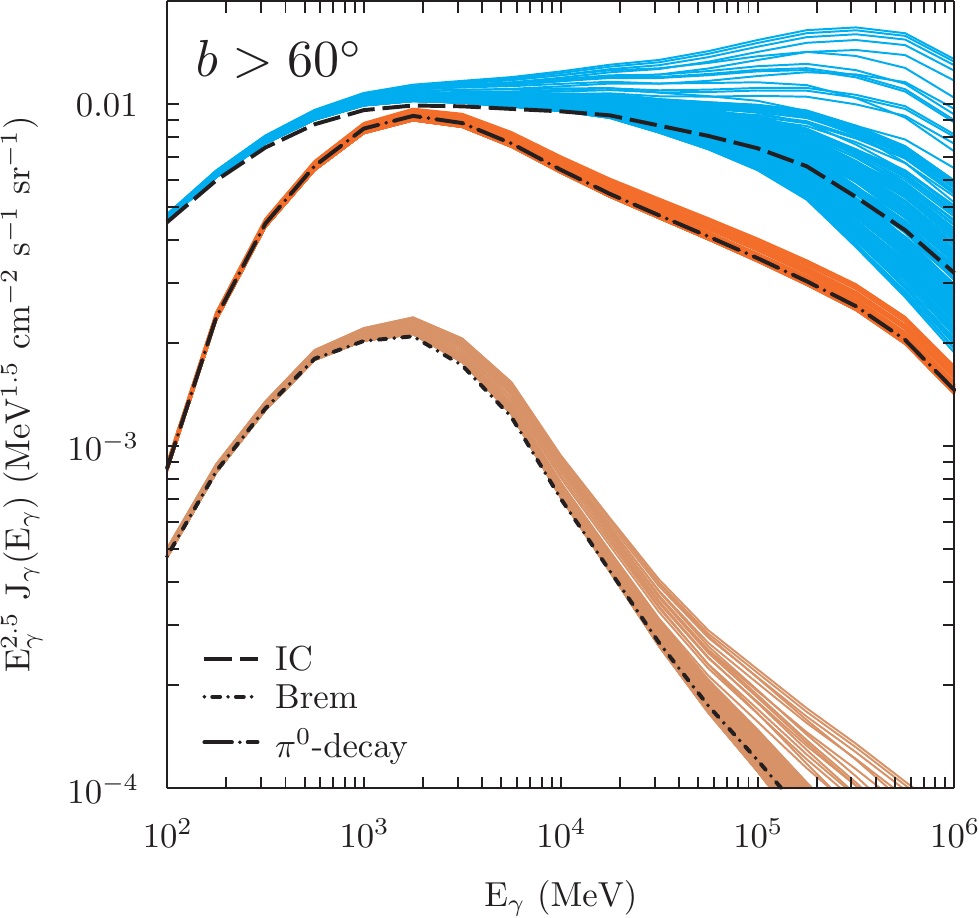}
    \includegraphics[scale=0.8]{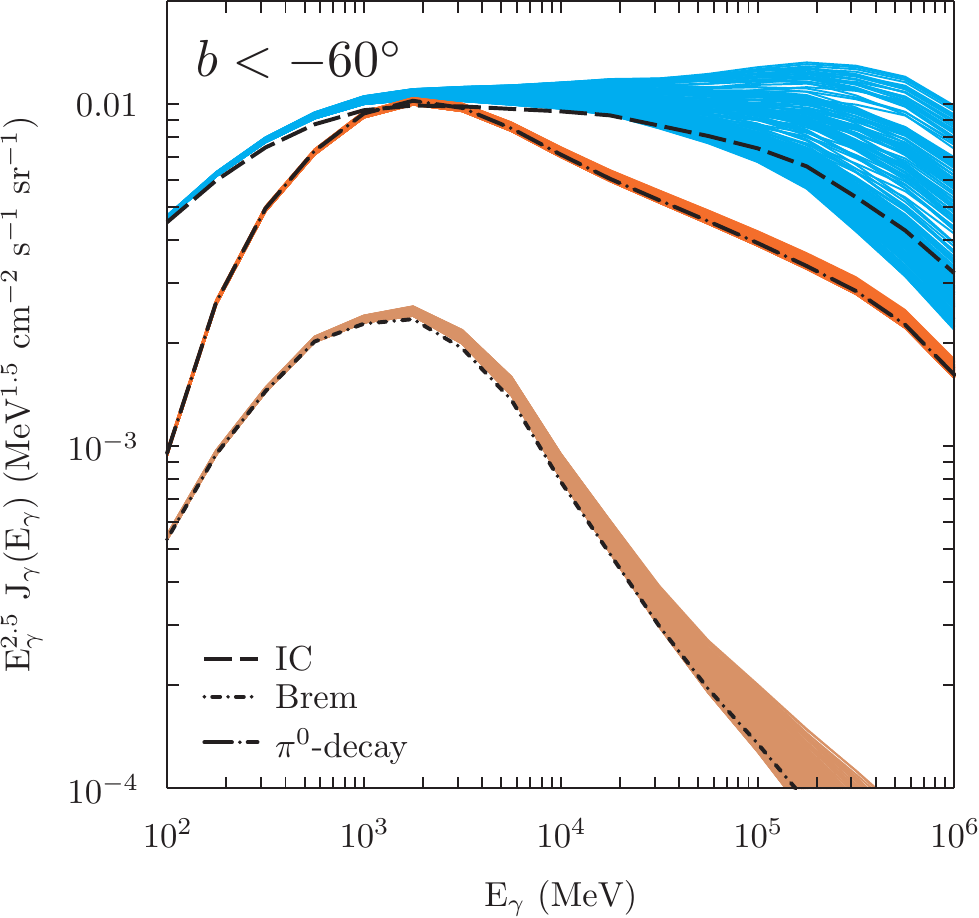}
  }
  \caption{Intensity spectra for \gray{} emission processes within $|b| < 10^\circ$ of the Galactic plane and toward the poles. Top row: quadrants 1 (left) and 4 (right). Middle row: quadrants 2 (left) and 3 (right). Bottow row: north polar (left) and south polar (right). Various black line styles show the SS$_{_{\rm SA50}}$ solution for respective sky regions: long dashed, IC; short dash-dotted, bremsstrahlung; long dash-dotted, $\pi^0$-decay.
    Coloured curves show the envelope of TDD intensity spectra over the last 5~Myr of the simulation run with 10~kyr sampling. The TDD intensity spectra colour coding is as follows: IC, cyan; bremsstrahlung, tan; $\pi^0$-decay, orange.
    \label{fig:gammaspec}
  }
\end{figure*}

\begin{figure*}
  \subfigure{
  \includegraphics[scale=0.6]{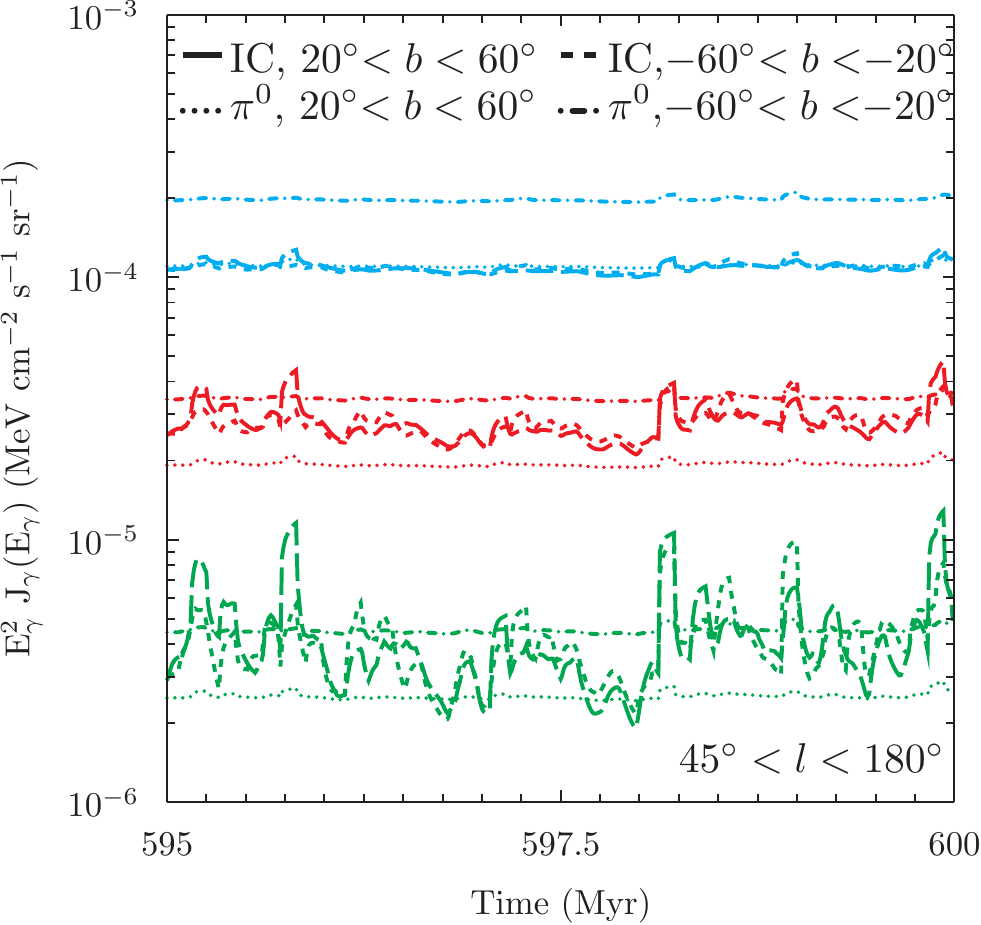}
  \includegraphics[scale=0.6]{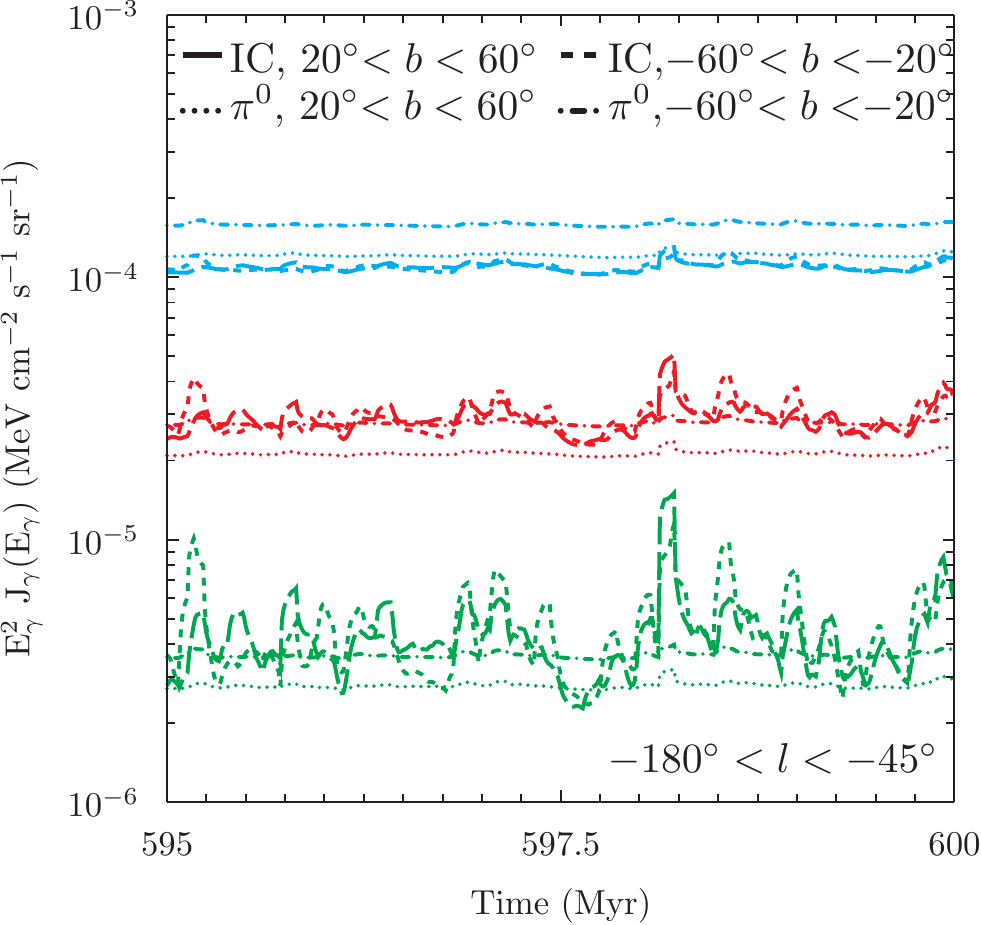}
  \includegraphics[scale=0.6]{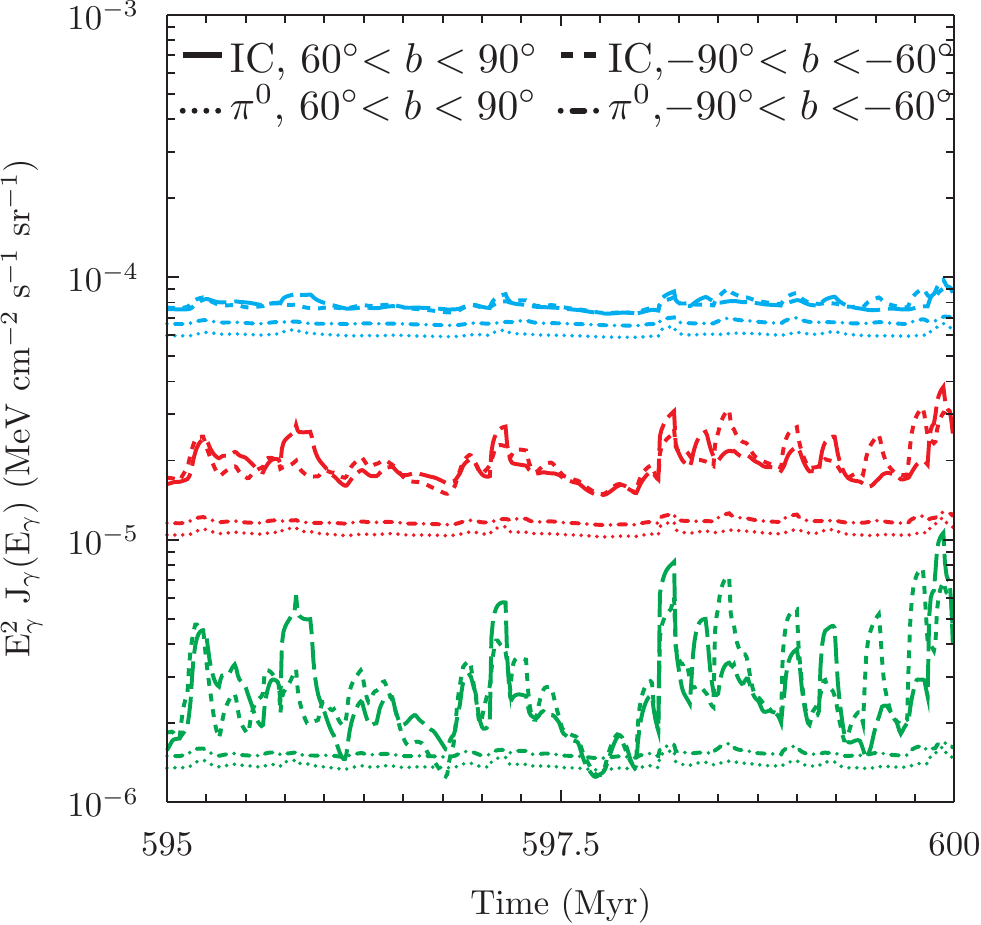}
  }
  \caption{Time series of \gray{} intensities at 10~GeV (cyan), 100~GeV (red), and 1~TeV (green) for IC and $\pi^0$-decay processes for (left) $45^\circ < l < 180^\circ$ and (middle) $-180^\circ < l < -45^\circ$ ranges for intermediate latitudes above/below plane and (right) north/south Galactic polar regions $|b| > 60^\circ$.
    The time range is for the last 5~Myr of the TDD simulation epoch and has 10~kyr sampling.
    This range and sampling is the same as shown by the right panels of Fig.~\ref{fig:crtimeseries}.
    \label{fig:gammaoutofplanetimeseries}
  }
\end{figure*}

The top panels of Fig.~\ref{fig:edenfracres} show the fractional residuals for the CR protons in the Galactic and the $X/Z$ planes for $Y=0$~kpc at 12~GeV (left) and 1.6~TeV (right) and the CR source regions that have been active within the last 100~kyr whose centres are within $\pm$50~pc of the zero coordinate for either planar slice.
For any such ``snapshot'', the discretised intensity distribution reflects the recently active and ongoing injection of CRs, together with the formerly active region remnant contributions that continue to propagate through the ISM.
This can be readily seen with the overlay, because not all discretised intensity enhancements over the steady state have an associated injection region.
Those enhancements without an associated injection region are due to CRs propagating from a currently active one outside of the $\pm$50~pc about the plane, or are from the remnant propagation from injection activity earlier than 100~kyr.

The fractional residuals for the protons clearly illustrate the non-smoothness caused by the time-/energy-dependent propagation and injection from the discretised regions, especially for the highest energies, where overdensities caused by individual regions stand out compared to the steady-state case.
These features are more prominent in the outer Galaxy, where the presence/absence of individual injection regions affects the fractional residuals more than toward the inner Galaxy.
The regional density is higher there, which produces a larger accumulation of particles with time that reduces fluctuations.

At smaller scales, e.g., for arm/inter-arm regions, the contrast in probabilities for individual injection regions to occur can also produce more prominent fluctuations away from the highest-density areas.
Interestingly, even for the $\sim$10~GeV energies for the protons, where it is typically assumed that the long mixing times erase the injection and propagation history, there are many small overdensities.
Although not shown, for even lower energies still ($\lesssim$1~GeV), small variations of the intensity distribution also occur due to the slow diffusion and ionization losses.
Meanwhile, looking at the $X/Z$ slice reveals asymmetrical residuals above/below the Galactic plane.
As for the plane slice, the distribution and magnitude of these asymmetries depends on the injection and propagation history.
Even though the modelling uses a $Z=0$~kpc symmetric injection region density distribution the finite statistics means that at any moment over the simulation epoch, there is no guarantee that individual regions will be distributed symmetrically about the Galactic plane. 

Figure~\ref{fig:edenfracres} (bottom row) shows the fractional residuals for CR electrons in the Galactic and $X/Z$ planes for $Y= 0$~kpc at the same energies as the protons, and, as for the proton intensity residuals, the CR injection regions active within the last 100~kyr.
The fractional residuals for the electrons exhibit much stronger spatial variations than the CR protons.
Across the Galactic plane, it is clear that at the highest energies there is no reasonable correspondence with the steady-state solution, with significant over- and under-densities spread throughout.

At lower energies, such density fluctuations are also evident, but the magnitude of their variations is lower than those $\gtrsim$1~TeV.
However, they are still numerous and widely distributed, indicating that a smooth CR sea for electrons even down to $\sim$~GeV energies, as might be defined by the steady-state solution, is not as easily justifiable as for the protons when the CR sources are discretised spatially and temporally.
Examining the $X/Z$ slice reveals asymmetries about the Galactic plane, and for the highest energies, the residuals are strong.
Some that are barely visible for the protons are clearly seen for the electrons (e.g., around $X\sim8.3$~kpc and $Z \sim -0.6$~kpc, and the regions about the plane for $X \sim 7$~and~$8.1$~kpc, respectively).

\subsection{$\gamma$-Rays}
\label{sec:gammaray}

The spatial grid used for the CR propagation calculations is also employed for determining \gray{} emissivities. 
The \gray{} intensity maps at the solar system location are obtained by line-of-sight integration of the \gray{} emissivities for the standard processes ($\pi^0$-decay, IC scattering, bremsstrahlung), where the emissivities are determined for a logarithmic grid from 100~MeV to 1~TeV using four bins per decade spacing.
All calculations of the IC component use the anisotropic scattering cross section \citep{2000ApJ...528..357M}, which accounts for the full directional intensity distribution for the R12 ISRF model.

The spatially averaged spectral intensities for $\pi^0$-decay, IC, and bremsstrahlung emission processes for the north and south polar regions and the four Galactic quadrants about the plane are shown in Fig.~\ref{fig:gammaspec}.
The intensity spectra for the TDD solution are shown for 10~kyr sampling over the last 5~Myr of the simulation as solid coloured lines, colour-coded according to the different processes (cyan for IC, brown for bremsstrahlung, and orange for $\pi^0$-decay).
This interval and sampling corresponds to that for the CR intensities at the solar system location shown in the right panels of Fig.~\ref{fig:crtimeseries}.
For the respective processes, the SS$_{_{\rm SA50}}$ solution spectral intensities are shown as the black curves.

About the Galactic plane, over all quadrants, there is essentially no difference between the TDD and SS$_{_{\rm SA50}}$ predictions for the $\pi^0$-decay emissions.
The modest fluctuations in the CR proton intensities seen in the upper panels of Fig.~\ref{fig:edenfracres} have little effect on the corresponding \gray{} emissions because of the smoothing when the line-of-sight integrated emissivities are averaged over sufficiently large regions of the sky.

For quadrants~1 and~4 ($0^\circ \leq l \leq 90^\circ$ and $-90^\circ \leq l \leq 0^\circ$, respectively) the IC and bremsstrahlung emissions are very similar for the TDD and SS$_{_{\rm SA50}}$ intensities.
The TDD intensities are slightly higher because of the integrated contributions per time epoch of the localised enhancements for the CR electron intensities toward the inner Galaxy.
The spectra of the IC and bremsstrahlung emissions also have a small envelope, $\gtrsim$100~GeV, that is generally higher compared to the SS$_{_{\rm SA50}}$ intensities.

For quadrants~2 and~3 ($90^\circ \leq l \leq 180^\circ$ and $-180^\circ \leq l \leq -90^\circ$, respectively), the TDD intensities are also slightly higher than the SS$_{_{\rm SA50}}$ solution around $\sim 1$~GeV.
But, at higher energies, there is a much larger spread of the spectral index, particularly for the IC emissions that show a variation of the intensity around $\sim$1~TeV of a factor of $\sim$3--5 depending on the quadrant.
The spread reflects the much stronger per time epoch fluctuations of the CR electron intensities (e.g., as seen in the lower panels of Fig.~\ref{fig:edenfracres}).
Because there are fewer injection regions toward the outer Galaxy, there is a stronger variation with respect to the steady-state intensities than toward the inner Galaxy, where the line-of-sight integration effectively samples much more of the Galactic disk and smoothes out the CR intensity fluctuations (see also discussion for the residual sky maps below).

Toward the north and south polar regions, the correspondence between TDD and SS$_{_{\rm SA50}}$  solution $\pi^0$-decay emissions is similar, as for the Galactic plane, but with larger fluctuations because of the smaller region averaged over in the line-of-sight integration.
The IC emissions display significant variations of overall intensity and spectral shape for the higher-latitude regions.
The spread of the IC intensities for the individual time samples is much larger than about the plane for quadrants~2 and~3, where the range of the IC spectral indices $\gtrsim$10~GeV is $\sim$0.5~dex over both polar regions.
For the polar regions, the harder spectra generally correspond to the presence of active CR injection regions, with the north/south variation with respect to the steady-state intensity a result of asymmetrical distribution of CR intensities from injection and propagation.

As seen for the $X/Z$~slices in Fig.~\ref{fig:edenfracres}, finite sampling from the smooth CR density model produces propagated CR intensity distributions that can be strongly asymmetric about the Galactic plane.
Because the high-latitude \gray{} emissions are produced by the relatively nearby ($\lesssim$1~kpc) regions, the asymmetrical CR intensity distributions also translate directly to north/south variation also for the \gray{} intensities.
There is some asymmetry for the bremsstrahlung emissions $\gtrsim$few GeV, but the intensity of this component is much lower than the other two processes; hence, the major variations at high latitudes are from the injection region history and its effects on the IC emissions.

The time series of the \gray{} intensities for $\pi^0$-decay and IC scattering at selected energies outside of the Galactic plane for the last 5~Myr of the simulation run is shown in Fig.~\ref{fig:gammaoutofplanetimeseries}.
The sampling is at 10~kyr intervals and also corresponds with the interval shown in Fig.~\ref{fig:gammaspec} and the right panels of Fig.~\ref{fig:crtimeseries}.
The time series gives a clearer view of the intensity variations with energy and time for the various processes over the sky. 
The left and middle panels show the intensities for $45^\circ < l < 180^\circ$ and $-180^\circ < l < -45^\circ$, respectively, for intermediate latitudes ($20^\circ < |b| < 60^\circ$). 
The right panel shows the intensities for the north and south polar regions ($|b| > 60^\circ$) averaged over all longitudes.

%\begin{figure*}
%  \subfigure{
%  \includegraphics[scale=0.6]{gamma_spec_intermediatelat_p45to180long_timeseries.pdf}
%  \includegraphics[scale=0.6]{gamma_spec_intermediatelat_m45to180long_timeseries.pdf}
%  \includegraphics[scale=0.6]{gamma_spec_polar_timeseries.pdf}
%  }
%  \caption{Time series of \gray{} intensities at 10~GeV (cyan), 100~GeV (red), and 1~TeV (green) for IC and $\pi^0$-decay processes for (left) $45^\circ < l < 180^\circ$ and (middle) $-180^\circ < l < -45^\circ$ ranges for intermediate latitudes above/below plane and (right) north/south Galactic polar regions $|b| > 60^\circ$.
%    The time range is for the last 5~Myr of the TDD simulation epoch and has 10~kyr sampling.
%    This range and sampling is the same as shown by the right panels of Fig.~\ref{fig:crtimeseries}.
%    \label{fig:gammaoutofplanetimeseries}
%  }
%\end{figure*}

\begin{figure*}[tb!]
 \includegraphics[scale=0.7]{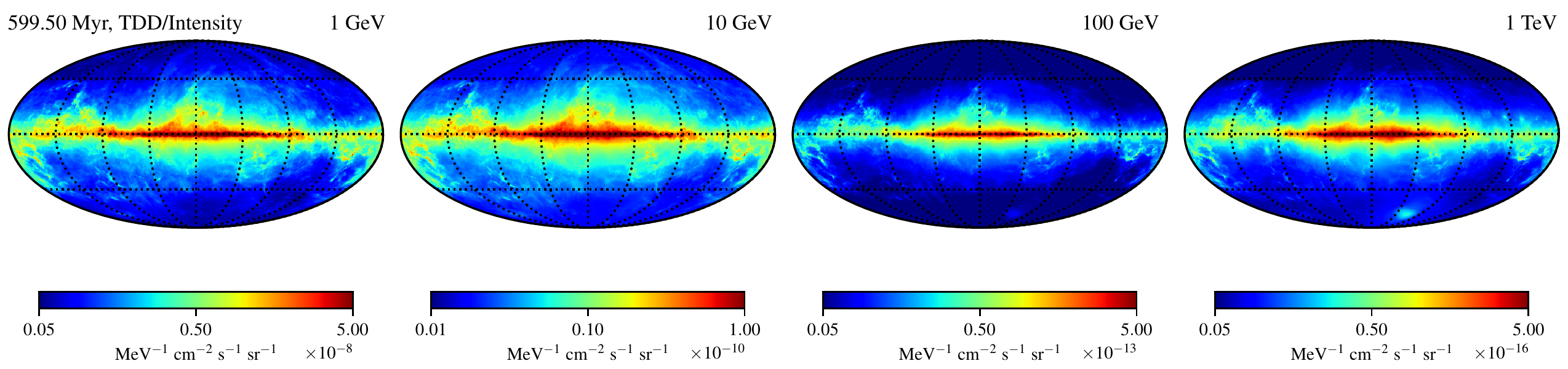}
  \includegraphics[scale=0.7]{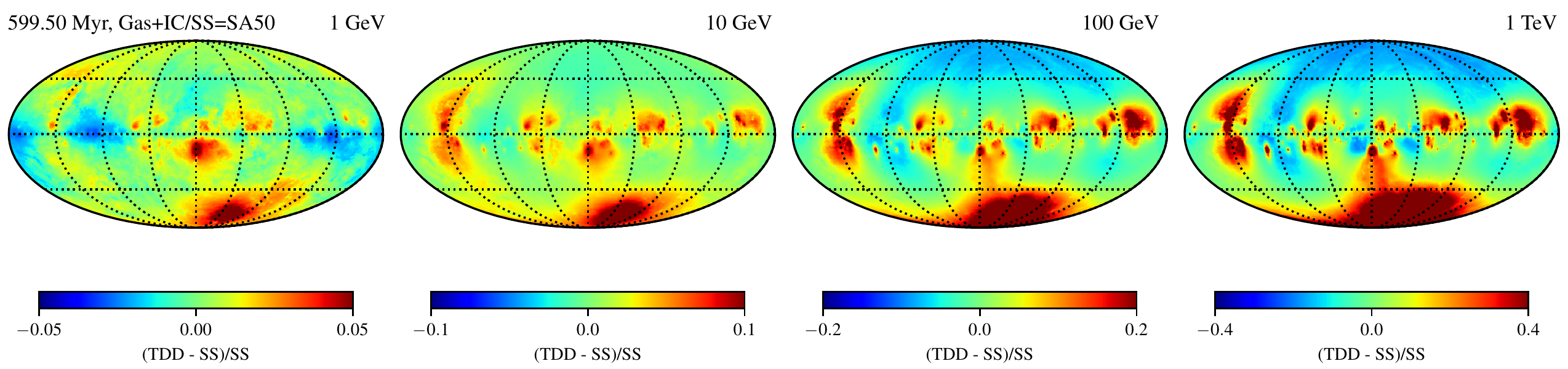}
  \includegraphics[scale=0.7]{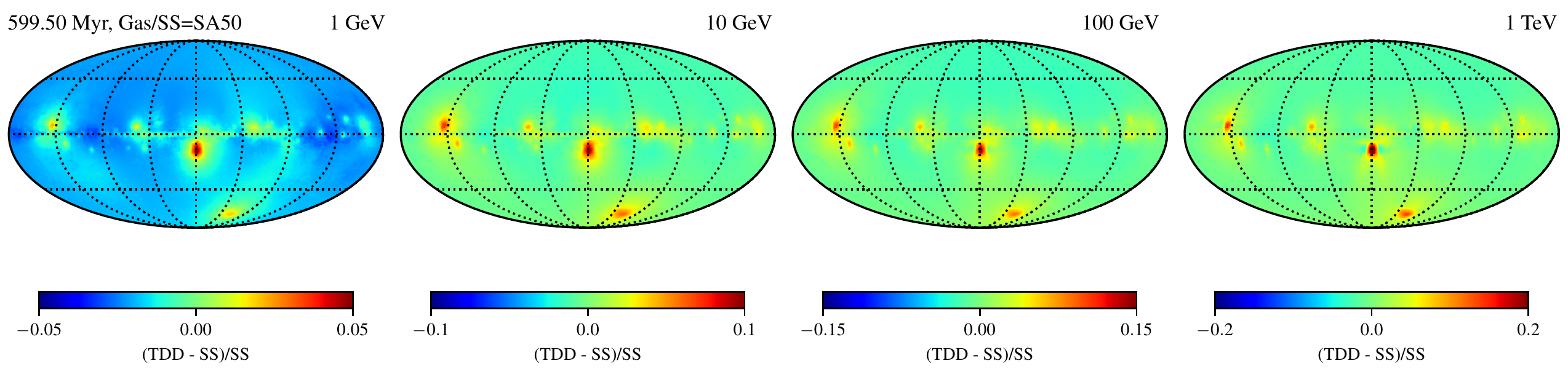}
  \includegraphics[scale=0.7]{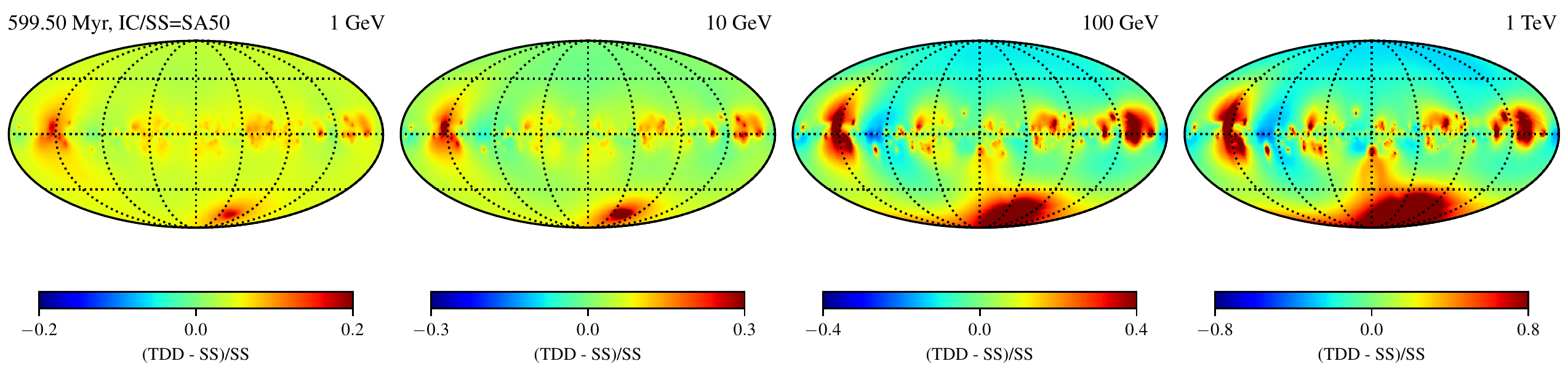}
  \caption{High-energy \gray{} emissions and fractional residuals for the TDD solution at selected energies.
    The first row shows the total all-sky intensities for the TDD solution at 599.5~Myr, corresponding to the CR intensity sample shown in Fig.~\ref{fig:edenfracres}.
    The second row shows fractional residuals at the same energies for total \gray{} intensities for the TDD solution using the SS$_{_{\rm SA50}}$ total \gray{} emissions as the baseline.
    The third and fourth rows show the corresponding fractional residuals for total gas-related ($\pi^0$-decay and bremsstrahlung) and IC emissions using the respective SS$_{_{\rm SA50}}$ predictions (gas-related, IC). The longitude meridians and latitude parallels have $45^\circ$ spacing.
    \label{fig:gammaintensityandfrac}
  }
\end{figure*}

\begin{figure*}[tb!]
  \includegraphics[scale=0.7]{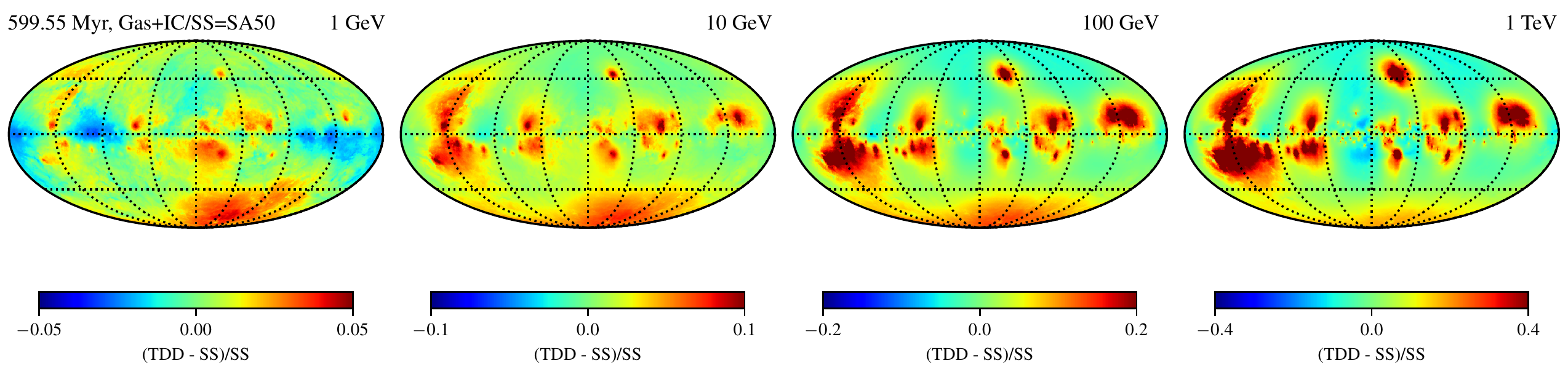}
  \includegraphics[scale=0.7]{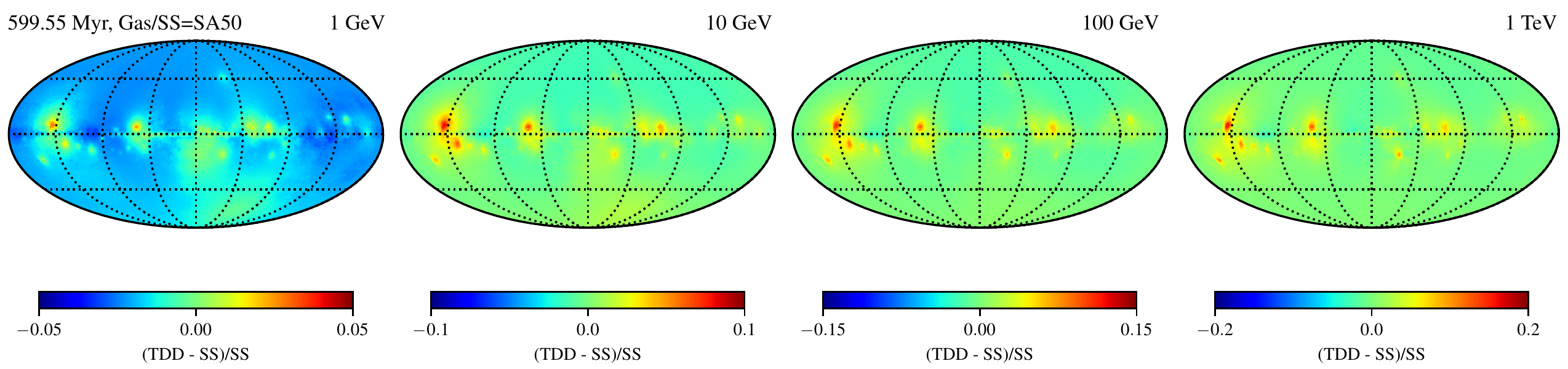}
  \includegraphics[scale=0.7]{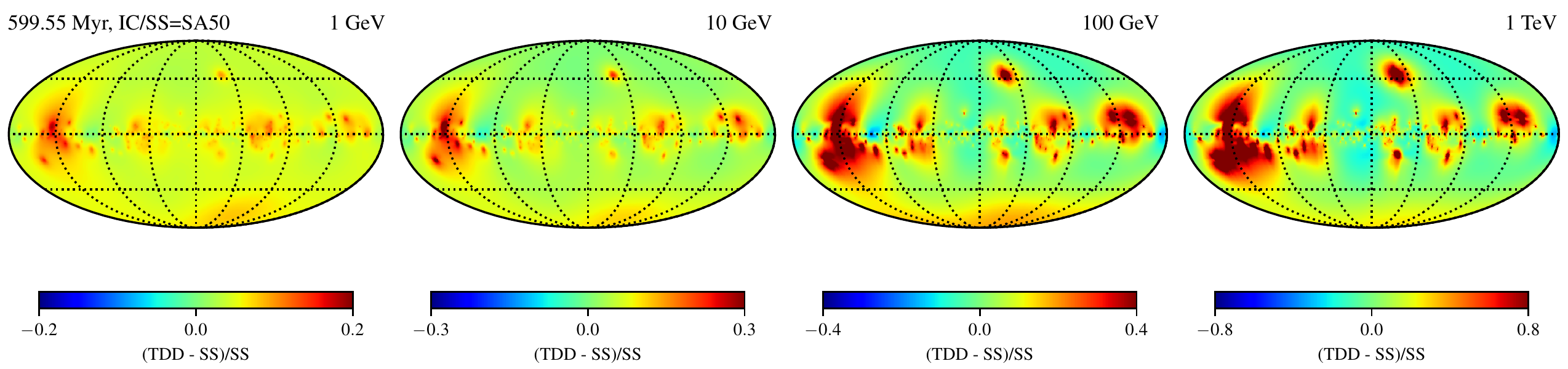}
  \caption{High-energy \gray{} emission fractional residuals for the TDD solution advanced by 50~kyr from those shown in Fig.~\ref{fig:gammaintensityandfrac} at selected energies. 
    The first row shows fractional residuals for total \gray{} intensities for the TDD solution using the SS$_{_{\rm SA50}}$ total \gray{} emissions as the baseline at time sample 599.55~Myr.
    The second and third rows show corresponding fractional residuals for total gas-related ($\pi^0$-decay and bremsstrahlung) and IC emissions using the respective SS$_{_{\rm SA50}}$ predictions (gas-related, IC). The longitude meridians and latitude parallels have $45^\circ$ spacing.
    \label{fig:gammafracadd50kyr}
  }
\end{figure*}

The $\pi^0$-decay intensities show variations with time of $\sim$5--10\% depending on region and energy (the general difference between north/south intensities is due to the distribution of the gas column density out of the Galactic plane).
The IC intensities display much stronger variability ranging from $\sim 10-20$\% around 10~GeV, to $\sim 50$\% at 100~GeV, and much higher at 1~TeV, again depending on the region of the sky averaged over.
The structure of the IC intensity time series is similar to that of the local CR electrons (Fig.~\ref{fig:crtimeseries}, lower right panel), but it is not exactly the same because the \gray{s} do not experience the CR propagation delay.

The asymmetry for the \gray{} emissions from both processes is directional and energy-dependent but is generally more prominent for the IC emissions because the rapid electron energy losses mean that contributions by the individual discretised regions show up more for this component.
An example can be seen for the intermediate-latitude regions left/right of the GC (Fig.~\ref{fig:gammaoutofplanetimeseries} left and middle panels), where the interval $\sim$598.2--599~Myr shows much stronger asymmetry for the north/south of the negative longitude region compared to that of the positive longitude region.
Toward the high-latitude regions, the asymmetry variations correlate to some degree, but not precisely the same as for either of the positive/negative intermediate-latitude regions.
Over all of the regions, the asymmetry can also change its north/south balance, with the \gray{} intensities reflecting the fluctuations about the plane for the CR intensity distributions from the finite sampling of the symmetric continuous source density, as shown in Sec.~\ref{sec:CR}.
This is a novel behaviour exhibited by the TDD solution that is entirely absent for the steady-state case.

The asymmetry due to the discretised injection region modelling is interesting because there has long been an acknowledged north/south difference in the \fermi-LAT data at high Galactic latitudes, where the north polar region is more intense than the south by $\sim$20\% around 1~GeV with the discrepancy increasing to $\sim$30\% at 10~GeV, and $\sim$50\% at 100~GeV albeit with large error bars for the latter \citep[see, e.g., Fig.~13 from][]{2012ApJ...750....3A}.
The \fermi-LAT asymmetry has the reverse sign to the difference between gas column density toward the respective polar regions, and an explanation has proved elusive since its initial finding.
The TDD solution produces broadly distributed hard spectral regions on the sky that are biased differently north/south depending on the CR injection and propagation history about the Galactic plane.

An explanation for the \fermi-LAT north/south asymmetry in terms of discretised CR injection activity/IC emission is an intriguing possibility, which would have additional implications for determination of the so-called `isotropic' \gray{} background.
These background emissions are comprised of extragalactic emissions that are too faint or diffuse to be resolved, as well as residual Galactic foreground emissions that are approximately isotropic.
Analyses of \fermi-LAT data \citep{2010PhRvL.104j1101A,2015ApJ...799...86A} have employed steady-state \GP-based foreground models that are symmetric about the Galactic plane (with the exception of the gas column density distribution) to determine the isotropic background.
The IC intensities at mid-to-high latitudes from these models are a major cause of systematic uncertainty for the Galactic foreground estimation.
Correspondingly, modelling and analysis that does not account for possible discretisation effects may also induce biases in the derived isotropic background properties.

Figure~\ref{fig:gammaintensityandfrac} shows the all-sky intensities and fractional residuals for the TDD solution with SS$_{_{\rm SA50}}$ baseline, corresponding to the sample of CR intensities shown by Fig.~\ref{fig:edenfracres} (599.5~Myr) and also the second-to-last time axis tick mark of the \gray{} intensity time series shown in Fig.~\ref{fig:gammaoutofplanetimeseries}.
Comparing the total intensity\footnote{Note that the IC intensities are relatively higher because, as noted earlier, the contribution by CR He nuclei is not included for the $\pi^0$-decay emissions.} panels, only modest differences with increasing energy are evident $\lesssim 100$~GeV.
For higher energies, the distribution about the plane for the emissions becomes narrower, with individual regions appearing more prominent, particularly at high latitudes.
A clearer picture of the effect of the discretised regions is given by the fractional residuals, which are shown for the total intensities by the second row of Fig.~\ref{fig:gammaintensityandfrac} at the same energies as the first row.
The fractional residuals also for gas-related ($\pi^0$-decay and bremsstrahlung) and IC scattering are also separated (third and fourth rows) to enable visualisation of the effect of the TDD solution for the different processes/ISM components.
 
The gas-related emission residuals generally display low-level variations about the Galactic plane of $\sim$5--10\%, with stronger fluctuations appearing at higher energies.
There are also highly localised areas on the sky with larger enhancements of the \gray{} intensity, e.g., $\sim$$5^\circ-20^\circ$ directly below the GC and at high latitudes toward the south polar region.
For the IC intensities, the TDD fractional residuals are much more highly structured and exhibit stronger variations compared to the steady-state emissions for all energies; meanwhile, they also have some similarity to the gas-related residuals.

The relationship of the individual process residuals to the differing 3D CR energy density distributions for the TDD and SS$_{_{\rm SA50}}$ solutions is non-trivial, particularly for directions about and through the GC where the integration path length through the Galaxy is maximised.
For example, the localised feature $\sim$$5^\circ-20^\circ$ below the GC seen for the gas-related residuals is mainly due to a nearby recently active region that is $\lesssim$1~kpc distant and located $\sim$100--200~pc below the plane (see $Z$ slices in panels of Fig.~\ref{fig:edenfracres}).
The low-energy spatial distribution of this gas-related residual reflects the contribution by enhanced bremsstrahlung and $\pi^0$-decay from this nearby region.
Because the contribution by bremsstrahlung decreases significantly for higher energies, the distribution on the sky becomes smaller, because only the $\pi^0$-decay \gray{s} are dominating the gas-related residuals.

Further beyond the nearby region along the same line of sight, and hence further from the Galactic plane, there are also some over- and under-densities for the CR intensities that result in corresponding variations for the \gray{} emissivity distributions (not shown).
However, the scale height of the neutral gas is $\sim$100--200~pc, causing a fairly rapid decrease of its density away from the plane, which significantly reduces the effects on the \gray{} intensity of $\pi^0$-decay and bremsstrahlung emissivity fluctuations at larger distances above/below the Galactic plane.

Meanwhile, for the same general direction, the IC residuals are structured but less dominated by any localised spot, except for the higher energies ($\sim 1$~TeV).
For a given \gray{} energy, the IC contribution is produced by electrons with energies of a factor $\sim$few to 10 times higher (because of the energy integration over the ISRF spectral intensity for the IC emissivity) than those contributing to the bremsstrahlung emissions component.
The higher-energy electrons of the TDD solution have stronger intensity fluctuations that produce an IC emissivity distribution with correspondingly higher variations.
In addition, the ISRF scale height is much larger than that of the gas \citep[see Sec.~3.2 of][]{2017ApJ...846...67P}, so a longer integration path length for a given line of sight effectively contributes to the IC intensity.
Consequently, for longitudes $\lesssim$$5^\circ-20^\circ$ about the GC there is a kind of averaging when integrating along the line of sight from directions close to the plane up to intermediate latitudes that produces a smoother IC intensity than might be expected from simple examination of the CR electron intensity $Z$~slices (Fig.~\ref{fig:edenfracres}).
The appearance of the nearby injection region in the highest-energy \gray{} IC residuals is due to the very strong over-density for the CRs from it compared to the steady-state solution, which is standing out over the line of sight smoothing that still occurs, even for directions out of the plane.

Elsewhere along and about the Galactic plane, the line-of-sight path lengths that prove challenging for interpreting the residuals for directions about the GC are reduced.
Enhancements by individual regions are more easily recognised, but their spatial characteristics are complicated because of the variable compensation by the over- and under-emissive regions, even with the reduced integration path lengths.
For example, about the Galactic plane toward $|l|$$\sim 45-60^\circ$, there are enhancements for both the gas-related and IC intensities.
They are due to relatively close ($\lesssim$2~kpc) injection regions that appear extended because of their proximity and the particle propagation.
%and the particle propagation, and appear significantly extended because of their proximity.

However, the spatial distribution on the sky of the residuals does not appear as a singular continuous excess.
The morphology is more complex because the individual injection regions are not all located at the same distance, and the integration path lengths near the plane are affected more by the line-of-sight smoothing than for directions that are toward higher latitudes.
This is also the case for the prominent residual features toward the outer Galaxy, which are also due to injection regions in the direction toward $|l|$$\sim 130^\circ-150^\circ$ that are also very close ($\lesssim$0.5--1.5~kpc, see Fig.~\ref{fig:edenfracres}).
For all of these residuals the sizes on the sky are a combination of the originating injection region proximity, the time since they became active, and the time-dependent extent of the corresponding CR intensity distribution from the propagation and energy losses.

To illustrate the time-dependent effects on the \gray{} emissions, Fig.~\ref{fig:gammafracadd50kyr} shows the fractional intensity residuals for the TDD solution advanced by 50~kyr compared to those shown by Fig.~\ref{fig:gammaintensityandfrac}; this time slice corresponds to a fraction 0.2 beyond the second-to-last tick mark in Fig.~\ref{fig:gammaoutofplanetimeseries}.
For the high-latitude southern directions, which were dominated by the recent activity by a nearby injection region for the 599.5~Myr time slice, the majority of the highest-energy emissions are now dissipated, with the lowest energies showing the strongest over-intensity compared to the steady-state solution.
(Indeed, comparison with the \gray{} intensity time series in the right-most panel of Fig.~\ref{fig:gammaoutofplanetimeseries} shows a dramatic drop of the southern region IC intensity.)
This shift of the most prominent residuals to lower energies is an indication of injection activity cessation and the evolution of the remnant CRs as they lose energy and propagate away from the formerly active region.

Interestingly, a high-latitude northern injection region has become active in the interval since the 599.5~Myr time slice, and its residuals are prominent for the total intensity and IC fractional residuals.
They are also present for the gas-related residuals, but to a lesser degree.
Examination of the time history for the injection regions shows that it became active $\sim$30~kyr earlier with its centre $\sim$0.7~kpc outside the Galactic plane, and the corresponding CR energy density distribution at 599.55~Myr (not shown) is still relatively closely distributed about the injection region.
Because the gas density decreases quickly, its effect on the gas-related (and their contribution to the total) intensity residuals is minor given its height above the plane.
Meanwhile, the other nearby injection regions that were producing the strongest over-intensities about the Galactic plane for the 599.5~Myr time slice remain active at 599.55~Myr.
Their corresponding residuals are still present and display, at this time slice, the characteristics that can be expected for ongoing particle injection and propagation near the active sites.

\subsection{Synchrotron Radiation}
\label{sec:synchrotron}

\begin{figure*}
  \subfigure{
  \includegraphics[scale=0.6]{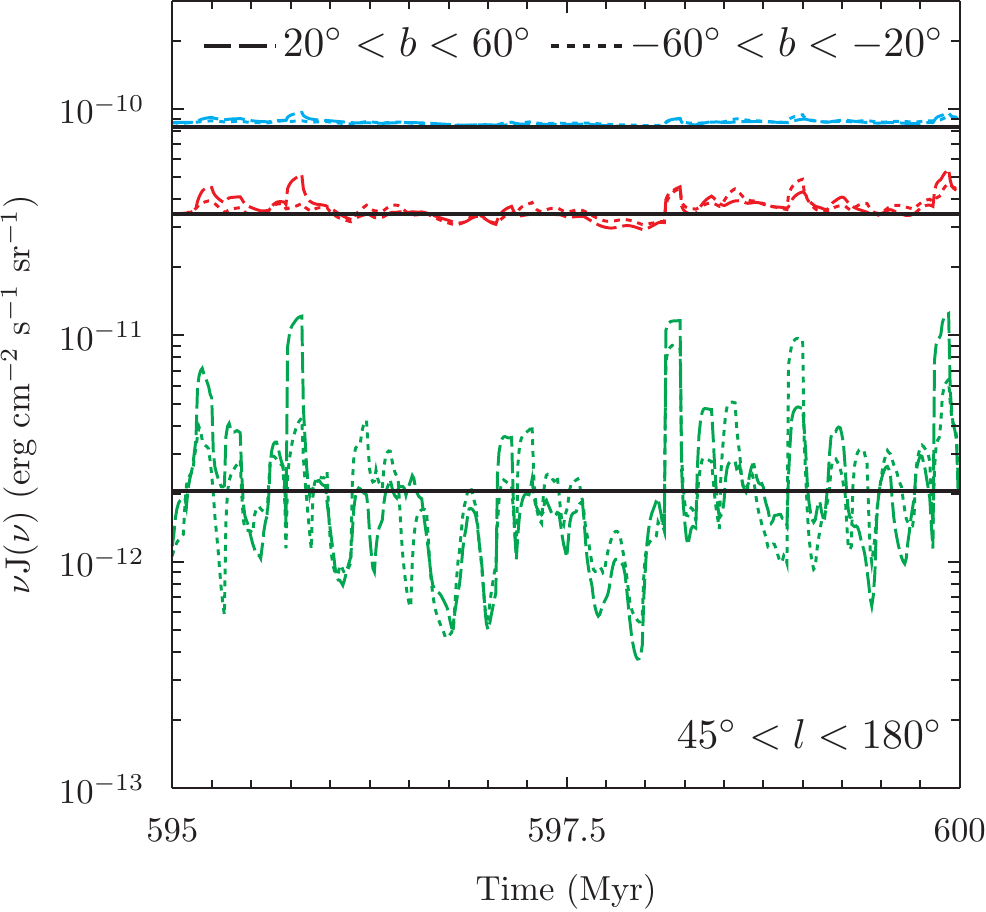}
  \includegraphics[scale=0.6]{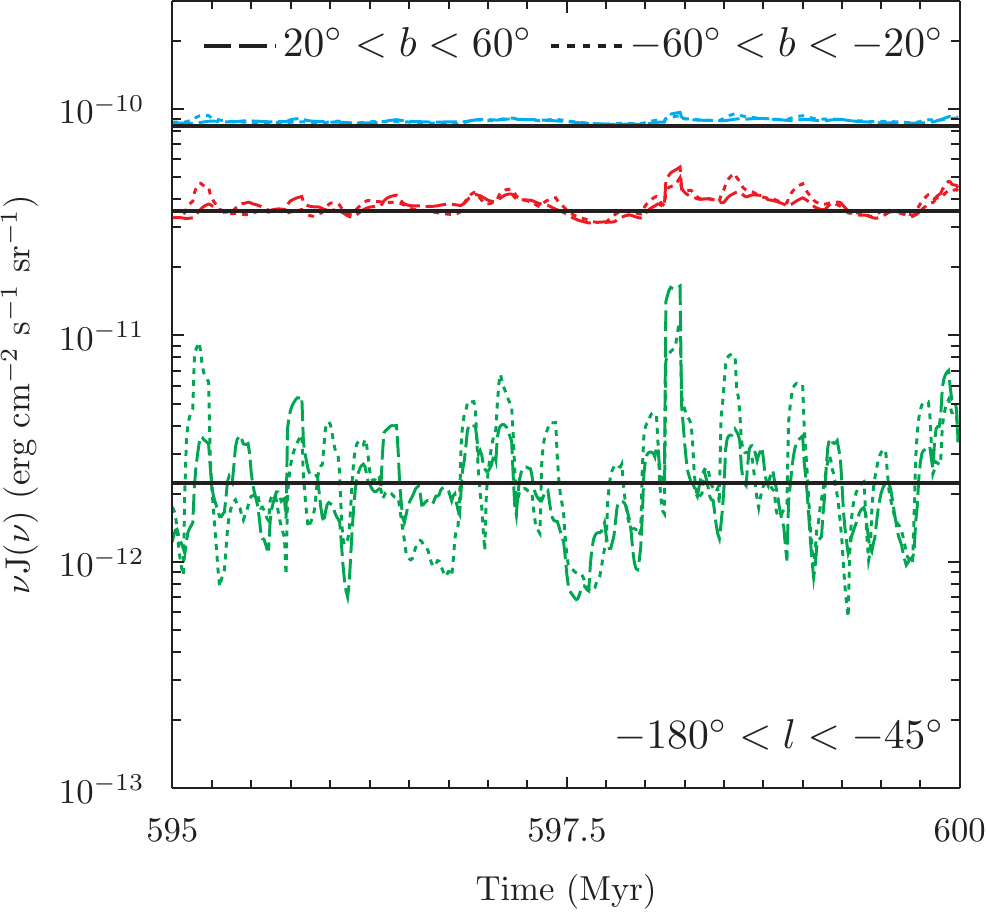}
  \includegraphics[scale=0.6]{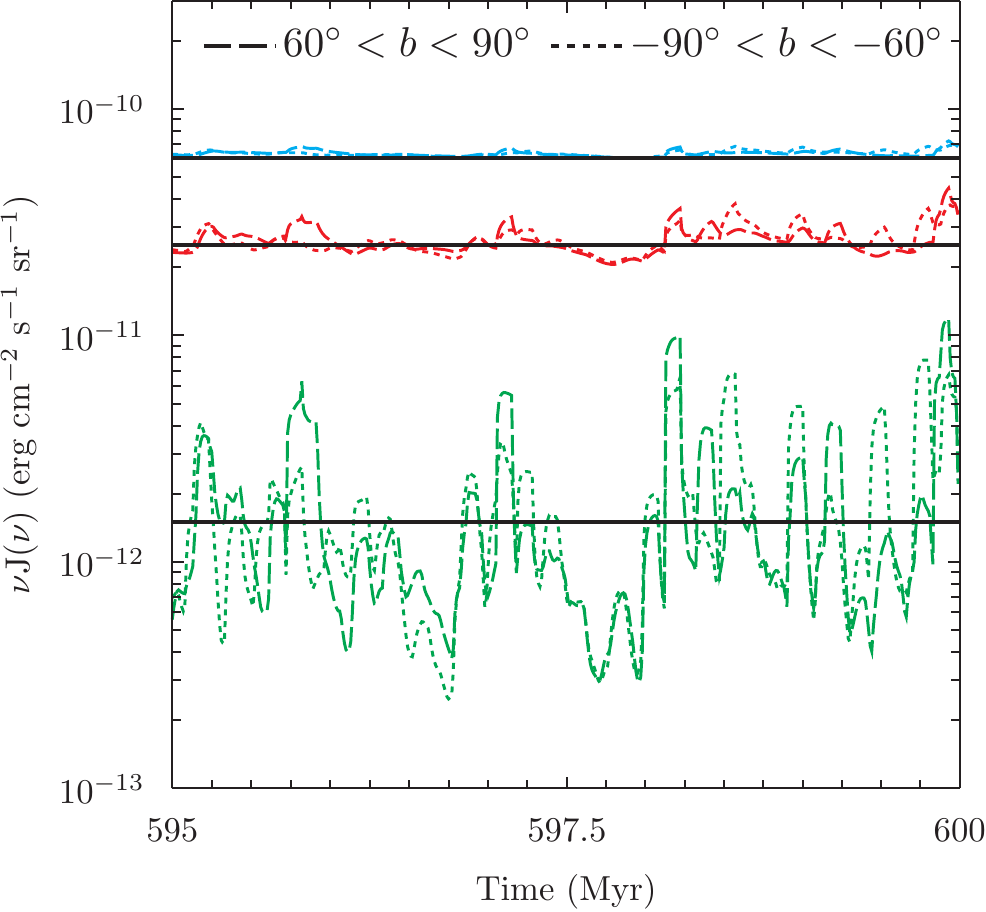}
  }
  \caption{Time series of synchrotron emissions at 102.4~GHz (cyan), 13.1~THz (red), and 6.7~PHz (green) for (left) $45^\circ < l < 180^\circ$ and (middle) $-180^\circ < l < -45^\circ$ ranges for intermediate latitudes above/below the plane, and (right) north/south Galactic polar regions $|b| > 60^\circ$.
    The time range is for the last 5~Myr of the TDD simulation epoch and has 10~kyr sampling.
    This range and sampling is the same as that shown by the Fig.~\ref{fig:gammaoutofplanetimeseries} and the right panels of Fig.~\ref{fig:crtimeseries}, respectively.
    The thin solid black line for each frequency shows the corresponding SS$_{_{\rm SA50}}$ intensity, which is symmetric above/below the Galactic plane.
  }
\label{fig:syncoutofplanetimeseries}
\end{figure*}

As for the \gray{s}, the spatial grid used for the CR propagation is also employed for the synchrotron emissivity calculations.
The synchrotron intensity maps at the solar system location are obtained by line-of-sight integration, where the emissivities are determined over a logarithmic frequency grid from 100~MHz to 241.5~PHz with 32 planes. 
Free--free absorption is accounted for in the line-of-sight integration following the method described by \citet{2013MNRAS.436.2127O} with an electron temperature $T_e = 7000$~K and the \hii{} distribution used for the propagation calculations (Sec.~\ref{sec:setup}).

Figure~\ref{fig:syncoutofplanetimeseries} shows the time series of the synchrotron intensities for the TDD solutions at 102.4~GHz, 13.1~THz (mid-infrared), and 6.7~PHz (extreme ultraviolet).
The corresponding SS$_{_{\rm SA50}}$ intensities at the same frequencies are shown as the solid black lines.
The frequencies are chosen so that the energies of the CR electrons\footnote{Using the approximation that the radiation is emitted about the synchrotron critical frequency for an $\sim$5~$\mu$G magnetic field in the ISM, which is a typical average field strength for the model used for this paper.} producing them are approximately the same as of those for the \gray{} intensity time series shown for the same regions of sky in Fig.~\ref{fig:gammaoutofplanetimeseries}.
It should be no surprise then, even though the frequency/energy correspondence is approximate, that there is a strong similarity between the synchrotron and IC time series.

The variability due to the CR injection region activity and propagation outside the Galactic plane appears for the emissions for both processes, providing a possible correlative test for high-energy asymmetries in the \gray{} data.
The predicted synchrotron intensities at soft X-ray ($\sim0.1$~keV) energies for the discretised model about the inner Galactic plane are $\sim 10-20$~keV cm$^{-2}$ s$^{-1}$ sr$^{-1}$, and decreasing toward higher Galactic latitudes.
Cosmic X-ray background estimates based on a combination of the hot Galactic plasma and a hard-spectrum extragalactic component \citep[e.g.,][]{2002A&A...389...93L,2018A&A...617A..92C} suggest fluxes $\sim$100~keV cm$^{-2}$ s$^{-1}$ sr$^{-1}$ at similar photon energies.
These are somewhat higher than obtained for the CR-induced emissions, but not orders of magnitude.
Because the predicted soft X-ray fluxes are produced by electrons with energies toward the upper bound of the spectral model used for the CR injection regions, the numerical cutoff can result in an artificially low prediction.
Improved modelling to higher energies and finer spatial resolutions than considered in this paper likely will result in larger intensities around $\sim$1~keV X-ray energies.

\begin{figure*}[htb!]
  \includegraphics[scale=0.7]{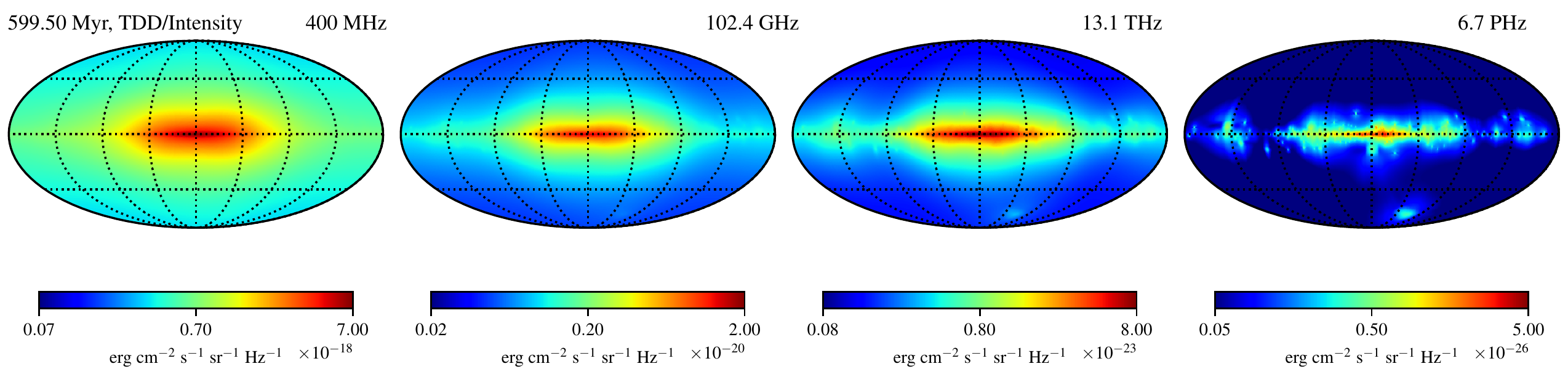}
  \includegraphics[scale=0.7]{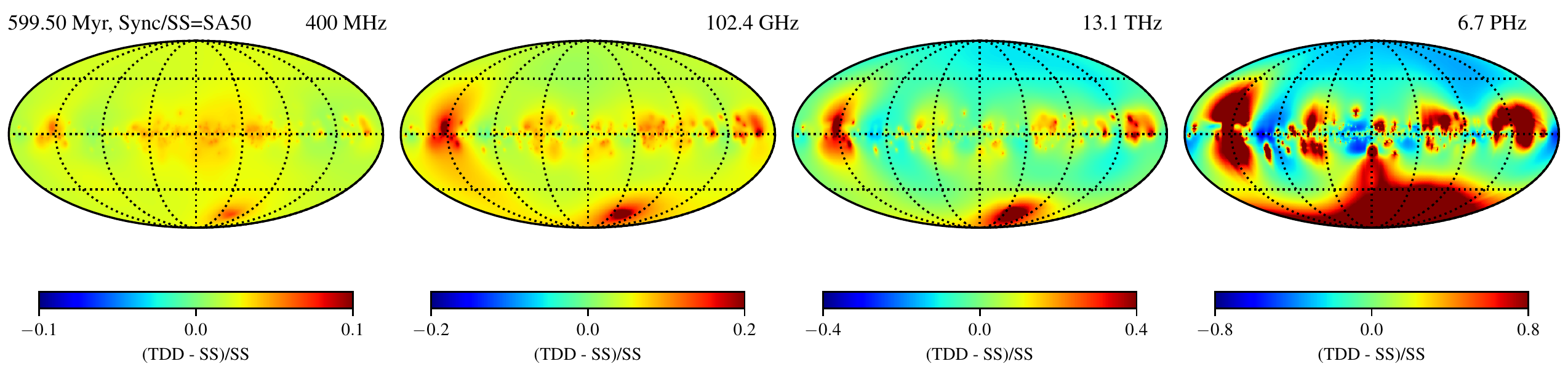}
\caption{Synchrotron radiation intensity and fractional residuals for the TDD solution at selected frequencies.
    The top row shows all-sky intensities for the TDD solution at 599.5~Myr, corresponding to the CR intensity sample shown in Fig.~\ref{fig:edenfracres}.
    The bottom row shows fractional residuals for the TDD solution using the SS$_{_{\rm SA50}}$ synchrotron emissions as the baseline.
    The longitude meridians and latitude parallels have $45^\circ$ spacing. 
    \label{fig:syncintensityandfrac}
  }
\end{figure*}

\begin{figure*}[htb!]
  \includegraphics[scale=0.7]{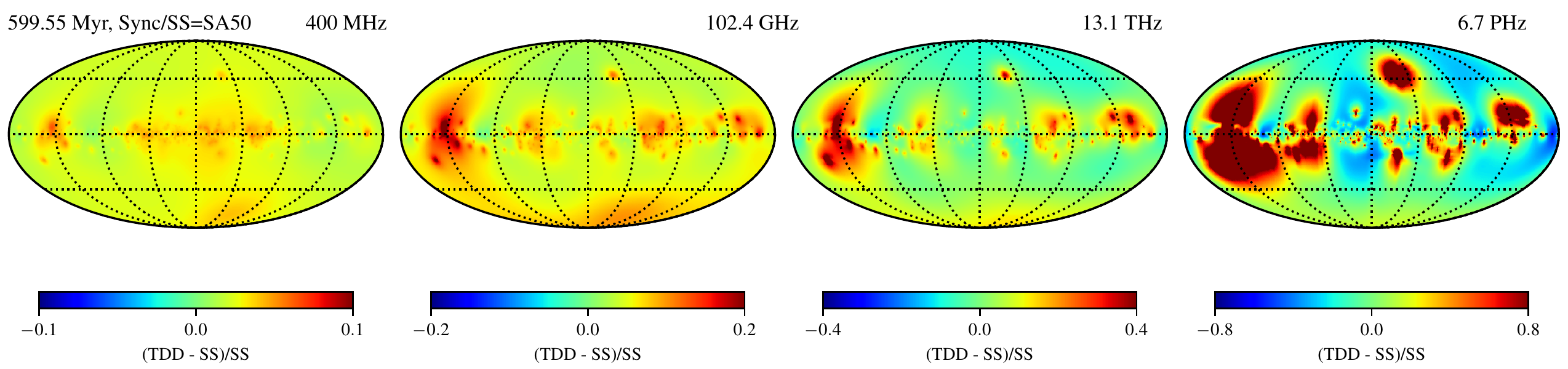}
  \caption{Synchrotron radiation fractional residuals for the TDD solution advanced by 50~kyr from those shown in Fig.~\ref{fig:syncintensityandfrac} at selected frequencies. 
    The longitude meridians and latitude parallels have $45^\circ$ spacing.
    \label{fig:syncfracadd50kyr}
  }
\end{figure*}

The all-sky view of the synchrotron emissions is given by Fig.~\ref{fig:syncintensityandfrac}.
The top row shows the predicted intensity for the TDD solution at selected frequencies for the same time sample as Fig.~\ref{fig:gammaintensityandfrac} (599.5~Myr).
The bottom row shows the corresponding fractional residuals with the SS$_{_{\rm SA50}}$ solution.
The lowest frequency shown in Fig.~\ref{fig:syncintensityandfrac} is chosen following the approximate relationship between CR electron energies\footnote{For the $\sim$1~GeV \gray{s} both bremsstrahlung and IC are contributing at approximately the same order of magnitude to the intensity, particularly about the Galactic plane (see top and middle row panels of Fig.~\ref{fig:gammaspec}).
Recall that for a given \gray{} energy the corresponding CR electron energies for bremsstrahlung are a factor of about a few higher. 
For IC emissions the CR electron energies are over a range factors of few to 10 higher because of the integration over the ISRF spectral energy density for the IC emissivity.} producing the $\sim$1~GeV IC \gray{s} and synchrotron emissions, as employed for the highest three frequencies whose time series were shown in Fig.~\ref{fig:syncoutofplanetimeseries}.

The intensity maps up to the 100s of GHz frequencies appear less featured than higher frequencies, because the line-of-sight integration effectively reduces any fluctuations of the synchrotron emissivity caused by the low-level intensity distribution spatial variations of the $\lesssim$10s of GeV CR electrons producing them (e.g., lower left panel of Fig.~\ref{fig:edenfracres}).
The higher-frequency all-sky intensities shown in the figure display apparent structure that is directly related to the current snapshot of discretised injection region activity. 
It is a clearer picture of the current (at 599.5~Myr) injection and propagation of the CR electrons than available through the approximate equivalent energy \gray{} intensities (Fig.~\ref{fig:gammaintensityandfrac}, first row), because the inclusion of interactions of CRs with the interstellar gas obscures the view obtained via the latter.
The higher-frequency synchrotron emissions, in fact, show a strong departure from the smooth intensity maps predicted for a steady-state model, being the result of the numerous recent and currently active injection regions.

The fractional residuals show, again not unexpectedly, similar features across most of the frequency range to those for the IC component shown earlier.
The correspondence is not precise because, as already mentioned, the frequency/energy mapping is not exact, and the spatial distributions of the ISM targets differ: the spiral pattern for the R12 ISRF is not the same as that for the PBSS magnetic field, and the ISRF energy density is generally higher toward the inner Galaxy than that of the large-scale magnetic field.
For the higher frequencies, these different ISM target distributions do not produce a strong effect because the line-of-sight integration is effectively smoothing under- and over-emissive region fluctuations, leaving only the over-intensities coming from the nearby injection region activity standing out in the residuals, as already discussed in Sec.~\ref{sec:gammaray}.

For the lowest frequency (400~MHz) shown, the emissions are coming from CR electrons with energies of $\sim$1 to a few~GeV energies, which are the particles producing mostly bremsstrahlung \gray{s} contributing to the lowest-energy \gray{} intensities/residuals shown above.
For the $\sim$few~GeV CR electrons, the TDD intensity distribution is not completely smooth, but the small fluctuations are effectively always higher than the SS$_{_{\rm SA50}}$ intensity distribution across the Galactic disk.
This produces the broadly distributed enhancement for the lower-frequency synchrotron residuals about the inner Galaxy.
There is some hint of similar structure for the gas-related residuals for the 1~GeV \gray{s} over the same region (Fig.~\ref{fig:gammaintensityandfrac}, third row), but it is difficult to definitively associate it with bremsstrahlung emission because of the much brighter $\pi^0$-decay component.

Figure~\ref{fig:syncfracadd50kyr} shows the fractional residuals for the TDD solution for synchrotron radiation advanced by 50~kyr beyond those shown by Fig.~\ref{fig:syncintensityandfrac}, as already shown for the evolution of the \gray{s} in Figs.~\ref{fig:gammaintensityandfrac} and~\ref{fig:gammafracadd50kyr}, respectively. 
The highest-frequency residuals show very similar evolution to the higher-energy \gray{s}, with the cessation of injection activity by the high-latitude southern region that formerly dominated that area of the sky at 599.5~Myr (with the attendant shifting of the remnant emissions to lower frequencies/energies), as well as the appearance of the newly active region at the high northern latitudes that was discussed earlier.

The existence of a number of loops and spurs observed in radio and polarised microwave emission that are covering large regions on the sky \citep{2015MNRAS.452..656V} is perhaps the most direct evidence of the presented picture.
These loops are likely very old shells of local SNRs or walls of the Local Bubble cavity that are essentially blended into the ISM.
Their angular sizes are very large and can extend up to $\sim80^\circ$ on the sky.
The fact that they are still observed via their synchrotron emissions means that they continue to accelerate electrons to moderately relativistic energies.
They were active injection regions $\gtrsim$50--100~kyr ago that at the earlier times may have produced more intense emissions than presently visible.
A similar picture could, in principle, be observed by taking any direction, but the contributions by individual old shells are not bright enough to be separately distinguished in the intensity maps \citep[][]{2013JCAP...06..041M}.
However, a number of such weak shells may still produce a low-intensity extended component of the radio-to-\gray{} interstellar emissions.
This component may be partially responsible for the residual emissions that are not accounted for by the standard steady-state models.

Toward the inner Galaxy, the relative invariance of the residuals is not unexpected because the time scale for the particle intensities to change for the corresponding energies is much longer than 50~kyr, and the injection lifetime (100~kyr) means that about half of the many regions contributing are still active.
Toward the outer Galaxy, the contributions by the activity of nearby injection regions are also evident, as for the \gray{s}, indicating that the broadband synchrotron emissions are an effective tracer of the time/space-localised injection activity, as well as the accumulation of CR electrons about the Galactic disk on long time scales.

The complementarity in particular of the highest-frequency synchrotron emissions and $\sim$1~TeV \gray{s} is consistent with the predictions of \citet{1997MNRAS.291..162A} of broadband ``halos'' produced by $\gtrsim$1~TeV energy electrons interacting with the interstellar magnetic and radiation fields about individual sources.
The so-called ``TeV haloes'' suggested as a new source class of very high energy \gray{} emitters \citep[e.g.,][]{2019PhRvD.100d3016S} could also have counterpart emissions detectable at X-ray energies with the new generation of all-sky observations.

\section{Discussion}
\label{sec:discussion}

\begin{figure*}[tb!]
  \includegraphics[scale=0.7]{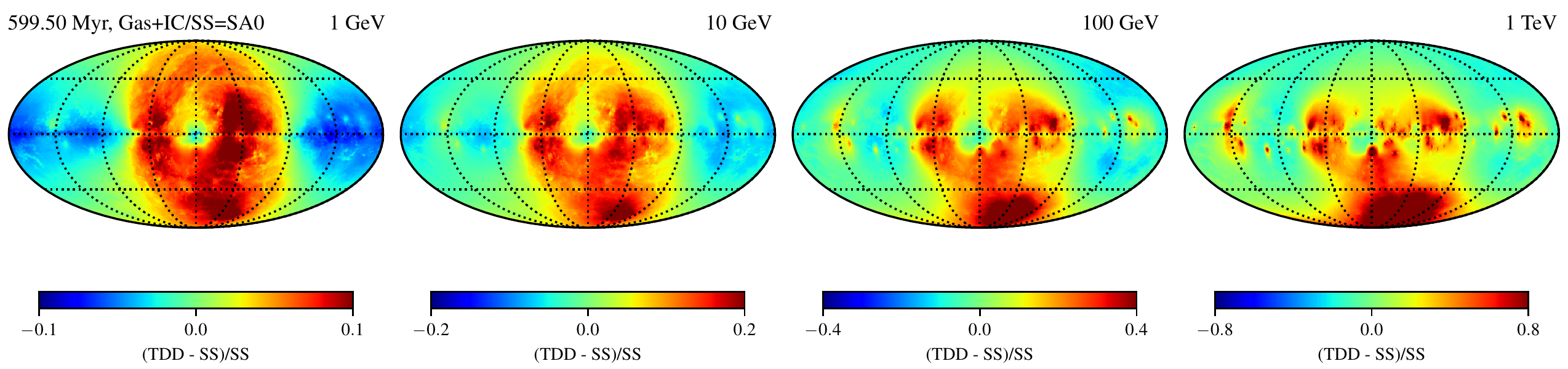}
  \includegraphics[scale=0.7]{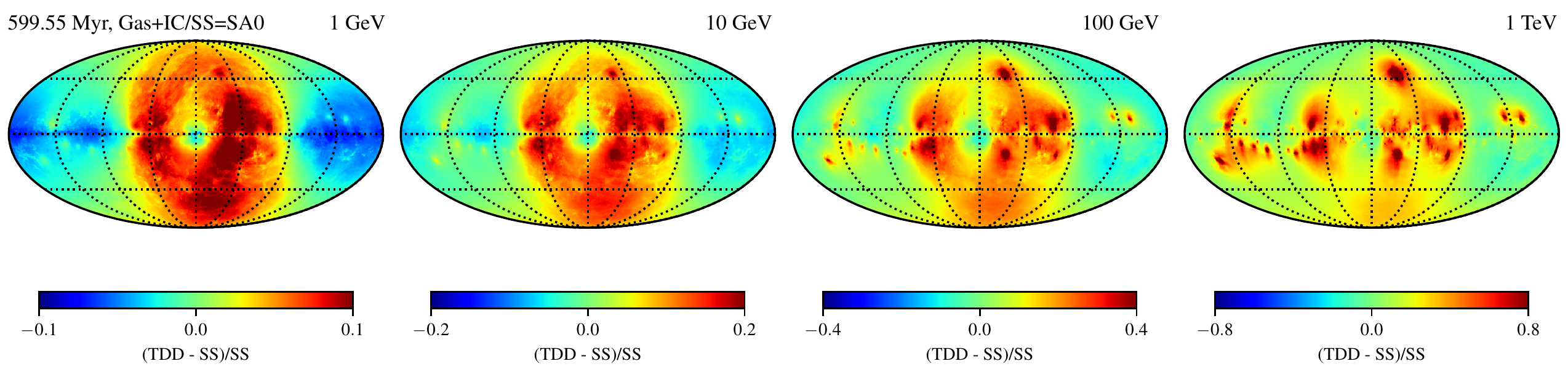}
  \caption{Total \gray{} fractional residuals at selected energies for the TDD solution at time samples 599.5 (top row) and 599.55 (bottom row) Myr using the SS$_{_{\rm SA0}}$ intensities for the baseline/reference prediction.
The longitude meridians and latitude parallels have $45^\circ$ spacing.}
  \label{fig:gammatotalfracsa0}
\end{figure*}

\begin{figure*}[tb!]
  \includegraphics[scale=0.7]{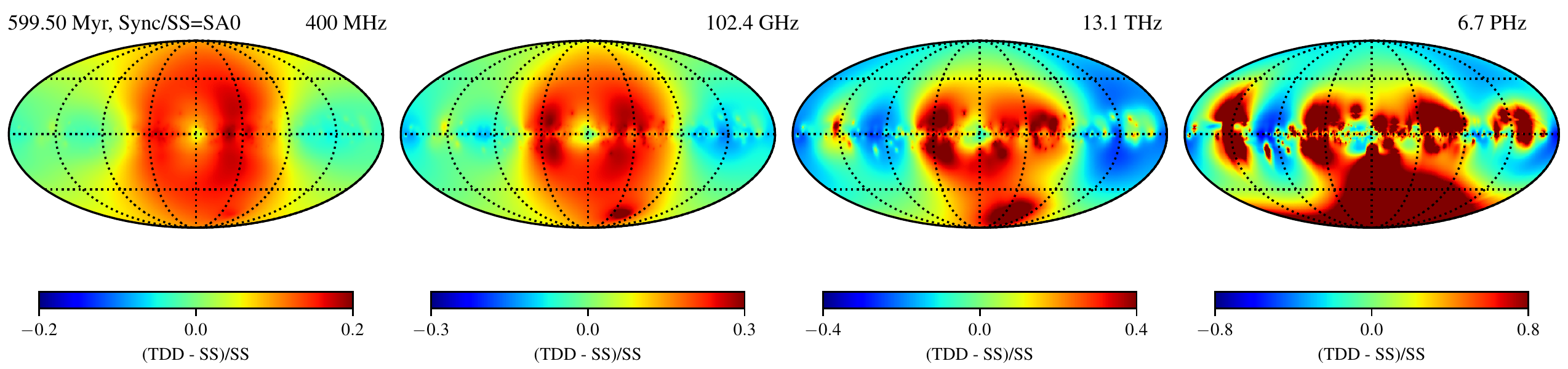}
  \includegraphics[scale=0.7]{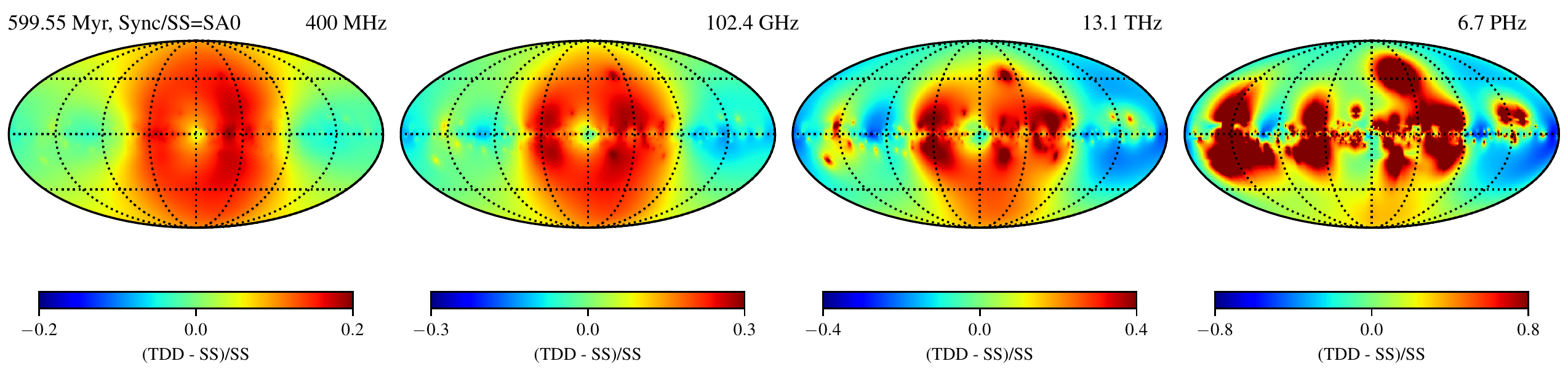}
  \caption{Synchrotron fractional residuals at selected frequencies for the TDD solution at time samples 599.5 (top row) and 599.55 (bottom row) Myr using the SS$_{_{\rm SA0}}$ intensities for the baseline/reference prediction.
The longitude meridians and latitude parallels have $45^\circ$ spacing.}
  \label{fig:syncfracsa0}
\end{figure*}

The differences between predictions made with the space/time-discretised and equivalent continuous/steady-state models are non-trivial.
The TDD solution produces CR intensities that display localised energy-dependent intensity fluctuations with time that extend over the entire Galactic disk and out of the plane.
For the local region to the solar system and energies $\lesssim$100~GeV, where the normalisation to CR data is made, the TDD solution intensities are close to those of steady state, apart from small upward fluctuations due to relatively nearby injection regions.
This shows a reasonable consistency with the prevailing picture that, at least for CR nuclei, the data $\lesssim$100~GeV energies are representative of the CR sea and not strongly dominated by local injection effects.

For higher energies, the discretised model intensities fluctuate about the corresponding steady-state solution, with the CR electron intensities having larger fluctuations than protons because they cool more rapidly.
In general, these results are qualitatively consistent with results obtained by other works employing a variety of stochastic source modelling configurations \citep[e.g.,][]{2005ApJ...619..314B,2006AdSpR..37.1909P,2012JCAP...01..010B,2018JCAP...11..045M}.
The new development made with this paper is the modelling of the associated non-thermal emissions (\gray{s}, synchrotron radiation) for space/time-discretised CR injection and propagation.

Deciphering the information contained in the non-thermal interstellar emissions, particularly for the residual sky maps, relies foremost on a correct determination of an equivalent steady-state/continuous source density approximation.
However, this is not known at all, and the 2D axisymmetric source density is often used as a basis for generating interstellar emission models (IEMs) for actual all-sky data analysis \citep[for examples for \gray{} analysis, see, e.g.,][]{2012ApJ...750....3A,2016ApJ...819...44A,2019ApJ...880...95K}.
If such an a priori  source density and corresponding steady-state solution is employed, the residual all-sky maps will encode multiple effects: signatures of the individual regions, together with the mismodelling between the assumed and actual (smooth) model that is the mathematical description for the large-scale discretised region distribution.

To illustrate for the multi-wavelength emissions, Fig.~\ref{fig:gammatotalfracsa0} shows total \gray{} fractional residuals for the 599.5 (top row) and 599.55 (bottom row) Myr snapshots at selected energies, and Fig.~\ref{fig:syncfracsa0} shows total synchrotron emissions fractional residuals for the 599.5 (top row) and 599.55 (bottom row) Myr snapshots at selected frequencies.
Both sets of figures use the SA0 CR source density distribution for the steady-state baseline.
The SA0 density model is a 2D galactocentric axisymmetric distribution of the Galactic pulsar population given by \citet{2004A&A...422..545Y} and has been used in the work described by \citet{2012ApJ...750....3A} and \citet{2016ApJ...819...44A}.
For this paper, the propagation parameters for the SA0 density distribution are obtained from the work by \citet{2019ApJ...879...91J}, who employed the same tuning procedure and data described in Sec.~\ref{sec:setup}.
The local CR spectra determined for the SA0 CR source density model agree with the data, as well as those for the SA50 model that was used for the discretised/steady-state comparison earlier in this paper. 

For the \gray{s} (Fig.~\ref{fig:gammatotalfracsa0}), it can be easily seen that the characteristics of the discretised solution identified for the same time epochs in Figs.~\ref{fig:gammaintensityandfrac} (second row) and~\ref{fig:gammafracadd50kyr} (top row) are significantly altered using the SA0 steady-state baseline.
For the lowest-energy (1~GeV) residuals, the ``incorrect'' steady-state baseline effectively erases any features of the discretised SA50 solution, particularly toward the GC and nearby Galactic plane, and instead there is a broad ``doughnut-like'' over-intensity band surrounding the inner Galaxy extending to high latitudes.
The origin of this feature has been discussed at length both by \citet{2017ApJ...846...67P} and \citet{2018ApJ...856...45J}, being due to the difference in number and intensities of the CRs injected in and about the spiral arms for the SA50 distribution, compared to the SA0 model.
Its spatial characteristics are also relatively invariant over the 50~kyr covered by the upper and lower rows of Fig.~\ref{fig:gammatotalfracsa0}, even though the corresponding total \gray{} residuals in Figs.~\ref{fig:gammaintensityandfrac} and~\ref{fig:gammafracadd50kyr} show features associated with localised emissions from individual injection regions.

The erasure of most features of the discretised solution persists up to $\sim$100~GeV \gray{} energies, with only the prominent high-latitude southern region at 599.5~Myr evident.
But it is still somewhat embedded in the residuals due to the mismatch of the density distributions.
Only for $\sim$1~TeV energies do at least spatial localisation features of the discretised solution for the different time slices become apparent.
However, the spatial distribution on the sky of the residuals is still significantly different due to the mismatched smooth density models.

For the synchrotron emissions (Fig.~\ref{fig:syncfracsa0}), the situation is similar where the use of the SA0 steady-state model for the baseline effectively erases features of the discretised solution.
Because the \gray{s} combine multiple processes and ISM target distributions, while the synchrotron emissions are tracing only the CR electrons and magnetic field, the latter have potential for probing the underlying large-scale distribution for the CR electron sources. 
The difference in the sky distribution of the residuals over the $\sim$100~MHz to $\sim$100~GHz range indicates that there is a sensitivity to both the spectral content and spatial distribution of the underlying CR electron source density model that is invariant over the 50~kyr time scale spanned by the upper and lower rows in the figure.
As for the \gray{s}, the higher-frequency emissions also begin to display features of the discretised solution but with the same issue, where the spatial distribution on the sky of the residuals has the additional contribution from the mismodelled smooth density distribution.

\begin{figure*}[tb!]
  \includegraphics[scale=0.7]{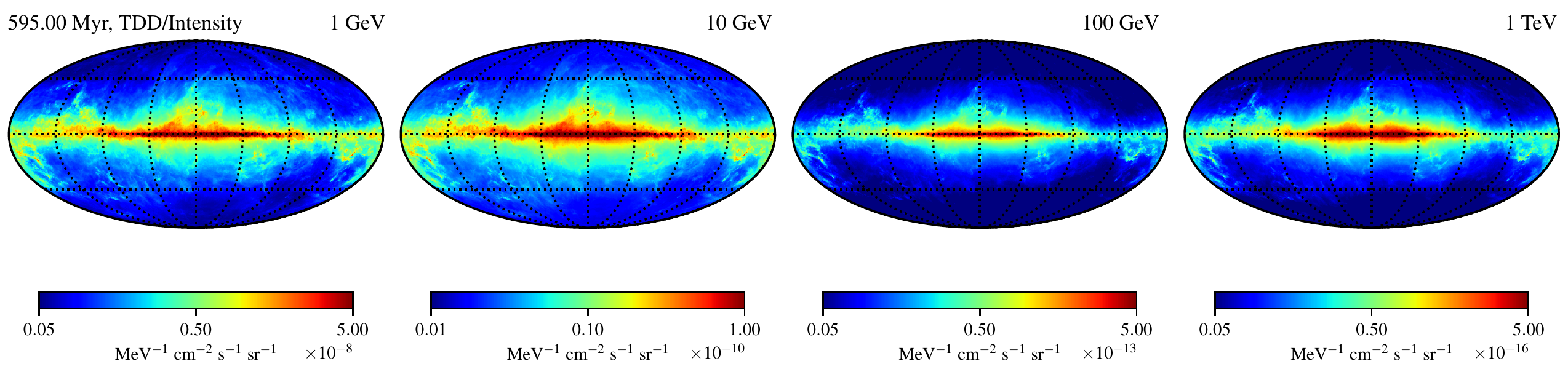}
  \includegraphics[scale=0.7]{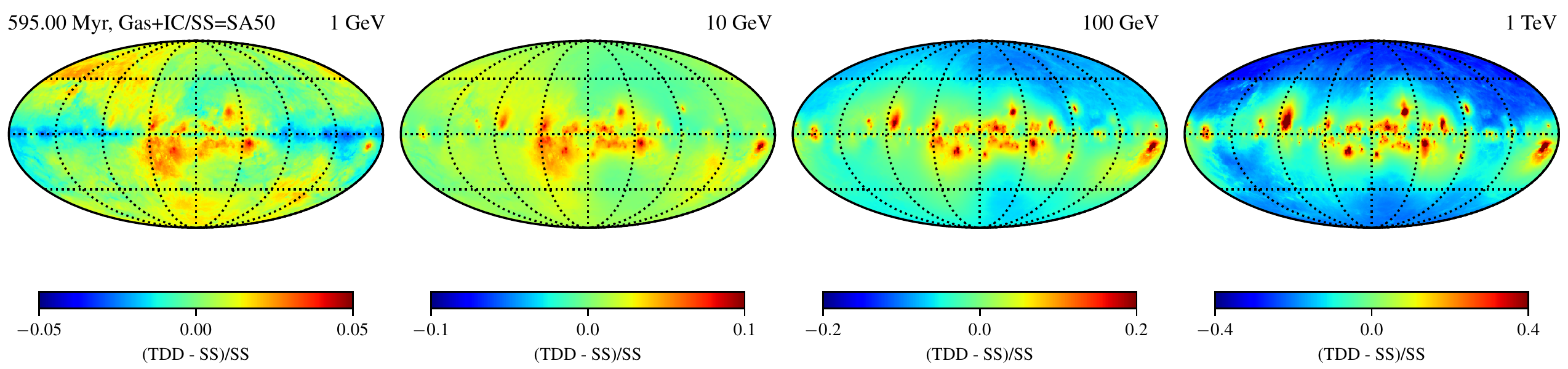}
  \includegraphics[scale=0.7]{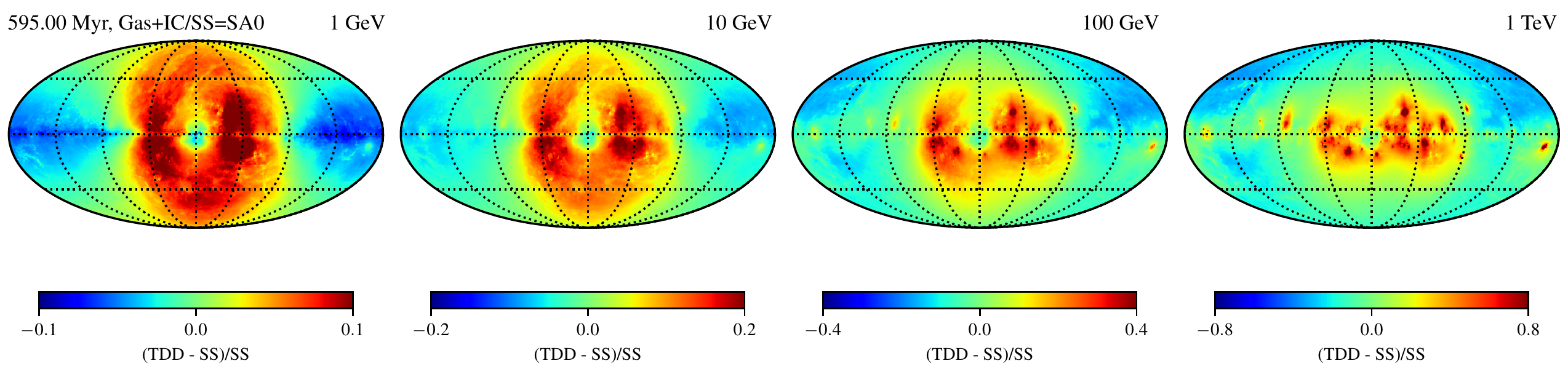}
  \caption{Snapshot at 599~Myr at selected energies for the TDD solution total \gray{} intensity (top) and the fractional residual using the SS$_{_{\rm SA50}}$ (middle) and SS$_{_{\rm SA0}}$ (bottom) intensities for the baseline/reference prediction.
    The longitude meridians and latitude parallels have $45^\circ$ spacing.
  An animation of this figure is available. The video begins at 595.00~Myr and ends at 599.99~Myr. The real-time duration of the video is 100 s.}
  \label{fig:gammamovie}
\end{figure*}

\begin{figure*}[tb!]
  \includegraphics[scale=0.7]{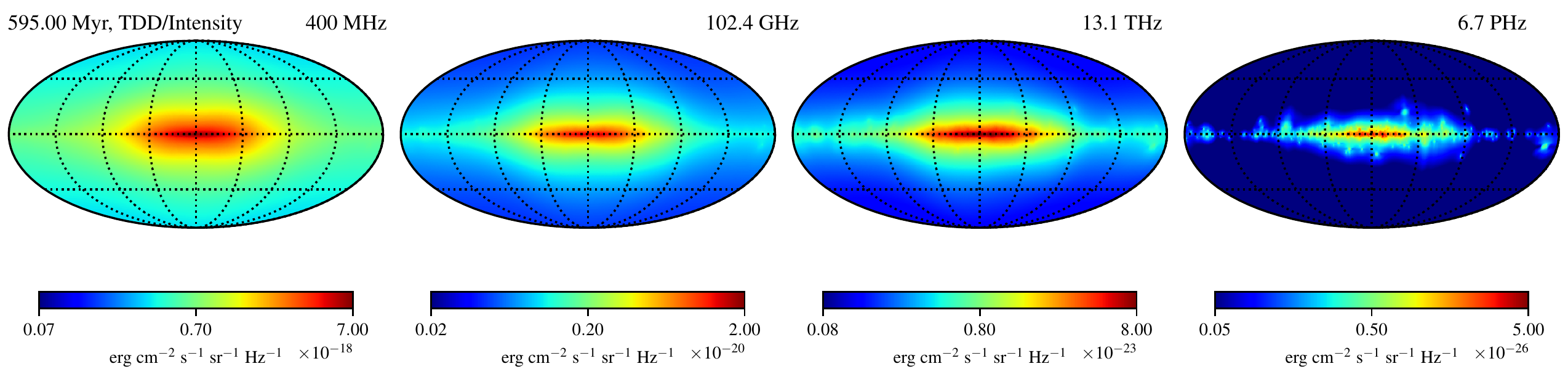}
  \includegraphics[scale=0.7]{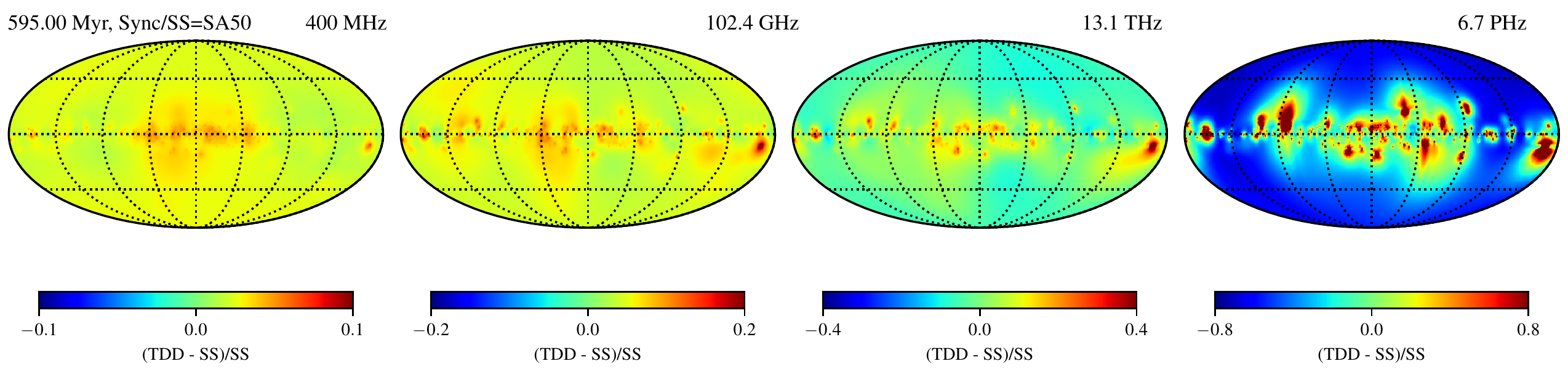}
  \includegraphics[scale=0.7]{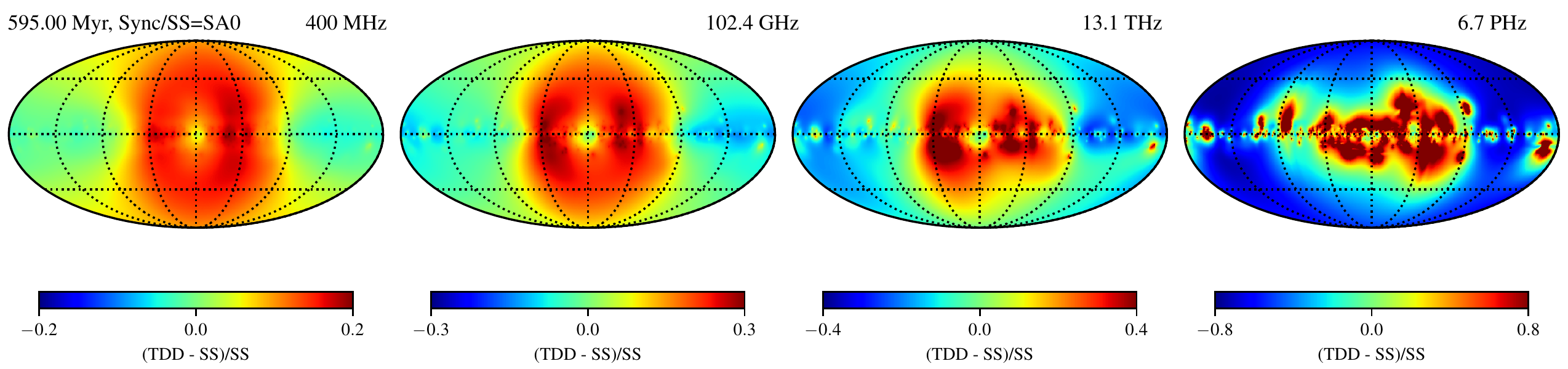}
  \caption{
    Snapshot at 599~Myr at selected frequencies for the TDD solution synchrotron intensity (top), and the fractional residual using the SS$_{_{\rm SA50}}$ (middle) and SS$_{_{\rm SA0}}$ (bottom) intensities for the baseline/reference prediction.
    The longitude meridians and latitude parallels have $45^\circ$ spacing.
  An animation of this figure is available. The video begins at 595.00~Myr and ends at 599.99~Myr. The real-time duration of the video is 100 s.}
  \label{fig:syncmovie}
\end{figure*}

The non-thermal emission time slices at 599.5 and 599.55~Myr necessarily show the evolution over a relatively short time span.
They are limited in their ability to show the range of features that may occur over millions of years but are sufficient to illustrate the general characteristics of the discretised injection model.
To provide a more complete visualisation and description over a longer time span is beyond the scope of the article text.
To facilitate, the online version provides synchronised movies of the energy/frequency-dependent \gray{} and synchrotron emission intensities and residuals using the SA50 and SA0 steady-state solutions, respectively, for the baselines.
Figures~\ref{fig:gammamovie} and~\ref{fig:syncmovie} show the 599~Myr first frame of the animations for \gray{} and synchrotron radiation, respectively.
The online movies are made for the last 5~Myr of the simulation epoch ($595-600$~Myr) with 10~kyr sampling, corresponding to the time span covered by the right panels of Fig.~\ref{fig:crtimeseries} and the non-thermal emission time series shown in Figs.~\ref{fig:gammaoutofplanetimeseries} and~\ref{fig:syncoutofplanetimeseries}, respectively.
Comparison between the 599.5~Myr snapshot and those at 599.5 and 599.55~Myr shows that the features in the residuals have variable spatial dimensions and relative enhancements over time that depend on the steady-state baseline model.
While it is not possible to elaborate on the entirety of the visible features over the whole 5~Myr time span covered by the movies, examination shows that there are many regions of extended enhanced intensity that have a degree of regularity in their features and are not completely amorphous ``blobs.''

Of the numerous interesting examples a few are described here.
First, around 597~Myr there is an over-intensity at intermediate northern latitudes near $l \sim -45^\circ$ associated with a single nearby region, while there are two such over-intense regions around $b\sim$$-(30-70)^\circ$ and $l \sim$$-(20-50)^\circ$ also from nearby injection activity (and as the time evolves, the emission residuals from the latter merge earlier for the higher energies than for the lower ones shown).
Second, around 599.05~Myr there are several over-intense regions at intermediate southern latitudes for $-45^\circ \gtrsim l \lesssim 45^\circ$ that are coming from nearby active regions, and, as for the first example, following the time evolution shows that the energy-dependent merging produces different residual structure.
Third, around 599.7~Myr, two unconnected regions toward the mid-to-high-latitude ($|b|$$\sim$$30--50^\circ$) regions within $\sim$$20^\circ-30^\circ$ of the $l = 0^\circ$ meridian are also active, with subsequent time evolution showing features similar to the other examples.
All of these examples appear ``bubble-like'' with a degree of symmetry and have hard spectra, somewhat similar to the proposed behaviour of the so-called ``Fermi bubbles'' \citep[e.g.,][]{2010ApJ...724.1044S}, although, of course, the latter are more closely aligned with the GC.

Moreover, there is some correspondence between the total high-energy \gray{} and synchrotron residuals supporting a common high-energy particle interaction origin and multiwavelength connection with the residuals in microwave data, perhaps consistent with a common origin of the Fermi bubbles and the so-called ``WMAP/{\it Planck} haze'' \citep[][]{2008ApJ...680.1222D,2013A&A...554A.139P}.
The microwave haze was obtained by subtraction of a template background model based on radio and other surveys extrapolated to higher frequencies, which introduces a collection of uncertainties into its physical interpretation \citep[e.g.,][]{2010JCAP...10..019M}, but its derived spectral characteristics are similar to those shown in this paper for such `bubble' over-intensities, e.g., hardening with frequency over the baseline model.

The residuals shown in the movies also exhibit large-scale features with some similarity to those obtained from analysis of \fermi-LAT data using a grid of \GP-based IEMs by \citet{2012ApJ...750....3A}.
That analysis employed 128 IEMs with multiple steady-state source density distributions, values for the \hi\ spin temperature, gas column density corrections from dust emission, and the size of the CR confinement volume as the fixed parameters.
The differences between the residuals for the collection of IEMs considered by \citet{2012ApJ...750....3A} produce excesses and deficits that are spatially colocated (see, e.g., Figs.~6 and~7 from that paper).
Some of the residuals are related to mismodelling of the ISM components, e.g., the ionised gas, the gas not traced by 21~cm and 2.6~mm surveys \citep[the so-called `dark neutral medium', e.g.,][]{2005Sci...307.1292G,2011A&A...536A..19P}; but such explanations cannot account for the very broadly distributed features.

All IEMs employed by \citet{2012ApJ...750....3A} use 2D axisymmetric CR source density models that have similar spatial distributions outside the inner Galaxy.
On the basis of the comparison between the discretised solution and SA0 (2D) steady-state model made above, a reasonable possibility is that the distributed residuals are coming from the mismatch between the 2D approximated CR source model and the actual discretisation of whatever the `true' 3D CR source distribution is across the Galaxy.

Some clues may also be found from IEMs developed to support generations of high-energy \gray{} source catalogues \citep[e.g.,][]{2012ApJS..199...31N,2015ApJS..218...23A}.
These IEMs have been developed with a primary criterion to flatten large-scale residual features as much as possible.
The established methodology for creating them employs iterative fitting to \gray{} data utilising templates based on gas surveys for \hi\ and CO-traced molecular hydrogen, together with dust tracers to model the highly structured emissions, models for the IC intensities across the sky, and additional large-scale ad hoc spatial templates \citep[e.g.,][]{2016ApJS..223...26A}.
The fitting procedure typically separates the ISM tracers into a finite number of galactocentric radial rings and fits their individual weights together with the ad hoc templates and other elements to obtain the IEM.
The criterion for the residual flatness is a choice made according to various considerations, with some including the number of fit iterations, as well as the characteristics of the ad hoc templates that are included to mitigate features that cannot be adequately modelled with the known ISM tracers.

For the IEM employed for the latest \fermi-LAT source catalogue~\citep[][]{2019arXiv190210045T}, the IEM is based on a \GP-generated model using a 2D axisymmetric CR source density with the corresponding \gray{} intensity templates decomposed over 10~galactocentric rings.
The residual flatness criterion is $\lesssim 3$~\% about the Galactic plane, with ad hoc templates used for the large-scale features with unknown physical origin.
These templates are of interest because they are broadly distributed and display similar characteristics to those shown in Fig.~\ref{fig:gammatotalfracsa0} (and, of course, for the total \gray{} fractional residual movies).
Combined with the observation of similarities with the residuals derived from the \citet{2012ApJ...750....3A}, where the fractional residual level can be $\sim$10--30\% (and larger for higher energies), these analyses may indicate that a component of the CR sources is associated with spiral arms.

Many of the clearest differences between the TDD and steady-state solution that appear in the residual maps in Figs.~\ref{fig:gammaintensityandfrac} and~\ref{fig:gammafracadd50kyr} occur at intermediate latitudes toward the outer Galaxy.
These residuals are mostly caused by nearby injection regions that stand out from the SS$_{_{\rm SA50}}$ background predictions.
It is therefore reasonable to look for such signatures in \fermi-LAT data analyses that focus on the nearby ISM.
There have been a number performed over the years \citep[e.g.,][]{2010ApJ...710..133A,2011Sci...334.1103A,2012ApJ...755...22A,2015A&A...582A..31P, 2016ApJ...833..278M,2017A&A...601A..78R}, 
but only \citet{2011Sci...334.1103A} found clear evidence for recent particle acceleration.
In fact, most of the analyses predict a fairly uniform CR intensity, even toward the outer Galaxy \citep[e.g.,][]{2010ApJ...710..133A}.

Figure~10 of \citet{2017A&A...601A..78R} provides a summary of the variation in gas emissivities for local clouds integrated in the energy range from 0.4 to 10 GeV.
The emissivities show a dispersion of $\sim$10\%, but with no clear trend in location or height about the plane.
Some of that dispersion could be an indication of discrete injection regions, but it could also be related to uncertainties in the estimates of the column density of gas.

A better indication of discrete injection activity could come from the IC emissions, which generally show stronger deviations from the steady-state predictions.
The local cloud analyses cited above all include a model for the IC emissions, but only $\sim 50$\% of them leave this component any freedom when fitting to the data, and the focus of discrepancies with models is always on the properties of the gas.
Some information can, however, be gleaned from the different analyses and the resulting scaling factors for the IC templates obtained for each.
The templates used are always created with GALPROP using a 2D symmetric CR source distribution, similar to the SA0 model used in this paper.

Depending on the region and the details of the individual analyses, the derived IC scaling factors vary from zero to 2.
The statistical error bars on the IC component are usually $\sim$10--50\%, and there is considerable correlation between the IC and the isotropic template, which is also included in all of the analyses.
Despite the correlation, these analyses seem to indicate significant variations in the IC emission at low-to-intermediate latitudes that could be a signature of recent nearby CR injection.
However, because of limited statistics, these templated-based analyses only reach to a few tens of GeV and are unfortunately not capable of measuring any hardening of the spectrum, a telltale signature of recent CR injection activity.

An interesting feature visible in the fractional maps is the symmetry of many of the residual structures above and below the Galactic plane.
Even though the discrete injection regions themselves are nearly spherically symmetric when projected onto the sky, calculating the residual as a fraction of the underlying emissions results in smaller residuals in and about bright regions.
This has implications for identification of such structures in \gray{} data analysis, because residual maps in units of standard deviation are most commonly used to identify regions requiring additional modelling components.

Such maps suffer from the same issues as the fractional residual maps, and any excess emission from a single recent CR injection region may be split into two or more (extended) components, depending on the structure of the emissions over the region being analysed.
Thus, adding components to the model based solely on the significance of the residual may lead to incorrect identification of the underlying physical contributions to the observations, as well as other components in the model being incorrectly determined.
It is thus very important to use as many data as possible to constrain the injection and propagation history of CRs in the Galaxy.

The discretised model employed in this paper is obviously simplified compared to the true physical description for CR sources in the Galaxy.
The injection regions employ a spectral model with a constant injection rate over a specified time interval for both CR protons and electrons.
This choice enables the comparison with the equivalent steady-state solution using the same spectral model continuous with time.
But the putative source classes for CR particles (SNRs, OB associations, etc.) very likely have different time-dependent injection spectral behaviours for the individual CR species.
Moreover, individual injection regions in this paper are effectively independent, determined solely according to the assumed source frequency and sampling from the smooth spatial distribution.

The formation of the massive stars that are likely progenitors for the CR sources is episodic and spatially localised, and, because of their short lifetimes, these stars have some proximity to the stellar nurseries where they are born.
The discrete CR sources associated with the massive stars would therefore also be clustered in time and space. 
%It is likely that CR sources associated with massive stars exhibit clustering in time and space, a consequence of the episodic and spatially localised nature of star formation.
%as a consequence of their short lifetimes and the stellar nurseries from where they are born, a consequence of the formation of massive stars that is episodic and spatially localised.
%Therefore, the discrete CR sources could be grouped in time and space.
The resulting stellar winds and SN explosion(s) would quench further star formation in the massive star birth region, resulting in a time-varying CR source density that follows active star formation at each epoch. 
Furthermore, the idealised picture that has been employed for this paper does not account for CR acceleration in larger structures, such as superbubbles.
However, the novel characteristics for the time- and space-discretised CR injection and propagation motivate consideration of more detailed models with future work.

Despite its simplified physical picture for the CR injection, the discretised modelling predicts noticeable differences from the equivalent steady-state model across the electromagnetic spectrum. 
In general, the all-sky intensity maps are less smooth where the dominant physical processes producing them are dominated by the CR particles propagating near the regions where they are freshly injected.
Even after cessation of injection about individual regions, there are noticeable remnant effects on the broadband interstellar emissions, as evinced by the differences between the 599.5 and 599.55~Myr intensity/residual maps discussed above.

Earlier attempts to model the very high energy interstellar emissions from the Galactic disk used the steady-state assumption \citep[e.g.,][]{2014PhRvD..90l2007A,2018PhRvD..98d3003L,2019arXiv190403894C}, which obviously cannot produce such behaviour.
\citet{2018PhRvD..98b3004N} and \citet{2019arXiv190706061N} gave estimates of the high-energy interstellar emissions $\gtrsim$300~GeV energies using \fermi-LAT data.
Both analyses show that outside the Galactic plane, there is directional dependence of the emissions, but they do not separate the northern/southern latitude regions.
The former work, in fact, suggests a correlation with IceCube neutrinos outside the Galactic plane as evidence for a nearby/recently active CR accelerator, although its location on the sky is not established apart from being outside of the plane at latitudes $|b| \gtrsim 10^\circ$. 
Given the myriad of features displayed by the discretised modelling for the relatively simplified case examined in this paper, a full data analysis to give definitive evidence for or against such an interpretation may require more sophistication than previously employed.

\section{Summary}
\label{summary}

We have made new calculations of the time-dependent injection and propagation through the ISM of CRs from discrete regions in the Galaxy using the \GP\ code.
These calculations are fully 3D for the CR source and ISM target density distributions.
Compared to the equivalent steady-state model, the TDD solution shows strong, energy-dependent fluctuations for the CR intensities that extend over the Galactic disk.
Because of the finite sampling, the TDD solution also shows asymmetrical distributions for the CR intensities about the plane over the Galaxy.
At the solar system location, where model predictions are normalised to the data, the TDD CR intensities generally attain very similar spectral shapes over the energy range $\sim$5--300~GeV after a long enough relaxation time so that they are comparable to those of the steady-state solution.
The effects of nearby injection activity and propagation generally appear at lower/higher energies, depending on particle species.

For the non-thermal emissions, comparison between the TDD and the equivalent steady-state solution reveals novel features in the intensity maps that are solely due to the discretised injection activity and propagation.
The features exhibited by the discretised model may explain puzzles in high-energy \gray{s}, such as the north/south asymmetry in \fermi-LAT data and residual features like the \fermi\ bubbles.
However, even with the ``known'' mathematical description for the spatial distribution for the CR injection regions the interpretation of multi-wavelength intensity sky maps is complicated.
Averaging of over- and under-emissive regions by the line-of-sight integration can give a limited picture of the general fluctuations across the Galaxy.
Identifying characteristics of the discrete injection regions depends on the directions being viewed, with particular difficulties occurring because of the averaging toward the inner Galaxy.
If the distribution of the CR injection regions is mismodelled, the features of the discretised CR injection and propagation are effectively erased, except for the highest energies or very intense nearby regions.
Even for the latter cases, their `true' features are incorrectly represented in residual maps because of the mismatch between the underlying smooth density distribution for the CR sources.

Our calculations show that the \fermi-LAT and current Cherenkov instruments, such as HESS, probe the energies from where there is dominance of \gray{} production by the large-scale CR sea to those where there is a major component due to injection and propagation of recently accelerated CRs about the sources.
Likewise, our modelling also shows that the microwave-to-X-ray emissions trace properties of the CR electrons over similar energy and spatial scales.
Forthcoming data from instruments such as {\it eROSITA} and CTA  and continued observations by HAWC have the potential to provide insights on the recent injection and propagation history about the CR sources.
But it is necessary to employ the data gathered already in the radio/microwave and high-energy ($\sim$GeV) domain to establish better knowledge of the large-scale spatial distribution of the CR sources.
Consideration of energy ranges in isolation provides an incomplete picture; hence, the analysis of multi-wavelength data is necessary to understand the evolution of CRs from injection about sources to their distribution throughout the Galaxy. 

\acknowledgements
Gavin Rowell is thanked for constructive comments. \galprop\ development is partially funded via NASA grant NNX17AB48G.
Some of the results in this paper have been derived using the 
HEALPix~\citep{2005ApJ...622..759G} package.

\bibliography{3d_cr}

\begin{thebibliography}{}
\expandafter\ifx\csname natexlab\endcsname\relax\def\natexlab#1{#1}\fi
\providecommand{\url}[1]{\href{#1}{#1}}
\providecommand{\dodoi}[1]{doi:~\href{http://doi.org/#1}{\nolinkurl{#1}}}
\providecommand{\doeprint}[1]{\href{http://ascl.net/#1}{\nolinkurl{http://ascl.net/#1}}}
\providecommand{\doarXiv}[1]{\href{https://arxiv.org/abs/#1}{\nolinkurl{https://arxiv.org/abs/#1}}}

\bibitem[{{Abdo} {et~al.}(2008){Abdo}, {Allen}, {Aune}, {Berley}, {Blaufuss},
  {Casanova}, {Chen}, {Dingus}, {Ellsworth}, {Fleysher}, {Fleysher},
  {Gonzalez}, {Goodman}, {Hoffman}, {H{\"u}ntemeyer}, {Kolterman}, {Lansdell},
  {Linnemann}, {McEnery}, {Mincer}, {Moskalenko}, {Nemethy}, {Noyes}, {Porter},
  {Pretz}, {Ryan}, {Saz Parkinson}, {Shoup}, {Sinnis}, {Smith}, {Strong},
  {Sullivan}, {Vasileiou}, {Walker}, {Williams}, \&
  {Yodh}}]{2008ApJ...688.1078A}
{Abdo}, A.~A., {Allen}, B., {Aune}, T., {et~al.} 2008, \apj, 688, 1078,
  \dodoi{10.1086/592213}

\bibitem[{{Abdo} {et~al.}(2010{\natexlab{a}}){Abdo}, {Ackermann}, {Ajello},
  {Atwood}, {Baldini}, {Ballet}, {Barbiellini}, {Bastieri}, {Baughman},
  {Bechtol}, {Bellazzini}, {Berenji}, {Blandford}, {Bloom}, {Bonamente},
  {Borgland}, {Bregeon}, {Brez}, {Brigida}, {Bruel}, {Burnett}, {Buson},
  {Caliandro}, {Cameron}, {Caraveo}, {Casandjian}, {Cavazzuti}, {Cecchi}, {{\c
  C}elik}, {Charles}, {Chekhtman}, {Cheung}, {Chiang}, {Ciprini}, {Claus},
  {Cohen-Tanugi}, {Cominsky}, {Conrad}, {Cutini}, {Dermer}, {de Angelis}, {de
  Palma}, {Digel}, {di Bernardo}, {do Couto e Silva}, {Drell}, {Drlica-Wagner},
  {Dubois}, {Dumora}, {Farnier}, {Favuzzi}, {Fegan}, {Focke}, {Fortin},
  {Frailis}, {Fukazawa}, {Funk}, {Fusco}, {Gaggero}, {Gargano}, {Gasparrini},
  {Gehrels}, {Germani}, {Giebels}, {Giglietto}, {Giommi}, {Giordano},
  {Glanzman}, {Godfrey}, {Grenier}, {Grondin}, {Grove}, {Guillemot}, {Guiriec},
  {Gustafsson}, {Hanabata}, {Harding}, {Hayashida}, {Hughes}, {Itoh},
  {Jackson}, {J{\'o}hannesson}, {Johnson}, {Johnson}, {Johnson}, {Johnson},
  {Kamae}, {Katagiri}, {Kataoka}, {Kawai}, {Kerr}, {Kn{\"o}dlseder}, {Kocian},
  {Kuehn}, {Kuss}, {Lande}, {Latronico}, {Lemoine-Goumard}, {Longo}, {Loparco},
  {Lott}, {Lovellette}, {Lubrano}, {Madejski}, {Makeev}, {Mazziotta},
  {McConville}, {McEnery}, {Meurer}, {Michelson}, {Mitthumsiri}, {Mizuno},
  {Moiseev}, {Monte}, {Monzani}, {Morselli}, {Moskalenko}, {Murgia}, {Nolan},
  {Norris}, {Nuss}, {Ohsugi}, {Omodei}, {Orlando}, {Ormes}, {Paneque},
  {Panetta}, {Parent}, {Pelassa}, {Pepe}, {Pesce-Rollins}, {Piron}, {Porter},
  {Rain{\`o}}, {Rando}, {Razzano}, {Reimer}, {Reimer}, {Reposeur}, {Ritz},
  {Rochester}, {Rodriguez}, {Roth}, {Ryde}, {Sadrozinski}, {Sanchez}, {Sander},
  {Parkinson}, {Scargle}, {Sellerholm}, {Sgr{\`o}}, {Shaw}, {Siskind}, {Smith},
  {Smith}, {Spandre}, {Spinelli}, {Starck}, {Strickman}, {Strong}, {Suson},
  {Tajima}, {Takahashi}, {Takahashi}, {Tanaka}, {Thayer}, {Thayer}, {Thompson},
  {Tibaldo}, {Torres}, {Tosti}, {Tramacere}, {Uchiyama}, {Usher}, {Vasileiou},
  {Vilchez}, {Vitale}, {Waite}, {Wang}, {Winer}, {Wood}, {Ylinen}, {Ziegler},
  \& {Fermi LAT Collaboration}}]{2010PhRvL.104j1101A}
{Abdo}, A.~A., {Ackermann}, M., {Ajello}, M., {et~al.} 2010{\natexlab{a}},
  Physical Review Letters, 104, 101101, \dodoi{10.1103/PhysRevLett.104.101101}

\bibitem[{{Abdo} {et~al.}(2010{\natexlab{b}}){Abdo}, {Ackermann}, {Ajello},
  {Baldini}, {Ballet}, {Barbiellini}, {Bastieri}, {Baughman}, {Bechtol},
  {Bellazzini}, {Berenji}, {Bloom}, {Bonamente}, {Borgland}, {Bregeon}, {Brez},
  {Brigida}, {Bruel}, {Burnett}, {Buson}, {Caliandro}, {Cameron}, {Caraveo},
  {Casandjian}, {Cecchi}, {{\c C}elik}, {Chekhtman}, {Cheung}, {Chiang},
  {Ciprini}, {Claus}, {Cohen-Tanugi}, {Cominsky}, {Conrad}, {Dermer}, {de
  Palma}, {Digel}, {Silva}, {Drell}, {Dubois}, {Dumora}, {Farnier}, {Favuzzi},
  {Fegan}, {Focke}, {Fortin}, {Frailis}, {Fukazawa}, {Funk}, {Fusco},
  {Gargano}, {Gehrels}, {Germani}, {Giavitto}, {Giebels}, {Giglietto},
  {Giordano}, {Glanzman}, {Godfrey}, {Grenier}, {Grondin}, {Grove},
  {Guillemot}, {Guiriec}, {Harding}, {Hayashida}, {Horan}, {Hughes}, {Jackson},
  {J{\'o}hannesson}, {Johnson}, {Johnson}, {Kamae}, {Katagiri}, {Kataoka},
  {Kawai}, {Kerr}, {Kn{\"o}dlseder}, {Kuss}, {Lande}, {Latronico},
  {Lemoine-Goumard}, {Longo}, {Loparco}, {Lott}, {Lovellette}, {Lubrano},
  {Makeev}, {Mazziotta}, {McEnery}, {Meurer}, {Michelson}, {Mitthumsiri},
  {Mizuno}, {Monte}, {Monzani}, {Morselli}, {Moskalenko}, {Murgia}, {Nolan},
  {Norris}, {Nuss}, {Ohsugi}, {Okumura}, {Omodei}, {Orlando}, {Ormes},
  {Paneque}, {Pelassa}, {Pepe}, {Pesce-Rollins}, {Piron}, {Porter},
  {Rain{\`o}}, {Rando}, {Razzano}, {Reimer}, {Reimer}, {Reposeur}, {Rodriguez},
  {Ryde}, {Sadrozinski}, {Sanchez}, {Sander}, {Saz Parkinson}, {Sgr{\`o}},
  {Siskind}, {Smith}, {Spandre}, {Spinelli}, {Starck}, {Strickman}, {Strong},
  {Suson}, {Takahashi}, {Tanaka}, {Thayer}, {Thayer}, {Thompson}, {Tibaldo},
  {Torres}, {Tosti}, {Tramacere}, {Uchiyama}, {Usher}, {Vasileiou}, {Vilchez},
  {Vitale}, {Waite}, {Wang}, {Winer}, {Wood}, {Ylinen}, {Ziegler}, \&
  {Fermi/LAT Collaboration}}]{2010ApJ...710..133A}
---. 2010{\natexlab{b}}, \apj, 710, 133, \dodoi{10.1088/0004-637X/710/1/133}

\bibitem[{{Abeysekara} {et~al.}(2017{\natexlab{a}}){Abeysekara}, {Albert},
  {Alfaro}, {Alvarez}, {{\'A}lvarez}, {Arceo}, {Arteaga-Vel{\'a}zquez}, {Ayala
  Solares}, {Barber}, {Baughman}, {Bautista-Elivar}, {Becerra Gonzalez},
  {Becerril}, {Belmont-Moreno}, {BenZvi}, {Berley}, {Bernal}, {Braun},
  {Brisbois}, {Caballero-Mora}, {Capistr{\'a}n}, {Carrami{\~n}ana}, {Casanova},
  {Castillo}, {Cotti}, {Cotzomi}, {Couti{\~n}o de Le{\'o}n}, {de la Fuente},
  {De Le{\'o}n}, {Diaz Hernandez}, {Dingus}, {DuVernois},
  {D{\'{\i}}az-V{\'e}lez}, {Ellsworth}, {Engel}, {Fiorino}, {Fraija},
  {Garc{\'{\i}}a-Gonz{\'a}lez}, {Garfias}, {Gerhardt}, {Gonz{\'a}lez
  Mu{\~n}oz}, {Gonz{\'a}lez}, {Goodman}, {Hampel-Arias}, {Harding},
  {Hernandez}, {Hernandez-Almada}, {Hinton}, {Hui}, {H{\"u}ntemeyer},
  {Iriarte}, {Jardin-Blicq}, {Joshi}, {Kaufmann}, {Kieda}, {Lara}, {Lauer},
  {Lee}, {Lennarz}, {Le{\'o}n Vargas}, {Linnemann}, {Longinotti}, {Raya},
  {Luna-Garc{\'{\i}}a}, {L{\'o}pez-Coto}, {Malone}, {Marinelli}, {Martinez},
  {Martinez-Castellanos}, {Mart{\'{\i}}nez-Castro}, {Mart{\'{\i}}nez-Huerta},
  {Matthews}, {Miranda-Romagnoli}, {Moreno}, {Mostaf{\'a}}, {Nellen},
  {Newbold}, {Nisa}, {Noriega-Papaqui}, {Pelayo}, {Pretz},
  {P{\'e}rez-P{\'e}rez}, {Ren}, {Rho}, {Rivi{\`e}re}, {Rosa-Gonz{\'a}lez},
  {Rosenberg}, {Ruiz-Velasco}, {Salazar}, {Salesa Greus}, {Sandoval},
  {Schneider}, {Schoorlemmer}, {Sinnis}, {Smith}, {Springer}, {Surajbali},
  {Taboada}, {Tibolla}, {Tollefson}, {Torres}, {Ukwatta}, {Vianello},
  {Villase{\~n}or}, {Weisgarber}, {Westerhoff}, {Wisher}, {Wood}, {Yapici},
  {Younk}, {Zepeda}, \& {Zhou}}]{2017ApJ...843...40A}
{Abeysekara}, A.~U., {Albert}, A., {Alfaro}, R., {et~al.} 2017{\natexlab{a}},
  \apj, 843, 40, \dodoi{10.3847/1538-4357/aa7556}

\bibitem[{{Abeysekara} {et~al.}(2017{\natexlab{b}}){Abeysekara}, {Albert},
  {Alfaro}, {Alvarez}, {{\'A}lvarez}, {Arceo}, {Arteaga-Vel{\'a}zquez}, {Avila
  Rojas}, {Ayala Solares}, {Barber}, {Bautista-Elivar}, {Becerril},
  {Belmont-Moreno}, {BenZvi}, {Berley}, {Bernal}, {Braun}, {Brisbois},
  {Caballero-Mora}, {Capistr{\'a}n}, {Carrami{\~n}ana}, {Casanova}, {Castillo},
  {Cotti}, {Cotzomi}, {Couti{\~n}o de Le{\'o}n}, {De Le{\'o}n}, {De la Fuente},
  {Dingus}, {DuVernois}, {D{\'{\i}}az-V{\'e}lez}, {Ellsworth}, {Engel},
  {Enr{\'{\i}}quez-Rivera}, {Fiorino}, {Fraija}, {Garc{\'{\i}}a-Gonz{\'a}lez},
  {Garfias}, {Gerhardt}, {Gonz{\'a}lez Mu{\~n}oz}, {Gonz{\'a}lez}, {Goodman},
  {Hampel-Arias}, {Harding}, {Hern{\'a}ndez}, {Hern{\'a}ndez-Almada}, {Hinton},
  {Hona}, {Hui}, {H{\"u}ntemeyer}, {Iriarte}, {Jardin-Blicq}, {Joshi},
  {Kaufmann}, {Kieda}, {Lara}, {Lauer}, {Lee}, {Lennarz}, {Vargas},
  {Linnemann}, {Longinotti}, {Luis Raya}, {Luna-Garc{\'{\i}}a},
  {L{\'o}pez-Coto}, {Malone}, {Marinelli}, {Martinez}, {Martinez-Castellanos},
  {Mart{\'{\i}}nez-Castro}, {Mart{\'{\i}}nez-Huerta}, {Matthews},
  {Miranda-Romagnoli}, {Moreno}, {Mostaf{\'a}}, {Nellen}, {Newbold}, {Nisa},
  {Noriega-Papaqui}, {Pelayo}, {Pretz}, {P{\'e}rez-P{\'e}rez}, {Ren}, {Rho},
  {Rivi{\`e}re}, {Rosa-Gonz{\'a}lez}, {Rosenberg}, {Ruiz-Velasco}, {Salazar},
  {Salesa Greus}, {Sandoval}, {Schneider}, {Schoorlemmer}, {Sinnis}, {Smith},
  {Springer}, {Surajbali}, {Taboada}, {Tibolla}, {Tollefson}, {Torres},
  {Ukwatta}, {Vianello}, {Weisgarber}, {Westerhoff}, {Wisher}, {Wood},
  {Yapici}, {Yodh}, {Younk}, {Zepeda}, {Zhou}, {Guo}, {Hahn}, {Li}, \&
  {Zhang}}]{2017Sci...358..911A}
---. 2017{\natexlab{b}}, Science, 358, 911, \dodoi{10.1126/science.aan4880}

\bibitem[{{Abeysekara} {et~al.}(2018){Abeysekara}, {Albert}, {Alfaro},
  {Alvarez}, {{\'A}lvarez}, {Arceo}, {Arteaga-Vel{\'a}zquez}, {Avila Rojas},
  {Ayala Solares}, {Belmont-Moreno}, {BenZvi}, {Brisbois}, {Caballero-Mora},
  {Capistr{\'a}n}, {Carrami{\~n}ana}, {Casanova}, {Castillo}, {Cotti},
  {Cotzomi}, {Couti{\~n}o de Le{\'o}n}, {De Le{\'o}n}, {De la Fuente},
  {D{\'{\i}}az-V{\'e}lez}, {Dichiara}, {Dingus}, {DuVernois}, {Ellsworth},
  {Engel}, {Espinoza}, {Fang}, {Fleischhack}, {Fraija}, {Galv{\'a}n-G{\'a}mez},
  {Garc{\'{\i}}a-Gonz{\'a}lez}, {Garfias}, {Gonz{\'a}lez-Mu{\~n}oz},
  {Gonz{\'a}lez}, {Goodman}, {Hampel-Arias}, {Harding}, {Hernandez}, {Hinton},
  {Hona}, {Hueyotl-Zahuantitla}, {Hui}, {H{\"u}ntemeyer}, {Iriarte},
  {Jardin-Blicq}, {Joshi}, {Kaufmann}, {Kar}, {Kunde}, {Lauer}, {Lee},
  {Le{\'o}n Vargas}, {Li}, {Linnemann}, {Longinotti}, {Luis-Raya},
  {L{\'o}pez-Coto}, {Malone}, {Marinelli}, {Martinez}, {Martinez-Castellanos},
  {Mart{\'{\i}}nez-Castro}, {Matthews}, {Miranda-Romagnoli}, {Moreno},
  {Mostaf{\'a}}, {Nayerhoda}, {Nellen}, {Newbold}, {Nisa}, {Noriega-Papaqui},
  {Pretz}, {P{\'e}rez-P{\'e}rez}, {Ren}, {Rho}, {Rivi{\`e}re},
  {Rosa-Gonz{\'a}lez}, {Rosenberg}, {Ruiz-Velasco}, {Salesa Greus}, {Sandoval},
  {Schneider}, {Schoorlemmer}, {Seglar Arroyo}, {Sinnis}, {Smith}, {Springer},
  {Surajbali}, {Taboada}, {Tibolla}, {Tollefson}, {Torres}, {Vianello},
  {Villase{\~n}or}, {Weisgarber}, {Werner}, {Westerhoff}, {Wood}, {Yapici},
  {Yodh}, {Zepeda}, {Zhang}, \& {Zhou}}]{2018Natur.562...82A}
---. 2018, \nat, 562, 82, \dodoi{10.1038/s41586-018-0565-5}

\bibitem[{{Abramowski} {et~al.}(2014){Abramowski}, {Aharonian}, {Ait Benkhali},
  {Akhperjanian}, {Ang{\"u}ner}, {Backes}, {Balenderan}, {Balzer}, {Barnacka},
  {Becherini}, {Becker Tjus}, {Berge}, {Bernhard}, {Bernl{\"o}hr}, {Birsin},
  {Biteau}, {B{\"o}ttcher}, {Boisson}, {Bolmont}, {Bordas}, {Bregeon}, {Brun},
  {Brun}, {Bryan}, {Bulik}, {Carrigan}, {Casanova}, {Chadwick}, {Chakraborty},
  {Chalme-Calvet}, {Chaves}, {Chr{\'e}tien}, {Colafrancesco}, {Cologna},
  {Conrad}, {Couturier}, {Cui}, {Davids}, {Degrange}, {Deil}, {deWilt},
  {Djannati-Ata{\"\i}}, {Domainko}, {Donath}, {Drury}, {Dubus}, {Dutson},
  {Dyks}, {Dyrda}, {Edwards}, {Egberts}, {Eger}, {Espigat}, {Farnier}, {Fegan},
  {Feinstein}, {Fernandes}, {Fernand ez}, {Fiasson}, {Fontaine}, {F{\"o}rster},
  {F{\"u}{\ss}ling}, {Gabici}, {Gajdus}, {Gallant}, {Garrigoux}, {Giavitto},
  {Giebels}, {Glicenstein}, {Gottschall}, {Grondin}, {Grudzi{\'n}ska},
  {Hadasch}, {H{\"a}ffner}, {Hahn}, {Harris}, {Heinzelmann}, {Henri},
  {Hermann}, {Hervet}, {Hillert}, {Hinton}, {Hofmann}, {Hofverberg}, {Holler},
  {Horns}, {Ivascenko}, {Jacholkowska}, {Jahn}, {Jamrozy}, {Janiak},
  {Jankowsky}, {Jung-Richardt}, {Kastendieck}, {Katarzy{\'n}ski}, {Katz},
  {Kaufmann}, {Kh{\'e}lifi}, {Kieffer}, {Klepser}, {Klochkov}, {Klu{\'z}niak},
  {Kolitzus}, {Komin}, {Kosack}, {Krakau}, {Krayzel}, {Kr{\"u}ger}, {Laffon},
  {Lamanna}, {Lefaucheur}, {Lefranc}, {Lemi{\`e}re}, {Lemoine-Goumard},
  {Lenain}, {Lohse}, {Lopatin}, {Lu}, {Marandon}, {Marcowith}, {Marx},
  {Maurin}, {Maxted}, {Mayer}, {McComb}, {M{\'e}hault}, {Meintjes}, {Menzler},
  {Meyer}, {Mitchell}, {Moderski}, {Mohamed}, {Mor{\^a}}, {Moulin}, {Murach},
  {de Naurois}, {Niemiec}, {Nolan}, {Oakes}, {Odaka}, {Ohm}, {Opitz},
  {Ostrowski}, {Oya}, {Panter}, {Parsons}, {Paz Arribas}, {Pekeur},
  {Pelletier}, {Petrucci}, {Peyaud}, {Pita}, {Poon}, {P{\"u}hlhofer}, {Punch},
  {Quirrenbach}, {Raab}, {Reichardt}, {Reimer}, {Reimer}, {Renaud}, {de los
  Reyes}, {Rieger}, {Romoli}, {Rosier-Lees}, {Rowell}, {Rudak}, {Rulten},
  {Sahakian}, {Salek}, {Sanchez}, {Santangelo}, {Schlickeiser},
  {Sch{\"u}ssler}, {Schulz}, {Schwanke}, {Schwarzburg}, {Schwemmer}, {Sol},
  {Spanier}, {Spengler}, {Spies}, {Stawarz}, {Steenkamp}, {Stegmann},
  {Stinzing}, {Stycz}, {Sushch}, {Tavernet}, {Tavernier}, {Taylor}, {Terrier},
  {Tluczykont}, {Trichard}, {Valerius}, {van Eldik}, {van Soelen},
  {Vasileiadis}, {Veh}, {Venter}, {Viana}, {Vincent}, {Vink}, {V{\"o}lk},
  {Volpe}, {Vorster}, {Vuillaume}, {Wagner}, {Wagner}, {Wagner}, {Ward},
  {Weidinger}, {Weitzel}, {White}, {Wierzcholska}, {Willmann}, {W{\"o}rnlein},
  {Wouters}, {Yang}, {Zabalza}, {Zaborov}, {Zacharias}, {Zdziarski}, {Zech},
  {Zechlin}, {Fukui}, \& {H.~E.~S.~S. Collaboration}}]{2014PhRvD..90l2007A}
{Abramowski}, A., {Aharonian}, F., {Ait Benkhali}, F., {et~al.} 2014, \prd, 90,
  122007, \dodoi{10.1103/PhysRevD.90.122007}

\bibitem[{{Acero} {et~al.}(2015){Acero}, {Ackermann}, {Ajello}, {Albert},
  {Atwood}, {Axelsson}, {Baldini}, {Ballet}, {Barbiellini}, {Bastieri},
  {Belfiore}, {Bellazzini}, {Bissaldi}, {Blandford}, {Bloom}, {Bogart},
  {Bonino}, {Bottacini}, {Bregeon}, {Britto}, {Bruel}, {Buehler}, {Burnett},
  {Buson}, {Caliandro}, {Cameron}, {Caputo}, {Caragiulo}, {Caraveo},
  {Casandjian}, {Cavazzuti}, {Charles}, {Chaves}, {Chekhtman}, {Cheung},
  {Chiang}, {Chiaro}, {Ciprini}, {Claus}, {Cohen-Tanugi}, {Cominsky}, {Conrad},
  {Cutini}, {D'Ammando}, {de Angelis}, {DeKlotz}, {de Palma}, {Desiante},
  {Digel}, {Di Venere}, {Drell}, {Dubois}, {Dumora}, {Favuzzi}, {Fegan},
  {Ferrara}, {Finke}, {Franckowiak}, {Fukazawa}, {Funk}, {Fusco}, {Gargano},
  {Gasparrini}, {Giebels}, {Giglietto}, {Giommi}, {Giordano}, {Giroletti},
  {Glanzman}, {Godfrey}, {Grenier}, {Grondin}, {Grove}, {Guillemot}, {Guiriec},
  {Hadasch}, {Harding}, {Hays}, {Hewitt}, {Hill}, {Horan}, {Iafrate}, {Jogler},
  {J{\'o}hannesson}, {Johnson}, {Johnson}, {Johnson}, {Johnson}, {Kamae},
  {Kataoka}, {Katsuta}, {Kuss}, {La Mura}, {Landriu}, {Larsson}, {Latronico},
  {Lemoine-Goumard}, {Li}, {Li}, {Longo}, {Loparco}, {Lott}, {Lovellette},
  {Lubrano}, {Madejski}, {Massaro}, {Mayer}, {Mazziotta}, {McEnery},
  {Michelson}, {Mirabal}, {Mizuno}, {Moiseev}, {Mongelli}, {Monzani},
  {Morselli}, {Moskalenko}, {Murgia}, {Nuss}, {Ohno}, {Ohsugi}, {Omodei},
  {Orienti}, {Orlando}, {Ormes}, {Paneque}, {Panetta}, {Perkins},
  {Pesce-Rollins}, {Piron}, {Pivato}, {Porter}, {Racusin}, {Rando}, {Razzano},
  {Razzaque}, {Reimer}, {Reimer}, {Reposeur}, {Rochester}, {Romani},
  {Salvetti}, {S{\'a}nchez-Conde}, {Saz Parkinson}, {Schulz}, {Siskind},
  {Smith}, {Spada}, {Spandre}, {Spinelli}, {Stephens}, {Strong}, {Suson},
  {Takahashi}, {Takahashi}, {Tanaka}, {Thayer}, {Thayer}, {Thompson},
  {Tibaldo}, {Tibolla}, {Torres}, {Torresi}, {Tosti}, {Troja}, {Van Klaveren},
  {Vianello}, {Winer}, {Wood}, {Wood}, {Zimmer}, \& {Fermi-LAT
  Collaboration}}]{2015ApJS..218...23A}
{Acero}, F., {Ackermann}, M., {Ajello}, M., {et~al.} 2015, \apjs, 218, 23,
  \dodoi{10.1088/0067-0049/218/2/23}

\bibitem[{{Acero} {et~al.}(2016){Acero}, {Ackermann}, {Ajello}, {Albert},
  {Baldini}, {Ballet}, {Barbiellini}, {Bastieri}, {Bellazzini}, {Bissaldi},
  {Bloom}, {Bonino}, {Bottacini}, {Brandt}, {Bregeon}, {Bruel}, {Buehler},
  {Buson}, {Caliandro}, {Cameron}, {Caragiulo}, {Caraveo}, {Casandjian},
  {Cavazzuti}, {Cecchi}, {Charles}, {Chekhtman}, {Chiang}, {Chiaro}, {Ciprini},
  {Claus}, {Cohen-Tanugi}, {Conrad}, {Cuoco}, {Cutini}, {D'Ammando}, {de
  Angelis}, {de Palma}, {Desiante}, {Digel}, {Di Venere}, {Drell}, {Favuzzi},
  {Fegan}, {Ferrara}, {Focke}, {Franckowiak}, {Funk}, {Fusco}, {Gargano},
  {Gasparrini}, {Giglietto}, {Giordano}, {Giroletti}, {Glanzman}, {Godfrey},
  {Grenier}, {Guiriec}, {Hadasch}, {Harding}, {Hayashi}, {Hays}, {Hewitt},
  {Hill}, {Horan}, {Hou}, {Jogler}, {J{\'o}hannesson}, {Kamae}, {Kuss},
  {Landriu}, {Larsson}, {Latronico}, {Li}, {Li}, {Longo}, {Loparco},
  {Lovellette}, {Lubrano}, {Maldera}, {Malyshev}, {Manfreda}, {Martin},
  {Mayer}, {Mazziotta}, {McEnery}, {Michelson}, {Mirabal}, {Mizuno}, {Monzani},
  {Morselli}, {Nuss}, {Ohsugi}, {Omodei}, {Orienti}, {Orlando}, {Ormes},
  {Paneque}, {Pesce-Rollins}, {Piron}, {Pivato}, {Rain{\`o}}, {Rando},
  {Razzano}, {Razzaque}, {Reimer}, {Reimer}, {Remy}, {Renault},
  {S{\'a}nchez-Conde}, {Schaal}, {Schulz}, {Sgr{\`o}}, {Siskind}, {Spada},
  {Spandre}, {Spinelli}, {Strong}, {Suson}, {Tajima}, {Takahashi}, {Thayer},
  {Thompson}, {Tibaldo}, {Tinivella}, {Torres}, {Tosti}, {Troja}, {Vianello},
  {Werner}, {Wood}, {Wood}, {Zaharijas}, \& {Zimmer}}]{2016ApJS..223...26A}
---. 2016, \apjs, 223, 26, \dodoi{10.3847/0067-0049/223/2/26}

\bibitem[{{Ackermann} {et~al.}(2011){Ackermann}, {Ajello}, {Allafort},
  {Baldini}, {Ballet}, {Barbiellini}, {Bastieri}, {Belfiore}, {Bellazzini},
  {Berenji}, {Bland ford}, {Bloom}, {Bonamente}, {Borgland }, {Bottacini},
  {Brigida}, {Bruel}, {Buehler}, {Buson}, {Caliandro}, {Cameron}, {Caraveo},
  {Casandjian}, {Cecchi}, {Chekhtman}, {Cheung}, {Chiang}, {Ciprini}, {Claus},
  {Cohen-Tanugi}, {de Angelis}, {de Palma}, {Dermer}, {do Couto e Silva},
  {Drell}, {Dumora}, {Favuzzi}, {Fegan}, {Focke}, {Fortin}, {Fukazawa},
  {Fusco}, {Gargano}, {Germani}, {Giglietto}, {Giordano}, {Giroletti},
  {Glanzman}, {Godfrey}, {Grenier}, {Guillemot}, {Guiriec}, {Hadasch},
  {Hanabata}, {Harding}, {Hayashida}, {Hayashi}, {Hays}, {J{\'o}hannesson},
  {Johnson}, {Kamae}, {Katagiri}, {Kataoka}, {Kerr}, {Kn{\"o}dlseder}, {Kuss},
  {Lande}, {Latronico}, {Lee}, {Longo}, {Loparco}, {Lott}, {Lovellette},
  {Lubrano}, {Martin}, {Mazziotta}, {McEnery}, {Mehault}, {Michelson},
  {Mitthumsiri}, {Mizuno}, {Monte}, {Monzani}, {Morselli}, {Moskalenko},
  {Murgia}, {Naumann-Godo}, {Nolan}, {Norris}, {Nuss}, {Ohsugi}, {Okumura},
  {Orlando}, {Ormes}, {Ozaki}, {Paneque}, {Parent}, {Pesce-Rollins},
  {Pierbattista}, {Piron}, {Pohl}, {Prokhorov}, {Rain{\`o}}, {Rando},
  {Razzano}, {Reposeur}, {Ritz}, {Parkinson}, {Sgr{\`o}}, {Siskind}, {Smith},
  {Spinelli}, {Strong}, {Takahashi}, {Tanaka}, {Thayer}, {Thayer}, {Thompson},
  {Tibaldo}, {Torres}, {Tosti}, {Tramacere}, {Troja}, {Uchiyama},
  {Vandenbroucke}, {Vasileiou}, {Vianello}, {Vitale}, {Waite}, {Wang}, {Winer},
  {Wood}, {Yang}, {Zimmer}, \& {Bontemps}}]{2011Sci...334.1103A}
{Ackermann}, M., {Ajello}, M., {Allafort}, A., {et~al.} 2011, Science, 334,
  1103, \dodoi{10.1126/science.1210311}

\bibitem[{{Ackermann} {et~al.}(2012{\natexlab{a}}){Ackermann}, {Ajello},
  {Atwood}, {Baldini}, {Ballet}, {Barbiellini}, {Bastieri}, {Bechtol},
  {Bellazzini}, {Berenji}, {Blandford}, {Bloom}, {Bonamente}, {Borgland},
  {Brandt}, {Bregeon}, {Brigida}, {Bruel}, {Buehler}, {Buson}, {Caliandro},
  {Cameron}, {Caraveo}, {Cavazzuti}, {Cecchi}, {Charles}, {Chekhtman},
  {Chiang}, {Ciprini}, {Claus}, {Cohen-Tanugi}, {Conrad}, {Cutini}, {de
  Angelis}, {de Palma}, {Dermer}, {Digel}, {Silva}, {Drell}, {Drlica-Wagner},
  {Falletti}, {Favuzzi}, {Fegan}, {Ferrara}, {Focke}, {Fortin}, {Fukazawa},
  {Funk}, {Fusco}, {Gaggero}, {Gargano}, {Germani}, {Giglietto}, {Giordano},
  {Giroletti}, {Glanzman}, {Godfrey}, {Grove}, {Guiriec}, {Gustafsson},
  {Hadasch}, {Hanabata}, {Harding}, {Hayashida}, {Hays}, {Horan}, {Hou},
  {Hughes}, {J{\'o}hannesson}, {Johnson}, {Johnson}, {Kamae}, {Katagiri},
  {Kataoka}, {Kn{\"o}dlseder}, {Kuss}, {Lande}, {Latronico}, {Lee},
  {Lemoine-Goumard}, {Longo}, {Loparco}, {Lott}, {Lovellette}, {Lubrano},
  {Mazziotta}, {McEnery}, {Michelson}, {Mitthumsiri}, {Mizuno}, {Monte},
  {Monzani}, {Morselli}, {Moskalenko}, {Murgia}, {Naumann-Godo}, {Norris},
  {Nuss}, {Ohsugi}, {Okumura}, {Omodei}, {Orlando}, {Ormes}, {Paneque},
  {Panetta}, {Parent}, {Pesce-Rollins}, {Pierbattista}, {Piron}, {Pivato},
  {Porter}, {Rain{\`o}}, {Rando}, {Razzano}, {Razzaque}, {Reimer}, {Reimer},
  {Sadrozinski}, {Sgr{\`o}}, {Siskind}, {Spandre}, {Spinelli}, {Strong},
  {Suson}, {Takahashi}, {Tanaka}, {Thayer}, {Thayer}, {Thompson}, {Tibaldo},
  {Tinivella}, {Torres}, {Tosti}, {Troja}, {Usher}, {Vandenbroucke},
  {Vasileiou}, {Vianello}, {Vitale}, {Waite}, {Wang}, {Winer}, {Wood}, {Wood},
  {Yang}, {Ziegler}, \& {Zimmer}}]{2012ApJ...750....3A}
{Ackermann}, M., {Ajello}, M., {Atwood}, W.~B., {et~al.} 2012{\natexlab{a}},
  \apj, 750, 3, \dodoi{10.1088/0004-637X/750/1/3}

\bibitem[{{Ackermann} {et~al.}(2012{\natexlab{b}}){Ackermann}, {Ajello},
  {Allafort}, {Baldini}, {Ballet}, {Barbiellini}, {Bastieri}, {Bechtol},
  {Bellazzini}, {Berenji}, {Bland ford}, {Bloom}, {Bonamente}, {Borgland },
  {Bottacini}, {Brandt}, {Bregeon}, {Brigida}, {Bruel}, {Buehler}, {Busetto},
  {Buson}, {Caliandro}, {Cameron}, {Caraveo}, {Casandjian}, {Cecchi},
  {Charles}, {Chekhtman}, {Chiang}, {Ciprini}, {Claus}, {Cohen-Tanugi},
  {Conrad}, {D'Ammand o}, {de Angelis}, {de Palma}, {Dermer}, {Digel}, {Silva},
  {Drell}, {Drlica-Wagner}, {Falletti}, {Favuzzi}, {Fegan}, {Ferrara}, {Focke},
  {Fukazawa}, {Fukui}, {Funk}, {Fusco}, {Gargano}, {Gasparrini}, {Germani},
  {Giglietto}, {Giordano}, {Giroletti}, {Glanzman}, {Godfrey}, {Grenier},
  {Grondin}, {Grove}, {Guiriec}, {Hadasch}, {Hanabata}, {Harding}, {Hayashi},
  {Horan}, {Hou}, {Hughes}, {Itoh}, {Jackson}, {J{\'o}hannesson}, {Johnson},
  {Kamae}, {Katagiri}, {Kataoka}, {Kn{\"o}dlseder}, {Kuss}, {Lande}, {Larsson},
  {Lee}, {Lemoine-Goumard}, {Longo}, {Loparco}, {Lovellette}, {Lubrano},
  {Martin}, {Mazziotta}, {McEnery}, {Mehault}, {Michelson}, {Mitthumsiri},
  {Mizuno}, {Moiseev}, {Monte}, {Monzani}, {Morselli}, {Moskalenko}, {Murgia},
  {Naumann-Godo}, {Nemmen}, {Nishino}, {Norris}, {Nuss}, {Ohno}, {Ohsugi},
  {Okumura}, {Omodei}, {Orlando}, {Ormes}, {Ozaki}, {Paneque}, {Panetta},
  {Parent}, {Pesce-Rollins}, {Pierbattista}, {Piron}, {Pivato}, {Porter},
  {Rain{\`o}}, {Rando}, {Razzano}, {Reimer}, {Reimer}, {Romoli}, {Roth},
  {Sada}, {Sadrozinski}, {Sanchez}, {Sbarra}, {Sgr{\`o}}, {Siskind}, {Spandre},
  {Spinelli}, {Strong}, {Suson}, {Takahashi}, {Takahashi}, {Tanaka}, {Thayer},
  {Thayer}, {Thompson}, {Tibaldo}, {Tibolla}, {Tinivella}, {Torres}, {Tosti},
  {Tramacere}, {Troja}, {Uchiyama}, {Uehara}, {Usher}, {Vandenbroucke},
  {Vasileiou}, {Vianello}, {Vitale}, {Waite}, {Wang}, {Winer}, {Wood},
  {Yamamoto}, {Yang}, \& {Zimmer}}]{2012ApJ...755...22A}
{Ackermann}, M., {Ajello}, M., {Allafort}, A., {et~al.} 2012{\natexlab{b}}, The
  Astrophysical Journal, 755, 22, \dodoi{10.1088/0004-637X/755/1/22}

\bibitem[{{Ackermann} {et~al.}(2015){Ackermann}, {Ajello}, {Albert}, {Atwood},
  {Baldini}, {Ballet}, {Barbiellini}, {Bastieri}, {Bechtol}, {Bellazzini},
  {Bissaldi}, {Blandford}, {Bloom}, {Bottacini}, {Brandt}, {Bregeon}, {Bruel},
  {Buehler}, {Buson}, {Caliandro}, {Cameron}, {Caragiulo}, {Caraveo},
  {Cavazzuti}, {Cecchi}, {Charles}, {Chekhtman}, {Chiang}, {Chiaro}, {Ciprini},
  {Claus}, {Cohen-Tanugi}, {Conrad}, {Cuoco}, {Cutini}, {D'Ammando}, {de
  Angelis}, {de Palma}, {Dermer}, {Digel}, {Silva}, {Drell}, {Favuzzi},
  {Ferrara}, {Focke}, {Franckowiak}, {Fukazawa}, {Funk}, {Fusco}, {Gargano},
  {Gasparrini}, {Germani}, {Giglietto}, {Giommi}, {Giordano}, {Giroletti},
  {Godfrey}, {Gomez-Vargas}, {Grenier}, {Guiriec}, {Gustafsson}, {Hadasch},
  {Hayashi}, {Hays}, {Hewitt}, {Ippoliti}, {Jogler}, {J{\'o}hannesson},
  {Johnson}, {Johnson}, {Kamae}, {Kataoka}, {Kn{\"o}dlseder}, {Kuss},
  {Larsson}, {Latronico}, {Li}, {Li}, {Longo}, {Loparco}, {Lott}, {Lovellette},
  {Lubrano}, {Madejski}, {Manfreda}, {Massaro}, {Mayer}, {Mazziotta},
  {McEnery}, {Michelson}, {Mitthumsiri}, {Mizuno}, {Moiseev}, {Monzani},
  {Morselli}, {Moskalenko}, {Murgia}, {Nemmen}, {Nuss}, {Ohsugi}, {Omodei},
  {Orlando}, {Ormes}, {Paneque}, {Panetta}, {Perkins}, {Pesce-Rollins},
  {Piron}, {Pivato}, {Porter}, {Rain{\`o}}, {Rando}, {Razzano}, {Razzaque},
  {Reimer}, {Reimer}, {Reposeur}, {Ritz}, {Romani}, {S{\'a}nchez-Conde},
  {Schaal}, {Schulz}, {Sgr{\`o}}, {Siskind}, {Spandre}, {Spinelli}, {Strong},
  {Suson}, {Takahashi}, {Thayer}, {Thayer}, {Tibaldo}, {Tinivella}, {Torres},
  {Tosti}, {Troja}, {Uchiyama}, {Vianello}, {Werner}, {Winer}, {Wood}, {Wood},
  {Zaharijas}, \& {Zimmer}}]{2015ApJ...799...86A}
{Ackermann}, M., {Ajello}, M., {Albert}, A., {et~al.} 2015, \apj, 799, 86,
  \dodoi{10.1088/0004-637X/799/1/86}

\bibitem[{{Aharonian} \& {Atoyan}(2000)}]{2000A&A...362..937A}
{Aharonian}, F.~A., \& {Atoyan}, A.~M. 2000, \aap, 362, 937

\bibitem[{{Aharonian} {et~al.}(1997){Aharonian}, {Atoyan}, \&
  {Kifune}}]{1997MNRAS.291..162A}
{Aharonian}, F.~A., {Atoyan}, A.~M., \& {Kifune}, T. 1997, \mnras, 291, 162,
  \dodoi{10.1093/mnras/291.1.162}

\bibitem[{{Ajello} {et~al.}(2016){Ajello}, {Albert}, {Atwood}, {Barbiellini},
  {Bastieri}, {Bechtol}, {Bellazzini}, {Bissaldi}, {Blandford}, {Bloom},
  {Bonino}, {Bottacini}, {Brandt}, {Bregeon}, {Bruel}, {Buehler}, {Buson},
  {Caliandro}, {Cameron}, {Caputo}, {Caragiulo}, {Caraveo}, {Cecchi},
  {Chekhtman}, {Chiang}, {Chiaro}, {Ciprini}, {Cohen-Tanugi}, {Cominsky},
  {Conrad}, {Cutini}, {D'Ammando}, {de Angelis}, {de Palma}, {Desiante}, {Di
  Venere}, {Drell}, {Favuzzi}, {Ferrara}, {Fusco}, {Gargano}, {Gasparrini},
  {Giglietto}, {Giommi}, {Giordano}, {Giroletti}, {Glanzman}, {Godfrey},
  {Gomez-Vargas}, {Grenier}, {Guiriec}, {Gustafsson}, {Harding}, {Hewitt},
  {Hill}, {Horan}, {Jogler}, {J{\'o}hannesson}, {Johnson}, {Kamae}, {Karwin},
  {Kn{\"o}dlseder}, {Kuss}, {Larsson}, {Latronico}, {Li}, {Li}, {Longo},
  {Loparco}, {Lovellette}, {Lubrano}, {Magill}, {Maldera}, {Malyshev},
  {Manfreda}, {Mayer}, {Mazziotta}, {Michelson}, {Mitthumsiri}, {Mizuno},
  {Moiseev}, {Monzani}, {Morselli}, {Moskalenko}, {Murgia}, {Nuss}, {Ohno},
  {Ohsugi}, {Omodei}, {Orlando}, {Ormes}, {Paneque}, {Pesce-Rollins}, {Piron},
  {Pivato}, {Porter}, {Rain{\`o}}, {Rando}, {Razzano}, {Reimer}, {Reimer},
  {Ritz}, {S{\'a}nchez-Conde}, {Saz Parkinson}, {Sgr{\`o}}, {Siskind}, {Smith},
  {Spada}, {Spandre}, {Spinelli}, {Suson}, {Tajima}, {Takahashi}, {Thayer},
  {Torres}, {Tosti}, {Troja}, {Uchiyama}, {Vianello}, {Winer}, {Wood},
  {Zaharijas}, \& {Zimmer}}]{2016ApJ...819...44A}
{Ajello}, M., {Albert}, A., {Atwood}, W.~B., {et~al.} 2016, \apj, 819, 44,
  \dodoi{10.3847/0004-637X/819/1/44}

\bibitem[{{Atoyan} {et~al.}(1995){Atoyan}, {Aharonian}, \&
  {V{\"o}lk}}]{1995PhRvD..52.3265A}
{Atoyan}, A.~M., {Aharonian}, F.~A., \& {V{\"o}lk}, H.~J. 1995, \prd, 52, 3265,
  \dodoi{10.1103/PhysRevD.52.3265}

\bibitem[{{Atwood} {et~al.}(2009){Atwood}, {Abdo}, {Ackermann}, {Althouse},
  {Anderson}, {Axelsson}, {Baldini}, {Ballet}, {Band}, {Barbiellini}, \&
  et~al.}]{2009ApJ...697.1071A}
{Atwood}, W.~B., {Abdo}, A.~A., {Ackermann}, M., {et~al.} 2009, \apj, 697,
  1071, \dodoi{10.1088/0004-637X/697/2/1071}

\bibitem[{{Bennett} {et~al.}(2013){Bennett}, {Larson}, {Weiland}, {Jarosik},
  {Hinshaw}, {Odegard}, {Smith}, {Hill}, {Gold}, {Halpern}, {Komatsu}, {Nolta},
  {Page}, {Spergel}, {Wollack}, {Dunkley}, {Kogut}, {Limon}, {Meyer}, {Tucker},
  \& {Wright}}]{2013ApJS..208...20B}
{Bennett}, C.~L., {Larson}, D., {Weiland}, J.~L., {et~al.} 2013, \apjs, 208,
  20, \dodoi{10.1088/0067-0049/208/2/20}

\bibitem[{{Bernard} {et~al.}(2012){Bernard}, {Delahaye}, {Salati}, \&
  {Taillet}}]{2012A&A...544A..92B}
{Bernard}, G., {Delahaye}, T., {Salati}, P., \& {Taillet}, R. 2012, \aap, 544,
  A92, \dodoi{10.1051/0004-6361/201219502}

\bibitem[{{Blasi} \& {Amato}(2012)}]{2012JCAP...01..010B}
{Blasi}, P., \& {Amato}, E. 2012, \jcap, 1, 010,
  \dodoi{10.1088/1475-7516/2012/01/010}

\bibitem[{{Blasi} \& {Amato}(2019)}]{2019PhRvL.122e1101B}
---. 2019, \prl, 122, 051101, \dodoi{10.1103/PhysRevLett.122.051101}

\bibitem[{{Boschini} {et~al.}(2018{\natexlab{a}}){Boschini}, {Della Torre},
  {Gervasi}, {La Vacca}, \& {Rancoita}}]{2018AdSpR..62.2859B}
{Boschini}, M.~J., {Della Torre}, S., {Gervasi}, M., {La Vacca}, G., \&
  {Rancoita}, P.~G. 2018{\natexlab{a}}, Advances in Space Research, 62, 2859,
  \dodoi{10.1016/j.asr.2017.04.017}

\bibitem[{{Boschini} {et~al.}(2017){Boschini}, {Della Torre}, {Gervasi},
  {Grandi}, {J{\'o}hannesson}, {Kachelriess}, {La Vacca}, {Masi}, {Moskalenko},
  {Orlando}, {Ostapchenko}, {Pensotti}, {Porter}, {Quadrani}, {Rancoita},
  {Rozza}, \& {Tacconi}}]{2017ApJ...840..115B}
{Boschini}, M.~J., {Della Torre}, S., {Gervasi}, M., {et~al.} 2017, \apj, 840,
  115, \dodoi{10.3847/1538-4357/aa6e4f}

\bibitem[{{Boschini} {et~al.}(2018{\natexlab{b}}){Boschini}, {Della Torre},
  {Gervasi}, {Grandi}, {J{\'o}hannesson}, {La Vacca}, {Masi}, {Moskalenko},
  {Pensotti}, {Porter}, {Quadrani}, {Rancoita}, {Rozza}, \&
  {Tacconi}}]{2018ApJ...854...94B}
---. 2018{\natexlab{b}}, \apj, 854, 94, \dodoi{10.3847/1538-4357/aaa75e}

\bibitem[{{Boschini} {et~al.}(2018{\natexlab{c}}){Boschini}, {Della Torre},
  {Gervasi}, {Grandi}, {J{\'o}hannesson}, {La Vacca}, {Masi}, {Moskalenko},
  {Pensotti}, {Porter}, {Quadrani}, {Rancoita}, {Rozza}, \&
  {Tacconi}}]{2018ApJ...858...61B}
---. 2018{\natexlab{c}}, \apj, 858, 61, \dodoi{10.3847/1538-4357/aabc54}

\bibitem[{{Boschini} {et~al.}(2019){Boschini}, {Della Torre}, {Gervasi}, {Grand
  i}, {Johannesson}, {La Vacca}, {Masi}, {Moskalenko}, {Pensotti}, {Porter},
  {Quadrani}, {Rancoita}, {Rozza}, \& {Tacconi}}]{2019arXiv191103108B}
---. 2019, arXiv e-prints, arXiv:1911.03108.
\newblock \doarXiv{1911.03108}

\bibitem[{{Brandenburg} \& {Dobler}(2002)}]{BrandenburgDobler:2002}
{Brandenburg}, A., \& {Dobler}, W. 2002, Computer Physics Communications, 147,
  471, \dodoi{10.1016/S0010-4655(02)00334-X}

\bibitem[{{B{\"u}sching} {et~al.}(2005){B{\"u}sching}, {Kopp}, {Pohl},
  {Schlickeiser}, {Perrot}, \& {Grenier}}]{2005ApJ...619..314B}
{B{\"u}sching}, I., {Kopp}, A., {Pohl}, M., {et~al.} 2005, \apj, 619, 314,
  \dodoi{10.1086/426537}

\bibitem[{{Cataldo} {et~al.}(2019){Cataldo}, {Pagliaroli}, {Vecchiotti}, \&
  {Villante}}]{2019arXiv190403894C}
{Cataldo}, M., {Pagliaroli}, G., {Vecchiotti}, V., \& {Villante}, F.~L. 2019,
  arXiv e-prints, arXiv:1904.03894.
\newblock \doarXiv{1904.03894}

\bibitem[{{Clerc} {et~al.}(2018){Clerc}, {Ramos-Ceja}, {Ridl}, {Lamer},
  {Brunner}, {Hofmann}, {Comparat}, {Pacaud}, {K{\"a}fer}, {Reiprich},
  {Merloni}, {Schmid}, {Brand}, {Wilms}, {Friedrich}, {Finoguenov}, {Dauser},
  \& {Kreykenbohm}}]{2018A&A...617A..92C}
{Clerc}, N., {Ramos-Ceja}, M.~E., {Ridl}, J., {et~al.} 2018, \aap, 617, A92,
  \dodoi{10.1051/0004-6361/201732119}

\bibitem[{{Cordes} \& {Lazio}(2002)}]{2002astro.ph..7156C}
{Cordes}, J.~M., \& {Lazio}, T.~J.~W. 2002, ArXiv Astrophysics e-prints

\bibitem[{{Cowsik} \& {Lee}(1979)}]{1979ApJ...228..297C}
{Cowsik}, R., \& {Lee}, M.~A. 1979, \apj, 228, 297, \dodoi{10.1086/156846}

\bibitem[{{Diehl} {et~al.}(2006){Diehl}, {Halloin}, {Kretschmer}, {Lichti},
  {Sch{\"o}nfelder}, {Strong}, {von Kienlin}, {Wang}, {Jean}, {Kn{\"o}dlseder},
  {Roques}, {Weidenspointner}, {Schanne}, {Hartmann}, {Winkler}, \&
  {Wunderer}}]{2006Natur.439...45D}
{Diehl}, R., {Halloin}, H., {Kretschmer}, K., {et~al.} 2006, \nat, 439, 45,
  \dodoi{10.1038/nature04364}

\bibitem[{{Dobler} \& {Finkbeiner}(2008)}]{2008ApJ...680.1222D}
{Dobler}, G., \& {Finkbeiner}, D.~P. 2008, \apj, 680, 1222,
  \dodoi{10.1086/587862}

\bibitem[{{Evoli} {et~al.}(2018{\natexlab{a}}){Evoli}, {Blasi}, {Morlino}, \&
  {Aloisio}}]{2018PhRvL.121b1102E}
{Evoli}, C., {Blasi}, P., {Morlino}, G., \& {Aloisio}, R. 2018{\natexlab{a}},
  \prl, 121, 021102, \dodoi{10.1103/PhysRevLett.121.021102}

\bibitem[{{Evoli} {et~al.}(2018{\natexlab{b}}){Evoli}, {Linden}, \&
  {Morlino}}]{2018PhRvD..98f3017E}
{Evoli}, C., {Linden}, T., \& {Morlino}, G. 2018{\natexlab{b}}, \prd, 98,
  063017, \dodoi{10.1103/PhysRevD.98.063017}

\bibitem[{{Florinski} {et~al.}(2003){Florinski}, {Zank}, \&
  {Pogorelov}}]{2003JGRA..108.1228F}
{Florinski}, V., {Zank}, G.~P., \& {Pogorelov}, N.~V. 2003, Journal of
  Geophysical Research (Space Physics), 108, \dodoi{10.1029/2002JA009695}

\bibitem[{{Gaensler} {et~al.}(2008){Gaensler}, {Madsen}, {Chatterjee}, \&
  {Mao}}]{2008PASA...25..184G}
{Gaensler}, B.~M., {Madsen}, G.~J., {Chatterjee}, S., \& {Mao}, S.~A. 2008,
  \pasa, 25, 184, \dodoi{10.1071/AS08004}

\bibitem[{{Genolini} {et~al.}(2017){Genolini}, {Salati}, {Serpico}, \&
  {Taillet}}]{2017A&A...600A..68G}
{Genolini}, Y., {Salati}, P., {Serpico}, P.~D., \& {Taillet}, R. 2017, \aap,
  600, A68, \dodoi{10.1051/0004-6361/201629903}

\bibitem[{{Gleeson} \& {Axford}(1968)}]{1968ApJ...154.1011G}
{Gleeson}, L.~J., \& {Axford}, W.~I. 1968, \apj, 154, 1011,
  \dodoi{10.1086/149822}

\bibitem[{{G{\'o}rski} {et~al.}(2005){G{\'o}rski}, {Hivon}, {Banday},
  {Wandelt}, {Hansen}, {Reinecke}, \& {Bartelmann}}]{2005ApJ...622..759G}
{G{\'o}rski}, K.~M., {Hivon}, E., {Banday}, A.~J., {et~al.} 2005, \apj, 622,
  759, \dodoi{10.1086/427976}

\bibitem[{{Grenier} {et~al.}(2015){Grenier}, {Black}, \&
  {Strong}}]{2015ARA&A..53..199G}
{Grenier}, I.~A., {Black}, J.~H., \& {Strong}, A.~W. 2015, \araa, 53, 199,
  \dodoi{10.1146/annurev-astro-082214-122457}

\bibitem[{{Grenier} {et~al.}(2005){Grenier}, {Casandjian}, \&
  {Terrier}}]{2005Sci...307.1292G}
{Grenier}, I.~A., {Casandjian}, J.-M., \& {Terrier}, R. 2005, Science, 307,
  1292, \dodoi{10.1126/science.1106924}

\bibitem[{{H.E.S.S.~Collaboration} {et~al.}(2018){H.E.S.S.~Collaboration},
  {Abdalla}, {Abramowski}, {Aharonian}, {Ait Benkhali}, {Ang{\"u}ner},
  {Arakawa}, {Arrieta}, {Aubert}, {Backes}, \& et~al.}]{2018A&A...612A...1H}
{H.E.S.S.~Collaboration}, {Abdalla}, H., {Abramowski}, A., {et~al.} 2018, \aap,
  612, A1, \dodoi{10.1051/0004-6361/201732098}

\bibitem[{{Higdon} \& {Lingenfelter}(2003)}]{2003ApJ...582..330H}
{Higdon}, J.~C., \& {Lingenfelter}, R.~E. 2003, \apj, 582, 330,
  \dodoi{10.1086/344510}

\bibitem[{{Jardin-Blicq} {et~al.}(2019){Jardin-Blicq}, {Marandon}, \&
  {Brun}}]{2019arXiv190806658J}
{Jardin-Blicq}, A., {Marandon}, V., \& {Brun}, F. 2019, ArXiv e-prints.
\newblock \doarXiv{1908.06658}

\bibitem[{{J{\'o}hannesson} {et~al.}(2018){J{\'o}hannesson}, {Porter}, \&
  {Moskalenko}}]{2018ApJ...856...45J}
{J{\'o}hannesson}, G., {Porter}, T.~A., \& {Moskalenko}, I.~V. 2018, \apj, 856,
  45, \dodoi{10.3847/1538-4357/aab26e}

\bibitem[{{J{\'o}hannesson} {et~al.}(2019){J{\'o}hannesson}, {Porter}, \&
  {Moskalenko}}]{2019ApJ...879...91J}
---. 2019, \apj, 879, 91, \dodoi{10.3847/1538-4357/ab258e}

\bibitem[{{Karwin} {et~al.}(2019){Karwin}, {Murgia}, {Campbell}, \&
  {Moskalenko}}]{2019ApJ...880...95K}
{Karwin}, C.~M., {Murgia}, S., {Campbell}, S., \& {Moskalenko}, I.~V. 2019,
  \apj, 880, 95, \dodoi{10.3847/1538-4357/ab2880}

\bibitem[{{Kerr} \& {Lynden-Bell}(1986)}]{1986MNRAS.221.1023K}
{Kerr}, F.~J., \& {Lynden-Bell}, D. 1986, \mnras, 221, 1023,
  \dodoi{10.1093/mnras/221.4.1023}

\bibitem[{{Kobayashi} {et~al.}(2004){Kobayashi}, {Komori}, {Yoshida}, \&
  {Nishimura}}]{2004ApJ...601..340K}
{Kobayashi}, T., {Komori}, Y., {Yoshida}, K., \& {Nishimura}, J. 2004, \apj,
  601, 340, \dodoi{10.1086/380431}

\bibitem[{{Langner} {et~al.}(2006){Langner}, {Potgieter}, {Fichtner}, \&
  {Borrmann}}]{2006ApJ...640.1119L}
{Langner}, U.~W., {Potgieter}, M.~S., {Fichtner}, H., \& {Borrmann}, T. 2006,
  \apj, 640, 1119, \dodoi{10.1086/500162}

\bibitem[{{Lee}(1979)}]{1979ApJ...229..424L}
{Lee}, M.~A. 1979, \apj, 229, 424, \dodoi{10.1086/156970}

\bibitem[{{Lingenfelter}(1969)}]{1969Natur.224.1182L}
{Lingenfelter}, R.~E. 1969, \nat, 224, 1182, \dodoi{10.1038/2241182a0}

\bibitem[{{Lingenfelter} \& {Higdon}(1973)}]{1973ICRC....1..621L}
{Lingenfelter}, R.~E., \& {Higdon}, J.~C. 1973, International Cosmic Ray
  Conference, 1, 621

\bibitem[{{Lingenfelter} \& {Ramaty}(1971)}]{1971ICRC....1..377L}
{Lingenfelter}, R.~E., \& {Ramaty}, R. 1971, International Cosmic Ray
  Conference, 1, 377

\bibitem[{{Lipari} \& {Vernetto}(2018)}]{2018PhRvD..98d3003L}
{Lipari}, P., \& {Vernetto}, S. 2018, \prd, 98, 043003,
  \dodoi{10.1103/PhysRevD.98.043003}

\bibitem[{{Liu} {et~al.}(2015){Liu}, {Salati}, \& {Chen}}]{2015RAA....15...15L}
{Liu}, W., {Salati}, P., \& {Chen}, X. 2015, Research in Astronomy and
  Astrophysics, 15, 15, \dodoi{10.1088/1674-4527/15/1/002}

\bibitem[{{Lumb} {et~al.}(2002){Lumb}, {Warwick}, {Page}, \& {De
  Luca}}]{2002A&A...389...93L}
{Lumb}, D.~H., {Warwick}, R.~S., {Page}, M., \& {De Luca}, A. 2002, \aap, 389,
  93, \dodoi{10.1051/0004-6361:20020531}

\bibitem[{{Malkov} {et~al.}(2013){Malkov}, {Diamond}, {Sagdeev}, {Aharonian},
  \& {Moskalenko}}]{2013ApJ...768...73M}
{Malkov}, M.~A., {Diamond}, P.~H., {Sagdeev}, R.~Z., {Aharonian}, F.~A., \&
  {Moskalenko}, I.~V. 2013, \apj, 768, 73, \dodoi{10.1088/0004-637X/768/1/73}

\bibitem[{{Mertsch}(2011)}]{2011JCAP...02..031M}
{Mertsch}, P. 2011, \jcap, 2, 031, \dodoi{10.1088/1475-7516/2011/02/031}

\bibitem[{{Mertsch}(2018)}]{2018JCAP...11..045M}
---. 2018, \jcap, 11, 045, \dodoi{10.1088/1475-7516/2018/11/045}

\bibitem[{{Mertsch} \& {Sarkar}(2010)}]{2010JCAP...10..019M}
{Mertsch}, P., \& {Sarkar}, S. 2010, \jcap, 2010, 019,
  \dodoi{10.1088/1475-7516/2010/10/019}

\bibitem[{{Mertsch} \& {Sarkar}(2013)}]{2013JCAP...06..041M}
---. 2013, \jcap, 2013, 041, \dodoi{10.1088/1475-7516/2013/06/041}

\bibitem[{{Miyake} {et~al.}(2015){Miyake}, {Muraishi}, \&
  {Yanagita}}]{2015A&A...573A.134M}
{Miyake}, S., {Muraishi}, H., \& {Yanagita}, S. 2015, \aap, 573, A134,
  \dodoi{10.1051/0004-6361/201424442}

\bibitem[{{Mizuno} {et~al.}(2016){Mizuno}, {Abdollahi}, {Fukui}, {Hayashi},
  {Okumura}, {Tajima}, \& {Yamamoto}}]{2016ApJ...833..278M}
{Mizuno}, T., {Abdollahi}, S., {Fukui}, Y., {et~al.} 2016, The Astrophysical
  Journal, 833, 278, \dodoi{10.3847/1538-4357/833/2/278}

\bibitem[{{Moskalenko} {et~al.}(2001){Moskalenko}, {Mashnik}, \&
  {Strong}}]{2001ICRC....5.1836M}
{Moskalenko}, I.~V., {Mashnik}, S.~G., \& {Strong}, A.~W. 2001, International
  Cosmic Ray Conference, 5, 1836

\bibitem[{{Moskalenko} {et~al.}(2006){Moskalenko}, {Porter}, \&
  {Strong}}]{2006ApJ...640L.155M}
{Moskalenko}, I.~V., {Porter}, T.~A., \& {Strong}, A.~W. 2006, \apjl, 640,
  L155, \dodoi{10.1086/503524}

\bibitem[{{Moskalenko} \& {Strong}(2000)}]{2000ApJ...528..357M}
{Moskalenko}, I.~V., \& {Strong}, A.~W. 2000, \apj, 528, 357,
  \dodoi{10.1086/308138}

\bibitem[{{Nava} {et~al.}(2016){Nava}, {Gabici}, {Marcowith}, {Morlino}, \&
  {Ptuskin}}]{2016MNRAS.461.3552N}
{Nava}, L., {Gabici}, S., {Marcowith}, A., {Morlino}, G., \& {Ptuskin}, V.~S.
  2016, \mnras, 461, 3552, \dodoi{10.1093/mnras/stw1592}

\bibitem[{{Neronov} {et~al.}(2018){Neronov}, {Kachelrie{\ss}}, \&
  {Semikoz}}]{2018PhRvD..98b3004N}
{Neronov}, A., {Kachelrie{\ss}}, M., \& {Semikoz}, D.~V. 2018, \prd, 98,
  023004, \dodoi{10.1103/PhysRevD.98.023004}

\bibitem[{{Neronov} \& {Semikoz}(2019)}]{2019arXiv190706061N}
{Neronov}, A., \& {Semikoz}, D.~V. 2019, arXiv e-prints, arXiv:1907.06061.
\newblock \doarXiv{1907.06061}

\bibitem[{{Nishimura} {et~al.}(1979){Nishimura}, {Fujii}, \&
  {Taira}}]{1979ICRC....1..488N}
{Nishimura}, J., {Fujii}, M., \& {Taira}, T. 1979, International Cosmic Ray
  Conference, 1, 488

\bibitem[{{Nolan} {et~al.}(2012){Nolan}, {Abdo}, {Ackermann}, {Ajello},
  {Allafort}, {Antolini}, {Atwood}, {Axelsson}, {Baldini}, {Ballet}, \&
  et~al.}]{2012ApJS..199...31N}
{Nolan}, P.~L., {Abdo}, A.~A., {Ackermann}, M., {et~al.} 2012, \apjs, 199, 31,
  \dodoi{10.1088/0067-0049/199/2/31}

\bibitem[{{Orlando} \& {Strong}(2013)}]{2013MNRAS.436.2127O}
{Orlando}, E., \& {Strong}, A. 2013, \mnras, 436, 2127,
  \dodoi{10.1093/mnras/stt1718}

\bibitem[{{Parker}(1965)}]{1965P&SS...13....9P}
{Parker}, E.~N. 1965, \planss, 13, 9, \dodoi{10.1016/0032-0633(65)90131-5}

\bibitem[{{Planck Collaboration} {et~al.}(2011){Planck Collaboration}, {Ade},
  {Aghanim}, {Arnaud}, {Ashdown}, {Aumont}, {Baccigalupi}, {Balbi}, {Banday},
  {Barreiro}, \& et~al.}]{2011A&A...536A..19P}
{Planck Collaboration}, {Ade}, P.~A.~R., {Aghanim}, N., {et~al.} 2011, \aap,
  536, A19, \dodoi{10.1051/0004-6361/201116479}

\bibitem[{{Planck Collaboration} {et~al.}(2013){Planck Collaboration}, {Ade},
  {Aghanim}, {Arnaud}, {Ashdown}, {Atrio-Barandela}, {Aumont}, {Baccigalupi},
  {Balbi}, \& {Banday}}]{2013A&A...554A.139P}
---. 2013, \aap, 554, A139, \dodoi{10.1051/0004-6361/201220271}

\bibitem[{{Planck Collaboration} {et~al.}(2015){Planck Collaboration}, {Fermi
  Collaboration}, {Ade}, {Aghanim}, {Aniano}, {Arnaud}, {Ashdown}, {Aumont},
  {Baccigalupi}, {Banday}, {Barreiro}, {Bartolo}, {Battaner}, {Benabed},
  {Benoit-L{\'e}vy}, {Bernard}, {Bersanelli}, {Bielewicz}, {Bonaldi},
  {Bonavera}, {Bond}, {Borrill}, {Bouchet}, {Boulanger}, {Burigana}, {Butler},
  {Calabrese}, {Cardoso}, {Casand jian}, {Catalano}, {Chamballu}, {Chiang},
  {Christensen}, {Colombo}, {Combet}, {Couchot}, {Crill}, {Curto}, {Cuttaia},
  {Danese}, {Davies}, {Davis}, {de Bernardis}, {de Rosa}, {de Zotti},
  {Delabrouille}, {D{\'e}sert}, {Dickinson}, {Diego}, {Digel}, {Dole},
  {Donzelli}, {Dor{\'e}}, {Douspis}, {Ducout}, {Dupac}, {Efstathiou}, {Elsner},
  {En{\ss}lin}, {Eriksen}, {Falgarone}, {Finelli}, {Forni}, {Frailis},
  {Fraisse}, {Franceschi}, {Frejsel}, {Fukui}, {Galeotta}, {Galli}, {Ganga},
  {Ghosh}, {Giard}, {Gjerl{\o}w}, {Gonz{\'a}lez-Nuevo}, {G{\'o}rski},
  {Gregorio}, {Grenier}, {Gruppuso}, {Hansen}, {Hanson}, {Harrison},
  {Henrot-Versill{\'e}}, {Hern{\'a}ndez-Monteagudo}, {Herranz}, {Hildebrand t},
  {Hivon}, {Hobson}, {Holmes}, {Hovest}, {Huffenberger}, {Hurier}, {Jaffe},
  {Jaffe}, {Jones}, {Juvela}, {Keih{\"a}nen}, {Keskitalo}, {Kisner}, {Kneissl},
  {Knoche}, {Kunz}, {Kurki-Suonio}, {Lagache}, {Lamarre}, {Lasenby},
  {Lattanzi}, {Lawrence}, {Leonardi}, {Levrier}, {Liguori}, {Lilje},
  {Linden-V{\o}rnle}, {L{\'o}pez-Caniego}, {Lubin}, {Mac{\'\i}as-P{\'e}rez},
  {Maffei}, {Maino}, {Mand olesi}, {Maris}, {Marshall}, {Martin},
  {Mart{\'\i}nez-Gonz{\'a}lez}, {Masi}, {Matarrese}, {Mazzotta}, {Melchiorri},
  {Mendes}, {Mennella}, {Migliaccio}, {Miville-Desch{\^e}nes}, {Moneti},
  {Montier}, {Morgante}, {Mortlock}, {Munshi}, {Murphy}, {Naselsky}, {Natoli},
  {N{\o}rgaard-Nielsen}, {Novikov}, {Novikov}, {Oxborrow}, {Pagano}, {Pajot},
  {Paladini}, {Paoletti}, {Pasian}, {Perdereau}, {Perotto}, {Perrotta},
  {Pettorino}, {Piacentini}, {Piat}, {Plaszczynski}, {Pointecouteau},
  {Polenta}, {Popa}, {Pratt}, {Prunet}, {Puget}, {Rachen}, {Reach}, {Rebolo},
  {Reinecke}, {Remazeilles}, {Renault}, {Ristorcelli}, {Rocha}, {Roudier},
  {Rusholme}, {Sandri}, {Santos}, {Scott}, {Spencer}, {Stolyarov}, {Strong},
  {Sudiwala}, {Sunyaev}, {Sutton}, {Suur-Uski}, {Sygnet}, {Tauber}, {Terenzi},
  {Tibaldo}, {Toffolatti}, {Tomasi}, {Tristram}, {Tucci}, {Umana},
  {Valenziano}, {Valiviita}, {Van Tent}, {Vielva}, {Villa}, {Wade}, {Wandelt},
  {Wehus}, {Yvon}, {Zacchei}, \& {Zonca}}]{2015A&A...582A..31P}
{Planck Collaboration}, {Fermi Collaboration}, {Ade}, P.~A.~R., {et~al.} 2015,
  Astronomy and Astrophysics, 582, A31, \dodoi{10.1051/0004-6361/201424955}

\bibitem[{{Planck Collaboration} {et~al.}(2018{\natexlab{a}}){Planck
  Collaboration}, {Akrami}, {Arg{\"u}eso}, {Ashdown}, {Aumont}, {Baccigalupi},
  {Ballardini}, {Banday}, {Barreiro}, {Bartolo}, {Basak}, {Benabed}, {Bernard},
  {Bersanelli}, {Bielewicz}, {Bonavera}, {Bond}, {Borrill}, {Bouchet},
  {Boulanger}, {Bucher}, {Burigana}, {Butler}, {Calabrese}, {Cardoso},
  {Colombo}, {Crill}, {Cuttaia}, {de Bernardis}, {de Rosa}, {de Zotti},
  {Delabrouille}, {Di Valentino}, {Dickinson}, {Diego}, {Donzelli}, {Ducout},
  {Dupac}, {Efstathiou}, {Elsner}, {En{\ss}lin}, {Eriksen}, {Fantaye},
  {Finelli}, {Frailis}, {Franceschi}, {Frolov}, {Galeotta}, {Galli}, {Ganga},
  {G{\'e}nova-Santos}, {Gerbino}, {Ghosh}, {Gonz{\'a}lez-Nuevo}, {G{\'o}rski},
  {Gratton}, {Gruppuso}, {Gudmundsson}, {Hand ley}, {Hansen}, {Herranz},
  {Hivon}, {Huang}, {Jaffe}, {Jones}, {Karakci}, {Keih{\"a}nen}, {Keskitalo},
  {Kiiveri}, {Kim}, {Kisner}, {Krachmalnicoff}, {Kunz}, {Kurki-Suonio},
  {Lamarre}, {Lasenby}, {Lattanzi}, {Lawrence}, {Leahy}, {Levrier}, {Liguori},
  {Lilje}, {Lindholm}, {L{\'o}pez-Caniego}, {Ma}, {Mac{\'\i}as-P{\'e}rez},
  {Maggio}, {Maino}, {Mand olesi}, {Mangilli}, {Maris}, {Martin},
  {Mart{\'\i}nez-Gonz{\'a}lez}, {Matarrese}, {Mauri}, {McEwen}, {Meinhold},
  {Melchiorri}, {Mennella}, {Migliaccio}, {Molinari}, {Montier}, {Morgante},
  {Moss}, {Natoli}, {Pagano}, {Paoletti}, {Partridge}, {Patanchon}, {Patrizii},
  {Peel}, {Perrotta}, {Pettorino}, {Piacentini}, {Polenta}, {Puget}, {Rachen},
  {Racine}, {Reinecke}, {Remazeilles}, {Renzi}, {Rocha}, {Roudier},
  {Rubi{\~n}o-Mart{\'\i}n}, {Salvati}, {Sandri}, {Savelainen}, {Scott},
  {Seljebotn}, {Sirignano}, {Sirri}, {Spencer}, {Suur-Uski}, {Tauber},
  {Tavagnacco}, {Tenti}, {Terenzi}, {Toffolatti}, {Tomasi}, {Trombetti},
  {Valiviita}, {Vansyngel}, {Van Tent}, {Vielva}, {Villa}, {Vittorio},
  {Wandelt}, {Watson}, {Wehus}, {Zacchei}, \& {Zonca}}]{2018arXiv180706206P}
{Planck Collaboration}, {Akrami}, Y., {Arg{\"u}eso}, F., {et~al.}
  2018{\natexlab{a}}, arXiv e-prints, arXiv:1807.06206.
\newblock \doarXiv{1807.06206}

\bibitem[{{Planck Collaboration} {et~al.}(2018{\natexlab{b}}){Planck
  Collaboration}, {Aghanim}, {Akrami}, {Ashdown}, {Aumont}, {Baccigalupi},
  {Ballardini}, {Banday}, {Barreiro}, {Bartolo}, {Basak}, {Benabed}, {Bernard},
  {Bersanelli}, {Bielewicz}, {Bond}, {Borrill}, {Bouchet}, {Boulanger},
  {Bucher}, {Burigana}, {Calabrese}, {Cardoso}, {Carron}, {Challinor},
  {Chiang}, {Colombo}, {Combet}, {Couchot}, {Crill}, {Cuttaia}, {de Bernardis},
  {de Rosa}, {de Zotti}, {Delabrouille}, {Delouis}, {Di Valentino}, {Diego},
  {Dor{\'e}}, {Douspis}, {Ducout}, {Dupac}, {Efstathiou}, {Elsner},
  {En{\ss}lin}, {Eriksen}, {Falgarone}, {Fantaye}, {Finelli}, {Frailis},
  {Fraisse}, {Franceschi}, {Frolov}, {Galeotta}, {Galli}, {Ganga},
  {G{\'e}nova-Santos}, {Gerbino}, {Ghosh}, {Gonz{\'a}lez-Nuevo}, {G{\'o}rski},
  {Gratton}, {Gruppuso}, {Gudmundsson}, {Hand ley}, {Hansen},
  {Henrot-Versill{\'e}}, {Herranz}, {Hivon}, {Huang}, {Jaffe}, {Jones},
  {Karakci}, {Keih{\"a}nen}, {Keskitalo}, {Kiiveri}, {Kim}, {Kisner},
  {Krachmalnicoff}, {Kunz}, {Kurki-Suonio}, {Lagache}, {Lamarre}, {Lasenby},
  {Lattanzi}, {Lawrence}, {Levrier}, {Liguori}, {Lilje}, {Lindholm},
  {L{\'o}pez-Caniego}, {Ma}, {Mac{\'\i}as-P{\'e}rez}, {Maggio}, {Maino}, {Mand
  olesi}, {Mangilli}, {Martin}, {Mart{\'\i}nez-Gonz{\'a}lez}, {Matarrese},
  {Mauri}, {McEwen}, {Melchiorri}, {Mennella}, {Migliaccio},
  {Miville-Desch{\^e}nes}, {Molinari}, {Moneti}, {Montier}, {Morgante}, {Moss},
  {Mottet}, {Natoli}, {Pagano}, {Paoletti}, {Partridge}, {Patanchon},
  {Patrizii}, {Perdereau}, {Perrotta}, {Pettorino}, {Piacentini}, {Puget},
  {Rachen}, {Reinecke}, {Remazeilles}, {Renzi}, {Rocha}, {Roudier}, {Salvati},
  {Sand ri}, {Savelainen}, {Scott}, {Sirignano}, {Sirri}, {Spencer}, {Sunyaev},
  {Suur-Uski}, {Tauber}, {Tavagnacco}, {Tenti}, {Toffolatti}, {Tomasi},
  {Tristram}, {Trombetti}, {Valiviita}, {Vansyngel}, {Van Tent}, {Vibert},
  {Vielva}, {Villa}, {Vittorio}, {Wandelt}, {Wehus}, \&
  {Zonca}}]{2018arXiv180706207P}
{Planck Collaboration}, {Aghanim}, N., {Akrami}, Y., {et~al.}
  2018{\natexlab{b}}, arXiv e-prints, arXiv:1807.06207.
\newblock \doarXiv{1807.06207}

\bibitem[{{Porter} {et~al.}(2017){Porter}, {J{\'o}hannesson}, \&
  {Moskalenko}}]{2017ApJ...846...67P}
{Porter}, T.~A., {J{\'o}hannesson}, G., \& {Moskalenko}, I.~V. 2017, \apj, 846,
  67, \dodoi{10.3847/1538-4357/aa844d}

\bibitem[{{Porter} \& {Protheroe}(1997)}]{1997JPhG...23.1765P}
{Porter}, T.~A., \& {Protheroe}, R.~J. 1997, Journal of Physics G Nuclear
  Physics, 23, 1765, \dodoi{10.1088/0954-3899/23/11/022}

\bibitem[{{Porter} {et~al.}(2018){Porter}, {Rowell}, {J{\'o}hannesson}, \&
  {Moskalenko}}]{2018PhRvD..98d1302P}
{Porter}, T.~A., {Rowell}, G.~P., {J{\'o}hannesson}, G., \& {Moskalenko}, I.~V.
  2018, \prd, 98, 041302, \dodoi{10.1103/PhysRevD.98.041302}

\bibitem[{{Potgieter} \& {Langner}(2004)}]{2004AnGeo..22.3729P}
{Potgieter}, M., \& {Langner}, U. 2004, Annales Geophysicae, 22, 3729,
  \dodoi{10.5194/angeo-22-3729-2004}

\bibitem[{{Predehl} {et~al.}(2016){Predehl}, {Andritschke}, {Babyshkin},
  {Becker}, {Bornemann}, {Br{\"a}uninger}, {Brunner}, {Boller}, {Burwitz},
  {Burkert}, {Clerc}, {Churazov}, {Coutinho}, {Dennerl}, {Dwelly}, {Eder},
  {Emberger}, {Freyberg}, {Friedrich}, {F{\"u}rmetz}, {Georgakakis},
  {Gilfanov}, {Grossberger}, {Haberl}, {H{\"a}lker}, {Hartner}, {Kienlin},
  {Kink}, {Kreykenbohm}, {Lamer}, {Lomakin}, {Lapshov}, {Meidinger}, {Merloni},
  {Mican}, {M{\"u}ller}, {Nandra}, {Pavlinsky}, {Pfeffermann}, {Pietschner},
  {Robrade}, {Salvato}, {Santangelo}, {Sasaki}, {Scheuerle}, {Schmitt},
  {Schwope}, {Sunyaev}, {Tenzer}, {Yaroshenko}, \&
  {Wilms}}]{2016SPIE.9905E..1KP}
{Predehl}, P., {Andritschke}, R., {Babyshkin}, V., {et~al.} 2016, in \procspie,
  Vol. 9905, Space Telescopes and Instrumentation 2016: Ultraviolet to Gamma
  Ray, 99051K

\bibitem[{{Protheroe} \& {Wolfendale}(1980)}]{1980A&A....92..175P}
{Protheroe}, R.~J., \& {Wolfendale}, A.~W. 1980, \aap, 92, 175

\bibitem[{{Pshirkov} {et~al.}(2011){Pshirkov}, {Tinyakov}, {Kronberg}, \&
  {Newton-McGee}}]{2011ApJ...738..192P}
{Pshirkov}, M.~S., {Tinyakov}, P.~G., {Kronberg}, P.~P., \& {Newton-McGee},
  K.~J. 2011, \apj, 738, 192, \dodoi{10.1088/0004-637X/738/2/192}

\bibitem[{{Ptuskin} {et~al.}(2006){Ptuskin}, {Jones}, {Seo}, \&
  {Sina}}]{2006AdSpR..37.1909P}
{Ptuskin}, V.~S., {Jones}, F.~C., {Seo}, E.~S., \& {Sina}, R. 2006, Advances in
  Space Research, 37, 1909, \dodoi{10.1016/j.asr.2005.08.036}

\bibitem[{{Ptuskin} \& {Soutoul}(1998)}]{1998A&A...337..859P}
{Ptuskin}, V.~S., \& {Soutoul}, A. 1998, \aap, 337, 859

\bibitem[{{Ptuskin} {et~al.}(2008){Ptuskin}, {Zirakashvili}, \&
  {Plesser}}]{2008AdSpR..42..486P}
{Ptuskin}, V.~S., {Zirakashvili}, V.~N., \& {Plesser}, A.~A. 2008, Advances in
  Space Research, 42, 486, \dodoi{10.1016/j.asr.2007.12.007}

\bibitem[{{Ramaty} {et~al.}(1970){Ramaty}, {Reames}, \&
  {Lingenfelter}}]{1970PhRvL..24..913R}
{Ramaty}, R., {Reames}, D.~V., \& {Lingenfelter}, R.~E. 1970, Physical Review
  Letters, 24, 913, \dodoi{10.1103/PhysRevLett.24.913}

\bibitem[{{Remy} {et~al.}(2017){Remy}, {Grenier}, {Marshall}, \& {Casand
  jian}}]{2017A&A...601A..78R}
{Remy}, Q., {Grenier}, I.~A., {Marshall}, D.~J., \& {Casand jian}, J.~M. 2017,
  Astronomy and Astrophysics, 601, A78, \dodoi{10.1051/0004-6361/201629632}

\bibitem[{{Robitaille} {et~al.}(2012){Robitaille}, {Churchwell}, {Benjamin},
  {Whitney}, {Wood}, {Babler}, \& {Meade}}]{2012A&A...545A..39R}
{Robitaille}, T.~P., {Churchwell}, E., {Benjamin}, R.~A., {et~al.} 2012, \aap,
  545, A39, \dodoi{10.1051/0004-6361/201219073}

\bibitem[{{Shen}(1970)}]{1970ApJ...162L.181S}
{Shen}, C.~S. 1970, \apjl, 162, L181, \dodoi{10.1086/180650}

\bibitem[{{Shen} \& {Mao}(1971)}]{1971ApL.....9..169S}
{Shen}, C.~S., \& {Mao}, C.~Y. 1971, \aplett, 9, 169

\bibitem[{{Strong} \& {Moskalenko}(1998)}]{1998ApJ...509..212S}
{Strong}, A.~W., \& {Moskalenko}, I.~V. 1998, \apj, 509, 212,
  \dodoi{10.1086/306470}

\bibitem[{{Strong} \& {Moskalenko}(2001{\natexlab{a}})}]{2001AIPC..587..533S}
{Strong}, A.~W., \& {Moskalenko}, I.~V. 2001{\natexlab{a}}, in American
  Institute of Physics Conference Series, Vol. 587, Gamma 2001: Gamma-Ray
  Astrophysics, ed. S.~{Ritz}, N.~{Gehrels}, \& C.~R. {Shrader}, 533--537

\bibitem[{{Strong} \& {Moskalenko}(2001{\natexlab{b}})}]{2001ICRC....5.1964S}
---. 2001{\natexlab{b}}, International Cosmic Ray Conference, 5, 1964

\bibitem[{{Strong} {et~al.}(2007){Strong}, {Moskalenko}, \&
  {Ptuskin}}]{2007ARNPS..57..285S}
{Strong}, A.~W., {Moskalenko}, I.~V., \& {Ptuskin}, V.~S. 2007, Annual Review
  of Nuclear and Particle Science, 57, 285,
  \dodoi{10.1146/annurev.nucl.57.090506.123011}

\bibitem[{{Su} {et~al.}(2010){Su}, {Slatyer}, \&
  {Finkbeiner}}]{2010ApJ...724.1044S}
{Su}, M., {Slatyer}, T.~R., \& {Finkbeiner}, D.~P. 2010, \apj, 724, 1044,
  \dodoi{10.1088/0004-637X/724/2/1044}

\bibitem[{{Sudoh} {et~al.}(2019){Sudoh}, {Linden}, \&
  {Beacom}}]{2019PhRvD.100d3016S}
{Sudoh}, T., {Linden}, T., \& {Beacom}, J.~F. 2019, \prd, 100, 043016,
  \dodoi{10.1103/PhysRevD.100.043016}

\bibitem[{{Swordy}(2003)}]{2003ICRC....4.1989S}
{Swordy}, S.~P. 2003, International Cosmic Ray Conference, 4, 1989

\bibitem[{{Taillet} {et~al.}(2004){Taillet}, {Salati}, {Maurin},
  {Vangioni-Flam}, \& {Cass{\'e}}}]{2004ApJ...609..173T}
{Taillet}, R., {Salati}, P., {Maurin}, D., {Vangioni-Flam}, E., \& {Cass{\'e}},
  M. 2004, \apj, 609, 173, \dodoi{10.1086/421059}

\bibitem[{{The Fermi-LAT collaboration}(2019)}]{2019arXiv190210045T}
{The Fermi-LAT collaboration}. 2019, arXiv e-prints.
\newblock \doarXiv{1902.10045}

\bibitem[{{Tomassetti}(2015)}]{2015PhRvD..92h1301T}
{Tomassetti}, N. 2015, \prd, 92, 081301, \dodoi{10.1103/PhysRevD.92.081301}

\bibitem[{{Vidal} {et~al.}(2015){Vidal}, {Dickinson}, {Davies}, \&
  {Leahy}}]{2015MNRAS.452..656V}
{Vidal}, M., {Dickinson}, C., {Davies}, R.~D., \& {Leahy}, J.~P. 2015, \mnras,
  452, 656, \dodoi{10.1093/mnras/stv1328}

\bibitem[{{Webber} \& {Soutoul}(1998)}]{1998ApJ...506..335W}
{Webber}, W.~R., \& {Soutoul}, A. 1998, \apj, 506, 335, \dodoi{10.1086/306224}

\bibitem[{{Yusifov} \& {K{\"u}{\c c}{\"u}k}(2004)}]{2004A&A...422..545Y}
{Yusifov}, I., \& {K{\"u}{\c c}{\"u}k}, I. 2004, \aap, 422, 545,
  \dodoi{10.1051/0004-6361:20040152}

\end{thebibliography}
  
\end{document}